\documentclass[a4paper,iop,twocolumn,twocolappendix,numberedappendix]{emulateapj}
\usepackage[breaklinks,colorlinks=true,linkcolor=blue,anchorcolor=blue,citecolor=blue,urlcolor=blue,draft=false]{hyperref}
\usepackage{color}
\usepackage{graphicx}
\usepackage{amsmath} 
\usepackage{amssymb}
\usepackage{mathrsfs}
\usepackage{placeins}
\usepackage{times}
\usepackage{xcolor}
\usepackage{xspace}

\hypersetup{pdfauthor={Keiichi Umetsu},pdftitle={HSCxXXL-WL}}
\usepackage{natbib}

\newcommand{\simgt}{\lower.5ex\hbox{$\; \buildrel > \over \sim \;$}}
\newcommand{\simlt}{\lower.5ex\hbox{$\; \buildrel < \over \sim \;$}}
\newcommand{\llangle}{\langle\!\langle}
\newcommand{\rrangle}{\rangle\!\rangle}
\newcommand{\Om}{\Omega_\mathrm{m}}
\newcommand{\OL}{\Omega_\Lambda}
\newcommand{\kpch}{h^{-1}\mathrm{kpc}}

\newcommand{\Mpch}{h^{-1}\mathrm{Mpc}}

\newcommand{\Msunh}{h^{-1}M_\odot}
\newcommand{\Msun}{M_\odot}
\newcommand{\zref}{z_\mathrm{ref}}
\newcommand{\zmed}{z}
\newcommand{\ngal}{n_\mathrm{gal}}
\newcommand{\Tx}{T_\mathrm{300\,kpc}}
\newcommand{\Mwl}{M_\mathrm{200,WL}}
\newcommand{\Mmt}{M_{\Delta,\mathrm{MT}}}
\newcommand{\Mmtfv}{M_\mathrm{500,MT}}
\newcommand{\Mmttw}{M_\mathrm{200,MT}}
\newcommand{\Mtrue}{M_\mathrm{200,true}}
\newcommand{\cwl}{c_\mathrm{200,WL}}
\newcommand{\ctrue}{c_\mathrm{200,true}}
\newcommand{\percent}{\ensuremath{\%}}
\newcommand{\CBI}{C_\mathrm{BI}}
\newcommand{\SBI}{S_\mathrm{BI}}
\newcommand{\Ncl}{N_\mathrm{cl}}
\newcommand{\XMM}{{\em XMM}}
\newcommand{\XMMNewton}{{\em XMM-Newton}}
\newcommand{\Chandra}{{\em Chandra}}
\newcommand{\WMAP}{{\em WMAP}}
\newcommand{\dfmcal}{5\percent}
\newcommand{\Cpivot}{4.8\pm 1.0 \, (\mathrm{stat})\pm 0.8 \, (\mathrm{syst})}

\newcommand{\betaCM}{-0.07\pm 0.28}
\newcommand{\gammaCM}{-0.03\pm 0.47}

\newcommand{\sigmalnC}{(5.3\pm 3.4)\percent}
\newcommand{\sigmalnCup}{24\percent}
\newcommand{\Tpivot}{2.78\pm 0.54}
\newcommand{\TpivotFS}{2.58\pm 0.27}
\newcommand{\Tbest}{2.29\pm 0.36}
\newcommand{\betaTM}{0.85\pm 0.31}
\newcommand{\betaTMalt}{0.75\pm 0.27} 
\newcommand{\gammaTM}{0.18\pm 0.66}
\newcommand{\muZTM}{-0.17\pm 0.07}
\newcommand{\sigmalnT}{(14\pm 11)\percent}
\newcommand{\sigmalnTFS}{(16\pm 10)\percent} 
\newcommand{\bXXL}{(34\pm 20)\percent}       
\newcommand{\bXXLtwo}{(41\pm 20)\percent}    
\newcommand{\bXXLexc}{(28\pm 18)\percent}    
\newcommand{\bXXLexctwo}{(35\pm 18)\percent} 

\def\bd{\mbox{\boldmath $d$}}

\def\bp{\mbox{\boldmath $p$}}

\def\singlebond{\@makechembond\@ne}
\def\doublebond{\@makechembond\tw@}
\def\triplebond{\@makechembond\thr@@}    

\def\le{\leqslant}

\defcitealias{2016AA...592A...1P}{XXL~Paper~I}      
\defcitealias{2016AA...592A...2P}{XXL~Paper~II}     
\defcitealias{2016AA...592A...3G}{XXL~Paper~III}
\defcitealias{2016AA...592A...4L}{XXL~Paper~IV}     
\defcitealias{2014ApJ...794..157M}{XXL~Paper~V}
\defcitealias{2016AA...592A...5F}{XXL~Paper~VI}
\defcitealias{2016AA...592A...6P}{XXL~Paper~VII}
\defcitealias{2016AA...592A...7A}{XXL~Paper~VIII}
\defcitealias{2016AA...592A...8B}{XXL~Paper~IX}
\defcitealias{2016AA...592A...9Z}{XXL~Paper~X}
\defcitealias{2016AA...592A..10S}{XXL~Paper~XI}
\defcitealias{2016AA...592A..11K}{XXL~Paper~XII}
\defcitealias{2016AA...592A..12E}{XXL~Paper~XIII}   
\defcitealias{2016PASA...33....1L}{XXL~Paper~XIV}
\defcitealias{2016MNRAS.462.4141L}{XXL~Paper~XV}    
\defcitealias{2018AA...620A...1M}{XXL~Paper~XVI}    
\defcitealias{2018AA...620A...2M}{XXL~Paper~XVII}
\defcitealias{2018AA...620A...3B}{XXL~Paper~XVIII}
\defcitealias{2018AA...620A...4K}{XXL~Paper~XIX}
\defcitealias{2018AA...620A...5A}{XXL~Paper~XX}     
\defcitealias{2018AA...620A...6M}{XXL~Paper~XXI}
\defcitealias{2018AA...620A...7G}{XXL~Paper~XXII}
\defcitealias{2018AA...620A...8F}{XXL~Paper~XXIII}  
\defcitealias{2018AA...620A...9F}{XXL~Paper~XXIV}
\defcitealias{2018AA...620A..10P}{XXL~Paper~XXV}    
\defcitealias{2018AA...620A..11C}{XXL~Paper~XXVI}
\defcitealias{2018AA...620A..12C}{XXL~Paper~XXVII}
\defcitealias{2018AA...620A..13R}{XXL~Paper~XXVIII}
\defcitealias{2018AA...620A..14S}{XXL~Paper~XXIX}
\defcitealias{2018AA...620A..15G}{XXL~Paper~XXX}
\defcitealias{2018AA...620A..16B}{XXL~Paper~XXXI}
\defcitealias{2018AA...620A..17P}{XXL~Paper~XXXII}
\defcitealias{2018AA...620A..18L}{XXL~Paper~XXXIII}
\defcitealias{2018AA...620A..19H}{XXL~Paper~XXXIV}
\defcitealias{2018AA...620A..20K}{XXL~Paper~XXXV}

\lefthead{Umetsu et al.}
\righthead{Subaru HSC Weak-lensing Analysis of X-ray-selected XXL Galaxy
Groups and Clusters}
\usepackage{datetime}
\usepackage{version}    

\usepackage{etoolbox}
\makeatletter
\patchcmd{\NAT@citex}
  {\@citea\NAT@hyper@{\NAT@nmfmt{\NAT@nm}\NAT@date}}
  {\@citea\NAT@nmfmt{\NAT@nm}\NAT@hyper@{\NAT@date}}
  {}
  {}
\patchcmd{\NAT@citex}
  {\@citea\NAT@hyper@{%
     \NAT@nmfmt{\NAT@nm}%
     \hyper@natlinkbreak{\NAT@aysep\NAT@spacechar}{\@citeb\@extra@b@citeb}%
     \NAT@date}}
  {\@citea\NAT@nmfmt{\NAT@nm}%
   \NAT@aysep\NAT@spacechar%
   \NAT@hyper@{\NAT@date}}
  {}
  {}
\patchcmd{\NAT@citex}
  {\@citea\NAT@hyper@{%
     \NAT@nmfmt{\NAT@nm}%
     \hyper@natlinkbreak{\NAT@spacechar\NAT@@open\if*#1*\else#1\NAT@spacechar\fi}%
       {\@citeb\@extra@b@citeb}%
     \NAT@date}}
  {\@citea\NAT@nmfmt{\NAT@nm}%
   \NAT@spacechar\NAT@@open\if*#1*\else#1\NAT@spacechar\fi%
   \NAT@hyper@{\NAT@date}}
  {}
  {}
\makeatother

\begin{document}
\title{Weak-lensing Analysis of X-Ray-selected XXL Galaxy Groups and Clusters with Subaru HSC Data} 
\author{Keiichi Umetsu\altaffilmark{1}}          
\author{Mauro Sereno\altaffilmark{2,3}} 
\author{Maggie Lieu\altaffilmark{4}}
\author{Hironao Miyatake\altaffilmark{5,6}}
\author{Elinor Medezinski\altaffilmark{7}}
\author{Atsushi J. Nishizawa\altaffilmark{5,6}}
\author{Paul Giles\altaffilmark{8}}
\author{Fabio Gastaldello\altaffilmark{9}}
\author{Ian G. McCarthy\altaffilmark{10}}
\author{Martin Kilbinger\altaffilmark{11}}
\author{Mark Birkinshaw\altaffilmark{12}}
\author{Stefano Ettori\altaffilmark{2,3}}
\author{Nobuhiro Okabe\altaffilmark{13,14,15}}
\author{I-Non Chiu\altaffilmark{1}}
\author{Jean Coupon\altaffilmark{16}}
\author{Dominique Eckert\altaffilmark{16}}
\author{Yutaka Fujita\altaffilmark{17}}
\author{Yuichi Higuchi\altaffilmark{18,1}}
\author{Elias Koulouridis\altaffilmark{19,11}}
\author{Ben Maughan\altaffilmark{12}}
\author{Satoshi Miyazaki\altaffilmark{20,21}}
\author{Masamune Oguri\altaffilmark{22,23,24}}
\author{Florian Pacaud\altaffilmark{25}}
\author{Marguerite Pierre\altaffilmark{11}}
\author{David Rapetti\altaffilmark{26,27,28}}
\author{Graham P. Smith\altaffilmark{29}}

\email{keiichi@asiaa.sinica.edu.tw}
\altaffiltext{1}
 {Academia Sinica Institute of Astronomy and Astrophysics (ASIAA),
 No. 1, Section 4, Roosevelt Road, Taipei 10617, Taiwan}
\altaffiltext{2}
 {INAF - Osservatorio di Astrofisica e Scienza dello Spazio di Bologna,
 via Piero Gobetti 93/3, I-40129 Bologna, Italy}
\altaffiltext{3}
 {INFN, Sezione di Bologna, viale Berti Pichat 6/2, I-40127 Bologna, Italy}
\altaffiltext{4}{European Space Astronomy Centre, ESA, Villanueva de la Ca$\tilde{\mathrm{n}}$ada, E-28691 Madrid, Spain}
\altaffiltext{5}
 {Institute for Advanced Research, Nagoya University, Nagoya 464-8601, Japan}
\altaffiltext{6}
 {Division of Particle and Astrophysical Science, Graduate School of Science, Nagoya University, Nagoya 464-8602, Japan}
\altaffiltext{7}{Department of Astrophysical Sciences, Princeton
University, Princeton, NJ 08544, USA}
\altaffiltext{8}{Department of Physics and Astronomy, Pevensey Building, University of Sussex, Brighton, BN1 9QH, UK}
\altaffiltext{9}{INAF - IASF Milano, via Bassini 15, I-20133 Milano, Italy}
\altaffiltext{10}{Astrophysics Research Institute, Liverpool John Moores University, Liverpool, L3 5RF, United Kingdom}
\altaffiltext{11}{AIM, CEA, CNRS, Universit\'e Paris-Saclay,
Universit\'e Paris Diderot, Sorbonne Paris Cit\'e, F-91191 Gif-sur-Yvette, France}
\altaffiltext{12}{H. H. Wills Physics Laboratory, University of Bristol, Tyndall Ave., Bristol BS8 1TL, UK}
\altaffiltext{13}{Department of Physical Science, Hiroshima University,
1-3-1 Kagamiyama, Higashi-Hiroshima, Hiroshima 739-8526, Japan}
\altaffiltext{14}{Hiroshima Astrophysical Science Center, Hiroshima University, 1-3-1 Kagamiyama, Higashi-Hiroshima,
Hiroshima 739-8526, Japan}
\altaffiltext{15}{Core Research for Energetic Universe, Hiroshima University, 1-3-1 Kagamiyama, Higashi-Hiroshima,
Hiroshima 739-8526, Japan}
\altaffiltext{16}{Department of Astronomy, University of Geneva, ch. d'\'Ecogia 16, 1290 Versoix, Switzerland}
\altaffiltext{17}{Department of Earth and Space Science, Graduate School
of Science, Osaka University, Toyonaka, Osaka 560-0043, Japan} 
\altaffiltext{18}{Faculty of Science and Engineering, Kindai University,
Higashi-Osaka, Osaka, 577-8502, Japan} 
\altaffiltext{19}{Institute for Astronomy \& Astrophysics, Space
Applications \& Remote Sensing, National Observatory of Athens, GR-15236
Palaia Penteli, Greece} 
\altaffiltext{20}{National Astronomical Observatory of Japan, 2-21-1
Osawa, Mitaka, Tokyo 181-8588, Japan}
\altaffiltext{21}{SOKENDAI (The Graduate University for Advanced
Studies), 2-21-1 Osawa, Mitaka, Tokyo 181-8588, Japan}
\altaffiltext{22}{Research Center for the Early Universe, University of Tokyo, Tokyo 113-0033, Japan}
\altaffiltext{23}{Department of Physics, The University of Tokyo, 7-3-1 Hongo, Bunkyo-ku, Tokyo 113-0033, Japan}
\altaffiltext{24}{Kavli Institute for the Physics and Mathematics of the
Universe (Kavli IPMU, WPI), University of Tokyo, Chiba 277-8582, Japan}
\altaffiltext{25}{Argelander Institut f\"ur Astronomie,  Universit\"at Bonn, D-53121 Bonn, Germany}
\altaffiltext{26}{Center for Astrophysics and Space Astronomy,
Department of Astrophysical and Planetary Science, University of
Colorado, Boulder, CO 80309, USA} 
\altaffiltext{27}{NASA Ames Research Center, Moffett Field, CA 94035, USA}
\altaffiltext{28}{Universities Space Research Association, Mountain
View, CA 94043, USA} 
\altaffiltext{29}{School of Physics and Astronomy, University of
Birmingham, Birmingham, B15 2TT, United Kingdom}

\begin{abstract}
We present a weak-lensing analysis of X-ray galaxy groups and clusters
 selected from the \XMM-XXL survey using the first-year data from the
 Hyper Suprime-Cam (HSC) Subaru Strategic Program.
 Our joint weak-lensing and X-ray analysis focuses on 136
 spectroscopically confirmed X-ray-selected systems at
 $0.031 \le z \le 1.033$
 detected in the 25\,deg$^2$ XXL-N region, which largely overlaps with
 the HSC-\XMM\ field. 
 With high-quality HSC weak-lensing data,
 we characterize the mass distributions of individual clusters and
 establish the concentration--mass ($c$--$M$) relation for the XXL
 sample,
 by accounting for selection bias and statistical effects
 and marginalizing over the remaining mass calibration uncertainty.
We find the mass-trend parameter of the $c$--$M$ relation to
 be $\beta=\betaCM$ and the normalization to be
 $c_{200}=\Cpivot$ at $M_{200}=10^{14}\Msunh$ and $z=0.3$. We find no
 statistical evidence for redshift evolution.
Our weak-lensing results are in excellent agreement with
 dark-matter-only $c$--$M$ relations calibrated
 for recent $\Lambda$CDM cosmologies.
The level of intrinsic scatter in $c_{200}$ is constrained as
 $\sigma(\ln{c_{200}}) < \sigmalnCup$ ($99.7\percent$ CL), which is
 smaller than predicted for the full population of $\Lambda$CDM halos. 
This is likely caused in part by the X-ray selection bias in
 terms of the cool-core or relaxation state.
 We determine the temperature--mass ($T_\mathrm{X}$--$M_{500}$) relation
 for a subset of 105 XXL clusters that have both measured HSC lensing masses
 and X-ray temperatures.  
 The resulting $T_\mathrm{X}$--$M_{500}$ relation is consistent with the 
 self-similar prediction.
 Our $T_\mathrm{X}$--$M_{500}$ relation agrees with the XXL DR1  
 results at group scales but has a slightly steeper mass
 trend, implying a smaller mass scale in the cluster regime. The overall
 offset in the $T_\mathrm{X}$--$M_{500}$ relation is at the
 $\sim 1.5\sigma$ level, corresponding to a mean mass offset of $\bXXL$.
We also provide bias-corrected, weak-lensing-calibrated
 $M_{200}$ and $M_{500}$ mass estimates of individual XXL
 clusters based on their measured X-ray temperatures.
\end{abstract}   
 
\keywords{cosmology: observations ---
dark matter ---
gravitational lensing: weak ---
X-rays: galaxies: clusters ---
galaxies: clusters: general
}


\section{Introduction}
\label{sec:intro}

Galaxy clusters represent the largest bound objects formed in
the universe. Since galaxy clusters are highly massive and dominated
by dark matter (DM), they offer fundamental tests on the  assumed properties
of DM. For example, the standard cold dark matter (CDM) paradigm assumes
that DM is effectively cold and collisionless on astrophysical scales
\citep{Bertone+Gianfranco2018}.
In this context, the standard CDM model and its variants can provide a
series of observationally testable predictions. A prime example is the
``Bullet Cluster'', a merging pair of galaxy clusters 
exhibiting a significant offset between the centers of the gravitational
lensing mass and the peaks of the collisional intracluster gas
\citep{2004ApJ...604..596C,Clowe2006Bullet}.
The data support that DM is effectively
collisionless, like galaxies, placing a robust upper limit on the
self-interacting DM cross section of
$\sigma_\mathrm{DM}/m<1.25$\,cm$^2$\,g$^{-1}$ 
\citep{Randall2008}.

The evolution of the abundance of clusters across cosmic time is
sensitive to the amplitude and growth rate of primordial density 
fluctuations, as well as to the cosmic volume--redshift relation.
This cosmological sensitivity mainly comes from the fact that
cluster halos populate the exponential tail of the cosmic mass
function 
\citep{2001ApJ...553..545H,Watson+2014}.
Hence, large samples of galaxy clusters spanning a wide range 
of masses and redshifts provide an independent means of examining any 
viable cosmological model \citep{Allen+2004,Vikhlinin+2009CCC3,Mantz2010,Pratt2019}.
In principle, galaxy clusters can thus complement other cosmological
probes, such as cosmic microwave background (CMB) anisotropy,
large-scale galaxy clustering, distant supernovae, and cosmic shear
observations.

Significant progress has been made in recent years in constructing 
large statistical samples of clusters thanks to dedicated wide-field surveys
\citep[e.g.,][]{Planck2014XX,Planck2015XXIV,SPT2015sze,Rykoff2016,Oguri2018camira,Miyazaki2018wl}.  
Cluster samples are often defined by X-ray, Sunyaev--Zel'dovich effect
(SZE), or optical imaging observables, so that the cluster masses are
statistically inferred from mass scaling relations.
Since the level of mass bias is likely cluster mass dependent
\citep{CoMaLit2,CoMaLit5} and sensitive to calibration systematics of
the instruments \citep{Donahue2014clash,Israel+2015}, a concerted effort
is required to enable an accurate calibration of
mass scaling
relations using direct weak-lensing mass measurements
\citep[e.g.,][]{WtG1,WtG3,Umetsu2014clash,Hoekstra2015CCCP,Melchior2015DES,Okabe+Smith2016,Sereno2017psz2lens,Schrabback2018spt}
and well-defined selection functions \citep[e.g.,][]{JPAS2014}.

The distribution and concentration of DM in quasi-equilibrium objects
depend fundamentally on the properties of DM.
Hierarchical CDM models predict that the structure of halos
characterized in terms of the spherically averaged density profile
$\rho(r)$ is approximately self-similar with a 
characteristic density cusp in their centers, $\rho(r)\propto 1/r$,
albeit with large variance associated with the assembly histories of
individual halos \citep{Jing+Suto2000}. 
They also predict that the density gradient $d\ln{\rho(r)/d\ln{r}}$ of
DM halos continuously steepens from the center out to diffuse
outskirts
\citep[][hereafter NFW]{1996ApJ...462..563N,1997ApJ...490..493N}.
Clusters are predicted to have lower central concentrations, in contrast to
individual galaxies that have denser central regions
\citep{Diemer+Kravtsov2015}. The shape of clusters is predicted to be
not spherical but triaxial, reflecting the collisionless nature of DM
\citep{2002ApJ...574..538J}.

Gravitational lensing offers a direct probe of the cosmic matter
distribution dominated by DM. While strong lensing leads to highly
distorted and/or multiple images in the densest regions of the universe 
\citep[e.g.,][]{1999PThPS.133....1H},
namely the central regions of massive 
halos, weak lensing provides a direct measure of the mass distribution
on larger scales \citep[e.g.,][]{2001PhR...340..291B}. Galaxy clusters
act as powerful gravitational lenses, producing both strong- and
weak-lensing features in the images of background source galaxies.
The unique advantage of weak gravitational lensing is its ability to
constrain the mass distribution of individual systems independently of
assumptions about their physical or dynamical state.

Weak-lensing observations in the cluster regime have established that
the total matter distribution within clusters in projection can be well 
described by cuspy, outward steepening density profiles   
\citep{Umetsu+2011,Umetsu2014clash,Umetsu2016clash,Newman+2013a,Okabe+2013}, 
with a near-universal shape \citep{Niikura2015,Umetsu+Diemer2017}, as
predicted for collisionless halos in quasi-gravitational equilibrium
\citep[e.g.,][]{1996ApJ...462..563N,1997ApJ...490..493N,Taylor+Navarro2001,Hjorth+2010DARKexp,DARKexp2}.
Subsequent cluster lensing studies targeting lensing-unbiased samples
\citep[e.g.,][]{Merten2015clash,Du2015,Umetsu2016clash,Okabe+Smith2016,Cibirka2017,Sereno2017psz2lens,Klein2019}
have found that the degree of mass concentration derived for these
clusters agrees well with theoretical models calibrated for recent
$\Lambda$CDM cosmologies 
\citep[e.g.,][]{Bhatt+2013,Dutton+Maccio2014,Meneghetti2014clash,Diemer+Kravtsov2015}.  
The three-dimensional shapes of galaxy clusters as constrained by
weak-lensing and multiwavelength data sets are found to be in agreement
with $\Lambda$CDM predictions
\citep[e.g.,][]{Oguri2005,Morandi2012,Sereno2013glszx,Sereno2018clump3d,Umetsu2015A1689}. 
These results are all in support of the standard explanation for DM as
effectively collisionless and nonrelativistic on sub-Mpc scales and
beyond, with an excellent match with standard $\Lambda$CDM predictions.

The XXL program
\citep[hereafter \citetalias{2016AA...592A...1P}]{2016AA...592A...1P}
represents one of the largest \XMMNewton\ programs to date.
The ultimate science goal of the XXL survey
is to provide independent and self-sufficient cosmological constraints
using X-ray-selected galaxy clusters
\citep[hereafter \citetalias{2016AA...592A...2P}]{2016AA...592A...2P}.
The XXL survey covers two sky regions of $\simeq 25$\,deg$^2$ each at
high galactic latitudes, namely, the XXL-N and XXL-S fields.
With the aid of multiwavelength follow-up observations, the survey has
uncovered nearly 400 galaxy groups and clusters
out to a redshift of $z\sim 2$
\citep[hereafter \citetalias{2018AA...620A...5A}]{2018AA...620A...5A},
spanning approximately two decades in mass
\citepalias{2016AA...592A...1P}.
This XXL 365 galaxy cluster catalog was made public as part of the XXL
second-year data release (DR2).

Hyper Suprime-Cam is an optical wide-field imager with a 1.77\,deg$^2$ 
field of view mounted on the prime focus of the 8.2\,m Subaru telescope
\citep{Miyazaki2018hsc,Komiyama2018hsc,Furusawa2018hsc,Kawanomoto2018hsc}.
The Hyper Suprime-Cam Subaru Strategic Program
\citep[HSC-SSP;][]{hsc2018ssp,hsc2018dr1} has been
conducting an optical imaging survey in five broad bands ($grizy$) in
three layers of survey depths and areas (Wide, Deep, and Ultradeep),
aiming at observing $1400$\,deg$^2$ on the sky in its Wide layer
\citep{hsc2018ssp}. 
The HSC survey is optimized for weak-lensing studies
\citep{Mandelbaum2018shear,Miyaoka2018,Medezinski2018src,Hikage2019wl,Hamana2020wl} 
and overlaps with the XXL survey in its HSC-\XMM\ field.
It is therefore possible to directly estimate the masses of  XXL
clusters using well-calibrated weak-lensing data available from the HSC 
survey.

In this paper, we carry out a weak-lensing analysis on a
statistical sample of X-ray groups and clusters drawn from the
XXL DR2 cluster catalog
\citepalias{2018AA...620A...5A}.
Our analysis uses wide-field multiband imaging
from the HSC survey to measure the weak-lensing signal for our XXL
sample. 
The main goal of this paper is to obtain cluster mass estimates for
individual XXL clusters and to achieve ensemble mass calibration with
sufficient accuracy for scaling relation analyses.
With direct mass measurements from weak lensing, 
we aim to characterize observable--mass scaling
relations of the XXL sample.
We focus on the concentration--mass ($c$--$M$) and temperature--mass
($T_\mathrm{X}$--$M$) relations in this work.
In our companion paper \citep{Sereno2020xxl},
we examine joint multivariate X-ray observable--mass scaling
relations for the XXL sample using the cluster mass measurements
presented in this paper.

This paper is organized as follows. Section \ref{sec:data}
 describes the XXL cluster catalog and the HSC-SSP data.
Section \ref{sec:wl} describes the weak-lensing
 measurements, the selection of background galaxies, and their
 associated uncertainties (see also Appendix \ref{appendix:test}). 
In Section \ref{sec:mass}, after describing the methodology used to
 infer the mass and concentration parameters from the lensing signal, we
 present the results of weak-lensing mass measurements of the XXL sample.
In Section \ref{sec:scaling} we examine observable--mass scaling
 relations of the XXL sample through Bayesian population
 modeling. Finally, a summary is given in Section 
 \ref{sec:summary}.   

Throughout this paper, we assume a spatially flat $\Lambda$CDM cosmology     
with $\Om=0.28$, $\OL=0.72$, and a Hubble constant of  
$H_0 = 100h$\,km\,s$^{-1}$\,Mpc$^{-1}$ with $h=0.7$.
We adopt $\sigma_8=0.817$ \citep{Hinshaw+2013WMAP9} for the fiducial
normalization of the matter power spectrum, with $\sigma_8$
the rms amplitude of linear mass fluctuations in a sphere of comoving
radius $8\Mpch$.
We denote the critical density of the universe at a particular redshift
$z$ as $\rho_\mathrm{c}(z)=3H^2(z)/(8\pi G)$, with $H(z)$ the
redshift-dependent Hubble parameter. We also define the dimensionless
expansion function as $E(z)=H(z)/H_0$.
We adopt the standard notation
$M_\Delta$ 
to denote the mass enclosed within a sphere of radius
$r_\Delta$
within which the mean overdensity equals 
$\Delta \times \rho_\mathrm{c}(z)$.
We denote three-dimensional cluster radii as $r$, and
reserve the symbol $R$ for projected cluster-centric distances. 

We use ``$\log$'' to denote the base-10 logarithm and ``$\ln$'' to
denote the natural logarithm. 
The fractional scatter in natural logarithm is quoted as a percent.  All
quoted errors are $1\sigma$ confidence limits unless otherwise stated.

\section{Cluster Sample and Data}
\label{sec:data}
 
\subsection{XXL Cluster Sample}
\label{subsec:xxl}

\begin{deluxetable*}{ccccccccccccc}
\tablecolumns{13}
\tablewidth{0pt}
\tabletypesize{\scriptsize}
\tablecaption{\label{tab:stack}
Characteristics of the XXL samples}
\tablehead{
\multicolumn{1}{c}{Sample} & 
\multicolumn{1}{c}{$N_\mathrm{cl}$\tablenotemark{a}} & 
\multicolumn{1}{c}{$N_{T}$\tablenotemark{b}}  & 
\multicolumn{1}{c}{$T_\mathrm{X}$\tablenotemark{c}} & 
\multicolumn{1}{c}{$\langle T_\mathrm{X}\rangle_\mathrm{wl}$} & 
\multicolumn{1}{c}{$\zmed$\tablenotemark{d}} & 
\multicolumn{1}{c}{$\langle z\rangle_\mathrm{wl}$} &  
\multicolumn{1}{c}{$c_{200}$} & 
\multicolumn{1}{c}{$M_{200}$} & 
\multicolumn{1}{c}{$\langle M_{200}\rangle_\mathrm{wl}$} & 
\multicolumn{1}{c}{$\langle M_{200}\rangle_\mathrm{g}$} & 
\multicolumn{1}{c}{SNR} & 
\multicolumn{1}{c}{$($SNR$)_\mathrm{q}$} \\
\colhead{} & 
\colhead{} & 
\colhead{} & 
\multicolumn{1}{c}{(keV)} & 
\multicolumn{1}{c}{(keV)} & 
\colhead{} & 
\colhead{} & 
\colhead{} & 
\multicolumn{1}{c}{($10^{13}h^{-1}M_\odot$)} & 
\multicolumn{1}{c}{($10^{13}h^{-1}M_\odot$)} & 
\multicolumn{1}{c}{($10^{13}h^{-1}M_\odot$)} & 
\colhead{} & 
\colhead{}  
}
\startdata
C1+C2 & $136$ & $105$ & $1.9$ & $2.0$ & $0.30$ & $0.25$ & $3.5\pm 0.9$ & $8.7\pm 0.8$ & $8.0\pm 0.8$ & $9.8\pm 0.8$ & $15.6$ & $20.5$\\
C1 & $83$ & $76$ & $2.1$ & $2.1$ & $0.29$ & $0.23$ & $3.6\pm 1.1$ & $9.7\pm 1.0$ & $9.0\pm 1.0$ & $11.6\pm 1.2$ & $14.0$ & $18.4$\\
C2 & $53$ & $29$ & $1.7$ & $1.6$ & $0.43$ & $0.29$ & $3.4\pm 1.8$ & $6.4\pm 1.2$ & $6.1\pm 1.1$ & $6.5\pm 1.0$ & $7.2$ & $9.5$
\enddata
\tablenotetext{a}{Number of clusters.}
\tablenotetext{b}{Number of clusters with measured X-ray temperatures.}
\tablenotetext{c}{Median X-ray temperature.}
\tablenotetext{d}{Median cluster redshift.}
\tablecomments{Quantities in brackets with subscript "wl" denote lensing-weighted sample means (Equation (\ref{eq:wlmean})), and those in brackets with subscript "g"e denote error-weighted geometric means (Equation (\ref{eq:geom})). The effective mass and concentration parameters ($M_{200}, c_{200}$) of each subsample are obtained from a single-mass-bin fit to the respective stacked $\Delta\Sigma$ profile assuming an NFW density profile. For each sample, the effective $M_{200}$ mass is consistent with the respective weighted sample averages from individual cluster weak-lensing measurements. We provide two different estimates of the weak-lensing signal-to-nose ratio integrated over the comoving radial range $R\in [0.3, 3]\,\Mpch$, one based on the linear estimator, SNR (Equation (\ref{eq:SNR})), and the other based on the quadratic estimator, $($SNR$)_\mathrm{q}$ (Equation (\ref{eq:SNRq})).}
\end{deluxetable*}


\begin{figure*}[!htb] 
  \begin{center}
   \includegraphics[scale=0.7, angle=0, clip]{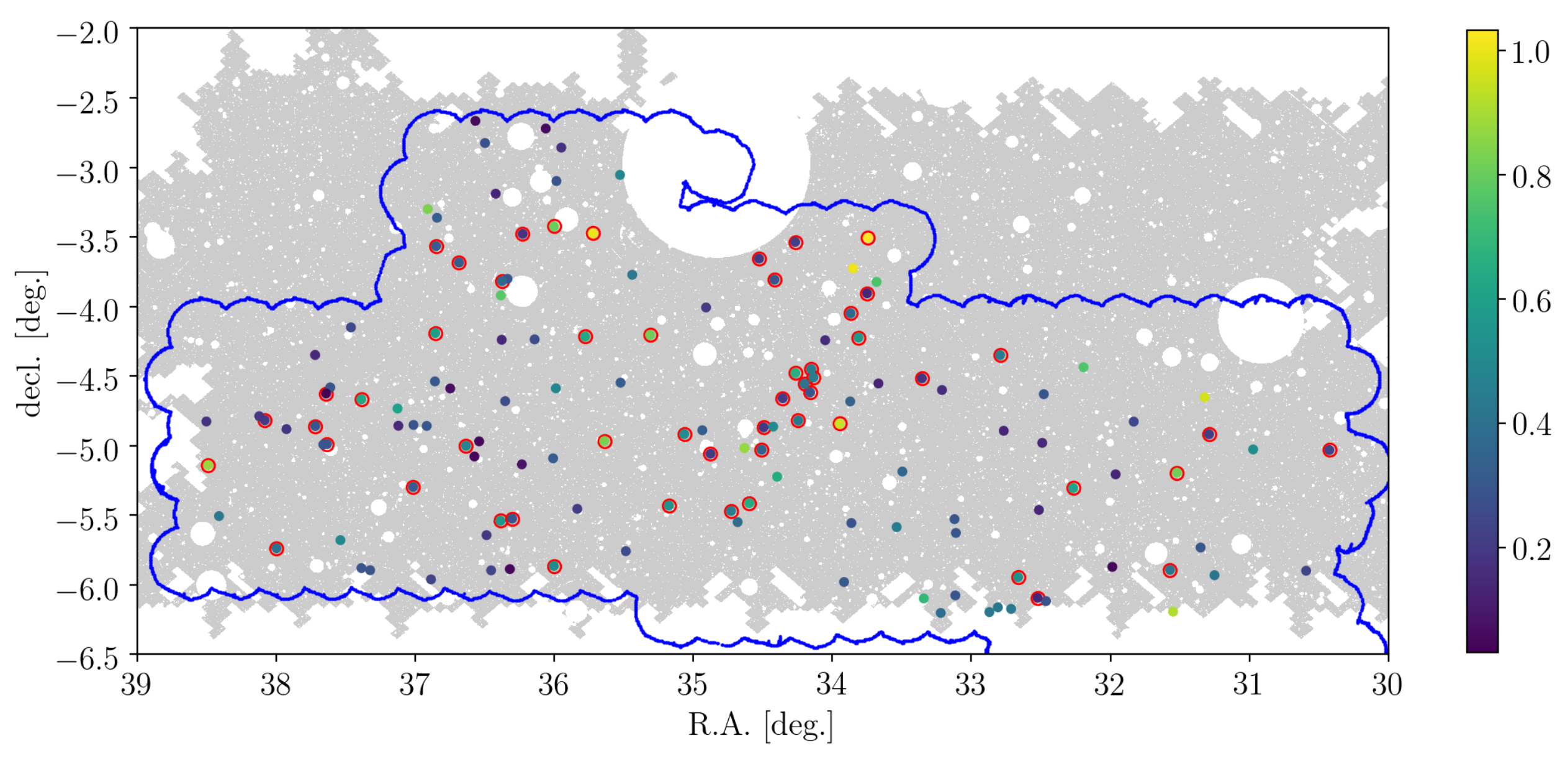} 
  \end{center}
\caption{
 \label{fig:sample}
Distribution of spectroscopically confirmed XXL-N C1+C2 groups and 
 clusters (filled circles) in the HSC-\XMM\ field. There are a total of
 136 XXL systems selected for our HSC weak-lensing analysis. The circles
 marked with red edges represent C2 clusters. 
 The gray shaded area shows the HSC survey footprint.
 The blue line shows the boundary of the combined exposure map of all
 \XMM\ pointings in the XXL-N field.
The area of the overlap region between the two surveys is
 $21.4$\,deg$^2$. The cluster redshift is color-coded according to the
 color bar on the right side. 
 }
\end{figure*}


\begin{figure}[!htb] 
  \begin{center}
   \includegraphics[scale=0.4, angle=0, clip]{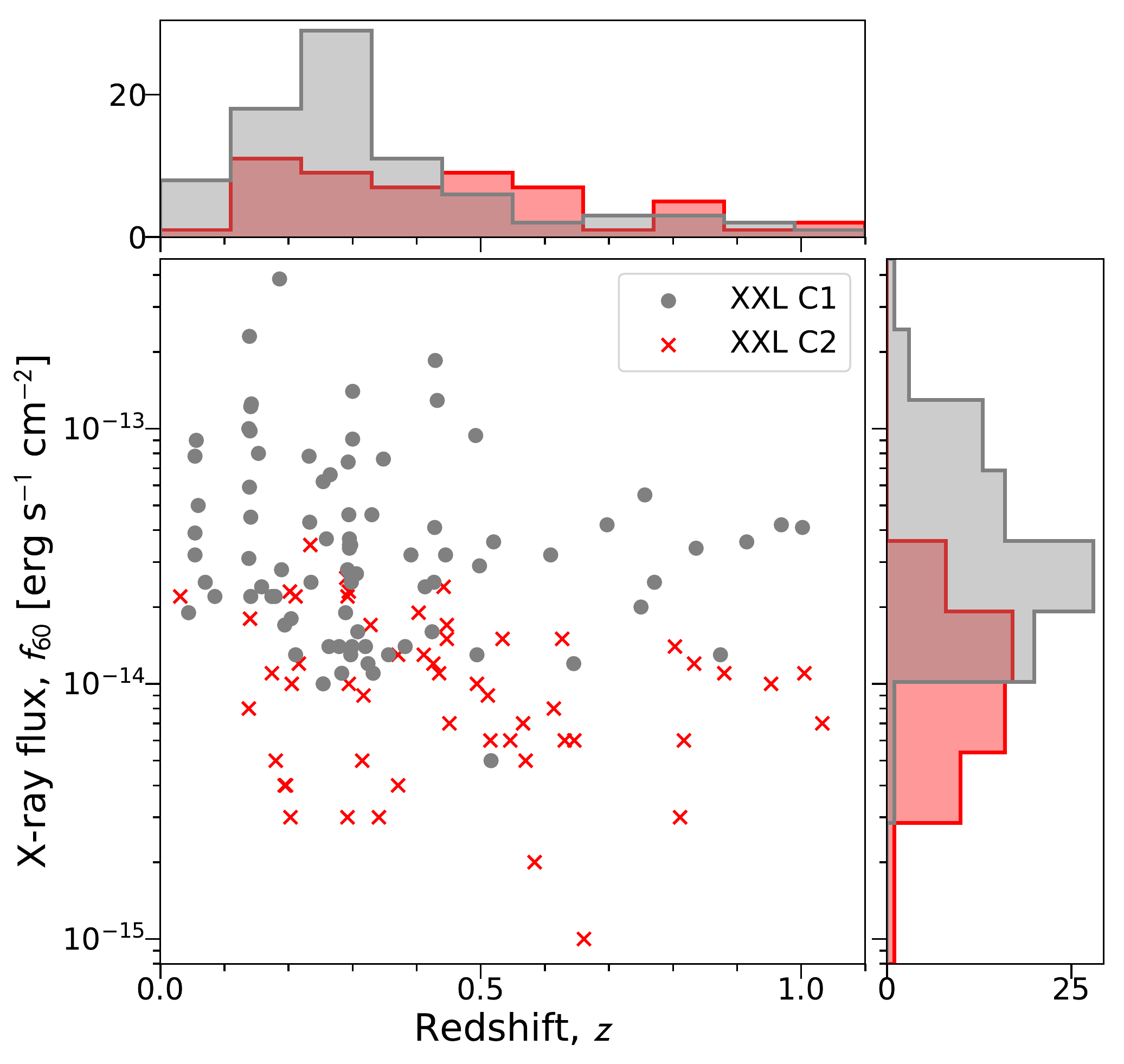} 
  \end{center}
\caption{
 \label{fig:XXL}
Distribution of our cluster sample in the X-ray flux ($f_{60}$)
 versus redshift ($z$) plane.  The gray circles and red crosses
 represent the C1 and C2 subsamples, respectively.
 }
\end{figure}

In the present study, we focus on spectroscopically confirmed
X-ray-selected systems of class C1 and C2 drawn from the XXL DR2 catalog  
presented in
\citetalias{2018AA...620A...5A}. 
The C1 population is designed to be free of contamination by spurious
detections or blended point sources, while
the C2 population is deeper but its initial selection is about
$50\percent$ contaminated
\citepalias{2016AA...592A...1P}.
Both populations of XXL clusters have been cleaned up a posteriori by
optical spectroscopic observations and from a detailed
comparison of X-ray and optical observations.

For our joint HSC-XXL analysis, we select XXL clusters that overlap with
the HSC survey footprint within a comoving transverse 
separation of $R_\mathrm{min}=0.3\Mpch$, which is the minimum
cluster-centric radius adopted in our HSC weak-lensing studies
\citep[Section \ref{subsec:DSigma}; see also][]{Medezinski2018src,Medezinski2018planck,Miyatake2019actpol}.  
These selection criteria leave us with
83 C1 clusters ($0.044\le z\le 1.002$)
 and
53 C2 clusters ($0.031\le z \le 1.033$), a total of 136 XXL clusters
with spectroscopic confirmation.
Of these, a subset of 105 clusters (76 C1 and 29 C2 clusters) have
X-ray temperatures $T_\mathrm{X}=\Tx$ measured in a fixed, core-included aperture of
$300$\,kpc, spanning the range $0.6\le \Tx/(\mathrm{keV}) \le 6.0$.
Here the X-ray temperatures $\Tx$ were measured with a spectral analysis
of the cluster single best pointing
\citepalias{2018AA...620A...5A}.
Spectra were extracted for each of the \XMMNewton\ cameras from the
region within a $300$\,kpc aperture and fitted in the 0.4--11.0\,keV
band with the absorbed Astrophysical Plasma Emission Code model 
(v2.0.2) in XSPEC \citep{XSPEC12}, with a fixed metal abundance of  
$Z=0.3Z_\odot$. The background was modeled following \citet{Eckert2014}.
X-ray temperatures could not be measured for all clusters, because
several cluster observations were at very low redshift with poor spatial 
coverage, were affected by flaring, were contaminated by point sources,
or had very low X-ray counts.  

\begin{deluxetable*}{ccccccccccccc}
\tablecolumns{13}
\tablewidth{0pt}
\tabletypesize{\scriptsize}
\tablecaption{\label{tab:clusters}
Cluster Properties and Weak-lensing Measurements}
\tablehead{
\multicolumn{1}{c}{ID\tablenotemark{a}} &
\multicolumn{1}{c}{R.A.\tablenotemark{b}} &
\multicolumn{1}{c}{Decl.\tablenotemark{b}} &
\multicolumn{1}{c}{$z$} & 
\multicolumn{1}{c}{Class} &
\multicolumn{1}{c}{$\Tx$} &
\multicolumn{1}{c}{$c_{200}$} &
\multicolumn{1}{c}{$M_\mathrm{200}$} &
\multicolumn{1}{c}{$M_\mathrm{500}$} &
\multicolumn{1}{c}{$\Mmttw$} &
\multicolumn{1}{c}{$\Mmtfv$} &
\multicolumn{1}{c}{SNR} & 
\multicolumn{1}{c}{$($SNR$)_\mathrm{q}$}\\
\colhead{} & \multicolumn{1}{c}{(deg)} & \multicolumn{1}{c}{(deg)} & \colhead{} & \colhead{} & \multicolumn{1}{c}{(keV)} & \colhead{} & \multicolumn{1}{c}{($10^{14}\Msun$)} & \multicolumn{1}{c}{($10^{14}\Msun$)} & \multicolumn{1}{c}{($10^{14}\Msun$)} & \multicolumn{1}{c}{($10^{14}\Msun$)}& \colhead{} & \colhead{} }
\startdata
002 & $36.384$ & $-3.920$ & $0.771$ & 1 & $2.50^{+0.18}_{-0.19}$ & $5.1 \pm 4.9$ ($4.0$) & $0.19 \pm 0.36$ ($0.36$) & $0.13 \pm 0.25$ ($0.24$) & $1.07^{+0.90}_{-0.53}$ & $0.72^{+0.65}_{-0.36}$ & $0.3$ & $1.7$\\
003 & $36.909$ & $-3.300$ & $0.836$ & 1 & $3.48^{+0.24}_{-0.25}$ & $4.9 \pm 4.8$ ($4.0$) & $0.43 \pm 0.92$ ($0.87$) & $0.29 \pm 0.62$ ($0.59$) & $1.64^{+1.39}_{-0.81}$ & $1.11^{+1.02}_{-0.56}$ & $0.7$ & $2.5$\\
006 & $35.439$ & $-3.772$ & $0.429$ & 1 & $4.24^{+0.61}_{-0.45}$ & $2.1 \pm 1.2$ ($2.0$) & $9.64 \pm 4.38$ ($9.45$) & $5.59 \pm 2.09$ ($5.53$) & $6.79^{+3.38}_{-2.20}$ & $4.72^{+1.92}_{-1.35}$ & $5.7$ & $6.6$\\
008 & $36.336$ & $-3.801$ & $0.299$ & 1 & $1.56^{+0.12}_{-0.10}$ & $2.4 \pm 2.0$ ($2.4$) & $1.65 \pm 1.67$ ($1.46$) & $1.01 \pm 0.95$ ($0.91$) & $1.07^{+0.81}_{-0.46}$ & $0.74^{+0.56}_{-0.32}$ & $3.2$ & $3.5$\\
009 & $36.685$ & $-3.684$ & $0.328$ & 2 & --                     & $5.8 \pm 5.1$ ($4.7$) & $0.42 \pm 0.53$ ($0.33$) & $0.29 \pm 0.37$ ($0.23$) & --                     & --                    & $1.0$ & $3.3$\\
010 & $36.843$ & $-3.362$ & $0.330$ & 1 & $2.36^{+0.35}_{-0.24}$ & $4.2 \pm 4.2$ ($3.7$) & $1.13 \pm 1.14$ ($1.01$) & $0.76 \pm 0.74$ ($0.68$) & $1.36^{+0.99}_{-0.59}$ & $0.92^{+0.70}_{-0.41}$ & $2.1$ & $4.1$\\
011 & $36.540$ & $-4.969$ & $0.054$ & 1 & $1.57^{+0.29}_{-0.11}$ & $3.0 \pm 3.0$ ($2.9$) & $1.15 \pm 1.38$ ($1.00$) & $0.76 \pm 0.84$ ($0.65$) & $0.90^{+0.72}_{-0.41}$ & $0.63^{+0.51}_{-0.29}$ & $5.0$ & $5.5$\\
013 & $36.858$ & $-4.538$ & $0.308$ & 1 & $1.30^{+0.22}_{-0.11}$ & $2.8 \pm 2.7$ ($2.7$) & $1.51 \pm 1.43$ ($1.40$) & $0.97 \pm 0.83$ ($0.90$) & $0.95^{+0.71}_{-0.40}$ & $0.67^{+0.50}_{-0.28}$ & $3.4$ & $3.8$\\
018 & $36.008$ & $-5.091$ & $0.324$ & 1 & $1.68^{+0.24}_{-0.16}$ & $5.9 \pm 5.1$ ($4.7$) & $0.29 \pm 0.40$ ($0.24$) & $0.20 \pm 0.27$ ($0.17$) & $0.77^{+0.63}_{-0.37}$ & $0.52^{+0.46}_{-0.26}$ & $-0.4$ & $2.8$\\
020 & $36.635$ & $-5.001$ & $0.494$ & 2 & $1.65^{+0.30}_{-0.23}$ & $2.1 \pm 1.6$ ($2.2$) & $4.36 \pm 3.39$ ($4.14$) & $2.60 \pm 1.80$ ($2.52$) & $1.89^{+1.39}_{-0.76}$ & $1.39^{+0.95}_{-0.55}$ & $3.6$ & $3.9$\\
021 & $36.233$ & $-5.134$ & $0.085$ & 1 & $0.79^{+0.06}_{-0.06}$ & $5.1 \pm 4.8$ ($4.0$) & $0.10 \pm 0.13$ ($0.10$) & $0.07 \pm 0.09$ ($0.07$) & $0.29^{+0.23}_{-0.14}$ & $0.19^{+0.17}_{-0.09}$ & $3.2$ & $4.5$\\
022 & $36.917$ & $-4.858$ & $0.293$ & 1 & $1.98^{+0.13}_{-0.12}$ & $4.9 \pm 4.5$ ($4.2$) & $2.15 \pm 1.34$ ($2.11$) & $1.48 \pm 0.87$ ($1.46$) & $1.67^{+0.98}_{-0.60}$ & $1.20^{+0.69}_{-0.43}$ & $2.6$ & $3.9$\\
025 & $36.353$ & $-4.680$ & $0.265$ & 1 & $2.23^{+0.24}_{-0.18}$ & $4.6 \pm 4.5$ ($3.9$) & $0.96 \pm 0.93$ ($0.86$) & $0.65 \pm 0.61$ ($0.58$) & $1.24^{+0.88}_{-0.54}$ & $0.84^{+0.62}_{-0.37}$ & $1.6$ & $3.5$\\
027 & $37.012$ & $-4.851$ & $0.295$ & 1 & $2.72^{+0.41}_{-0.40}$ & $2.8 \pm 2.7$ ($2.7$) & $0.91 \pm 1.04$ ($0.77$) & $0.58 \pm 0.63$ ($0.49$) & $1.45^{+1.13}_{-0.67}$ & $0.95^{+0.77}_{-0.44}$ & $2.6$ & $3.5$\\
028 & $35.984$ & $-3.098$ & $0.297$ & 1 & $1.53^{+0.27}_{-0.17}$ & $5.2 \pm 4.9$ ($4.1$) & $0.24 \pm 0.35$ ($0.21$) & $0.16 \pm 0.24$ ($0.14$) & $0.71^{+0.59}_{-0.34}$ & $0.48^{+0.43}_{-0.24}$ & $0.2$ & $2.6$\\
030 & $35.778$ & $-4.216$ & $0.631$ & 2 & --                     & $5.5 \pm 5.0$ ($4.3$) & $0.09 \pm 0.12$ ($0.10$) & $0.06 \pm 0.08$ ($0.07$) & --                     & --                    & $-1.2$ & $2.3$\\
032 & $36.002$ & $-3.424$ & $0.803$ & 2 & $2.16^{+0.49}_{-0.42}$ & $5.2 \pm 4.9$ ($4.1$) & $0.24 \pm 0.47$ ($0.48$) & $0.16 \pm 0.32$ ($0.32$) & $1.01^{+0.93}_{-0.51}$ & $0.69^{+0.67}_{-0.35}$ & $0.3$ & $1.5$\\
035 & $35.949$ & $-2.858$ & $0.174$ & 1 & $1.26^{+0.08}_{-0.08}$ & $5.5 \pm 5.0$ ($4.3$) & $0.11 \pm 0.14$ ($0.10$) & $0.07 \pm 0.10$ ($0.07$) & $0.47^{+0.38}_{-0.23}$ & $0.32^{+0.27}_{-0.15}$ & $-0.0$ & $3.4$\\
036 & $35.527$ & $-3.054$ & $0.492$ & 1 & $3.53^{+0.53}_{-0.43}$ & $6.4 \pm 5.3$ ($5.3$) & $0.52 \pm 0.70$ ($0.41$) & $0.36 \pm 0.50$ ($0.28$) & $1.63^{+1.33}_{-0.79}$ & $1.10^{+0.98}_{-0.54}$ & $0.5$ & $3.9$\\
038 & $36.856$ & $-4.190$ & $0.584$ & 2 & $1.67^{+0.30}_{-0.28}$ & $5.5 \pm 5.0$ ($4.3$) & $0.09 \pm 0.12$ ($0.09$) & $0.06 \pm 0.08$ ($0.06$) & $0.60^{+0.52}_{-0.30}$ & $0.39^{+0.36}_{-0.20}$ & $-0.3$ & $2.0$\\
040 & $35.523$ & $-4.547$ & $0.320$ & 1 & $1.95^{+0.26}_{-0.24}$ & $5.4 \pm 5.1$ ($4.2$) & $0.78 \pm 0.85$ ($0.66$) & $0.54 \pm 0.58$ ($0.45$) & $1.06^{+0.80}_{-0.47}$ & $0.72^{+0.57}_{-0.33}$ & $2.1$ & $3.9$\\
041 & $36.378$ & $-4.239$ & $0.142$ & 1 & $1.68^{+0.22}_{-0.08}$ & $7.3 \pm 5.1$ ($6.5$) & $2.47 \pm 1.11$ ($2.45$) & $1.82 \pm 0.76$ ($1.82$) & $1.94^{+0.92}_{-0.60}$ & $1.49^{+0.66}_{-0.45}$ & $2.9$ & $5.9$\\
044 & $36.141$ & $-4.236$ & $0.263$ & 1 & $1.21^{+0.11}_{-0.14}$ & $2.9 \pm 2.6$ ($2.8$) & $2.07 \pm 1.40$ ($2.00$) & $1.32 \pm 0.81$ ($1.30$) & $1.20^{+0.79}_{-0.45}$ & $0.87^{+0.54}_{-0.32}$ & $3.6$ & $4.8$\\
048 & $35.722$ & $-3.473$ & $1.005$ & 2 & $2.72^{+0.18}_{-0.16}$ & $5.7 \pm 5.1$ ($4.6$) & $1.04 \pm 2.61$ ($2.52$) & $0.71 \pm 1.78$ ($1.71$) & $1.28^{+1.12}_{-0.63}$ & $0.88^{+0.83}_{-0.44}$ & $1.1$ & $1.1$\\
049 & $35.988$ & $-4.588$ & $0.494$ & 1 & $2.13^{+0.12}_{-0.14}$ & $6.0 \pm 5.1$ ($4.9$) & $1.90 \pm 1.84$ ($1.73$) & $1.35 \pm 1.29$ ($1.21$) & $1.42^{+1.03}_{-0.60}$ & $1.00^{+0.76}_{-0.43}$ & $1.3$ & $3.2$\\
050 & $36.421$ & $-3.189$ & $0.140$ & 1 & $3.07^{+0.26}_{-0.25}$ & $6.3 \pm 5.3$ ($5.1$) & $1.10 \pm 0.96$ ($1.04$) & $0.78 \pm 0.68$ ($0.72$) & $1.65^{+1.14}_{-0.71}$ & $1.13^{+0.82}_{-0.49}$ & $1.0$ & $3.9$\\
051 & $36.498$ & $-2.825$ & $0.279$ & 1 & $1.34^{+0.09}_{-0.09}$ & $6.1 \pm 5.2$ ($4.9$) & $0.22 \pm 0.32$ ($0.19$) & $0.15 \pm 0.22$ ($0.13$) & $0.60^{+0.49}_{-0.29}$ & $0.40^{+0.35}_{-0.20}$ & $-0.3$ & $3.2$\\
052 & $36.567$ & $-2.666$ & $0.056$ & 1 & $0.63^{+0.04}_{-0.03}$ & $5.1 \pm 4.9$ ($4.0$) & $0.12 \pm 0.17$ ($0.12$) & $0.08 \pm 0.11$ ($0.08$) & $0.23^{+0.19}_{-0.11}$ & $0.16^{+0.14}_{-0.08}$ & $0.8$ & $2.5$\\
054 & $36.319$ & $-5.887$ & $0.054$ & 1 & $1.54^{+0.09}_{-0.08}$ & $5.9 \pm 5.2$ ($4.7$) & $0.60 \pm 0.76$ ($0.48$) & $0.42 \pm 0.54$ ($0.33$) & $0.77^{+0.60}_{-0.35}$ & $0.53^{+0.44}_{-0.25}$ & $-0.3$ & $3.6$\\
055 & $36.454$ & $-5.896$ & $0.232$ & 1 & $3.15^{+0.32}_{-0.51}$ & $8.2 \pm 5.4$ ($7.5$) & $3.19 \pm 1.49$ ($3.20$) & $2.39 \pm 1.07$ ($2.39$) & $2.92^{+1.37}_{-0.92}$ & $2.19^{+1.01}_{-0.69}$ & $3.1$ & $4.8$\\
056 & $33.871$ & $-4.682$ & $0.348$ & 1 & $2.99^{+0.50}_{-0.39}$ & $4.0 \pm 4.0$ ($3.5$) & $0.93 \pm 1.04$ ($0.79$) & $0.62 \pm 0.68$ ($0.52$) & $1.56^{+1.21}_{-0.72}$ & $1.04^{+0.84}_{-0.48}$ & $1.6$ & $3.1$\\
057 & $34.051$ & $-4.242$ & $0.153$ & 1 & $2.05^{+0.26}_{-0.18}$ & $4.6 \pm 4.6$ ($3.8$) & $0.41 \pm 0.53$ ($0.34$) & $0.28 \pm 0.36$ ($0.22$) & $0.97^{+0.78}_{-0.46}$ & $0.65^{+0.55}_{-0.31}$ & $1.8$ & $2.4$\\
058 & $34.935$ & $-4.889$ & $0.332$ & 1 & $2.19^{+0.27}_{-0.26}$ & $4.6 \pm 4.6$ ($3.7$) & $0.19 \pm 0.29$ ($0.18$) & $0.13 \pm 0.19$ ($0.12$) & $0.95^{+0.79}_{-0.47}$ & $0.63^{+0.56}_{-0.32}$ & $1.4$ & $3.6$\\
059 & $34.397$ & $-5.223$ & $0.645$ & 1 & $2.92^{+0.49}_{-0.37}$ & $6.3 \pm 5.4$ ($5.0$) & $0.33 \pm 0.61$ ($0.40$) & $0.22 \pm 0.40$ ($0.27$) & $1.37^{+1.19}_{-0.68}$ & $0.92^{+0.86}_{-0.47}$ & $0.2$ & $4.2$\\
060 & $33.668$ & $-4.553$ & $0.139$ & 1 & $4.70^{+0.26}_{-0.26}$ & $5.1 \pm 4.3$ ($4.5$) & $3.45 \pm 1.67$ ($3.38$) & $2.39 \pm 1.04$ ($2.38$) & $3.66^{+1.76}_{-1.21}$ & $2.56^{+1.15}_{-0.80}$ & $3.8$ & $6.0$\\
061 & $35.485$ & $-5.758$ & $0.259$ & 1 & $1.93^{+0.29}_{-0.22}$ & $4.4 \pm 4.5$ ($3.6$) & $0.29 \pm 0.41$ ($0.25$) & $0.20 \pm 0.28$ ($0.16$) & $0.90^{+0.74}_{-0.43}$ & $0.60^{+0.52}_{-0.29}$ & $1.4$ & $2.2$\\
062 & $36.061$ & $-2.721$ & $0.059$ & 1 & $0.77^{+0.12}_{-0.08}$ & $5.6 \pm 5.1$ ($4.5$) & $0.24 \pm 0.36$ ($0.22$) & $0.16 \pm 0.24$ ($0.15$) & $0.34^{+0.29}_{-0.16}$ & $0.23^{+0.21}_{-0.11}$ & $1.3$ & $4.1$\\
064 & $34.632$ & $-5.017$ & $0.874$ & 1 & --                     & $4.6 \pm 4.7$ ($3.7$) & $0.34 \pm 0.72$ ($0.71$) & $0.23 \pm 0.48$ ($0.48$) & --                     & --                    & $1.1$ & $2.1$\\
065 & $34.245$ & $-4.819$ & $0.435$ & 2 & --                     & $5.9 \pm 5.2$ ($4.7$) & $0.09 \pm 0.12$ ($0.09$) & $0.06 \pm 0.08$ ($0.06$) & --                     & --                    & $-1.3$ & $2.8$\\
067 & $34.681$ & $-5.549$ & $0.382$ & 1 & $1.22^{+0.13}_{-0.17}$ & $6.7 \pm 5.4$ ($5.5$) & $0.43 \pm 0.59$ ($0.34$) & $0.29 \pm 0.42$ ($0.23$) & $0.63^{+0.53}_{-0.30}$ & $0.43^{+0.39}_{-0.21}$ & $-0.1$ & $3.7$\\
071 & $35.640$ & $-4.967$ & $0.833$ & 2 & $2.18^{+0.13}_{-0.15}$ & $5.6 \pm 5.1$ ($4.4$) & $0.25 \pm 0.51$ ($0.51$) & $0.17 \pm 0.35$ ($0.35$) & $0.95^{+0.82}_{-0.47}$ & $0.64^{+0.59}_{-0.32}$ & $-0.2$ & $2.2$\\
072 & $33.850$ & $-3.726$ & $1.002$ & 1 & $2.00^{+0.27}_{-0.31}$ & $5.7 \pm 5.1$ ($4.6$) & $1.32 \pm 3.44$ ($3.26$) & $0.88 \pm 2.31$ ($2.18$) & $1.01^{+0.94}_{-0.50}$ & $0.69^{+0.69}_{-0.35}$ & $1.4$ & $1.4$\\
073 & $33.744$ & $-3.506$ & $1.033$ & 2 & $1.72^{+0.41}_{-0.33}$ & $5.7 \pm 5.1$ ($4.4$) & $1.08 \pm 2.77$ ($2.70$) & $0.72 \pm 1.84$ ($1.79$) & $0.94^{+0.91}_{-0.48}$ & $0.64^{+0.67}_{-0.34}$ & $1.4$ & $1.4$\\
075 & $35.834$ & $-5.454$ & $0.211$ & 1 & --                     & $5.8 \pm 5.1$ ($4.6$) & $0.08 \pm 0.10$ ($0.08$) & $0.06 \pm 0.07$ ($0.05$) & --                     & --                    & $-1.2$ & $2.4$\\
076 & $33.682$ & $-3.823$ & $0.750$ & 1 & --                     & $4.1 \pm 4.2$ ($3.5$) & $3.26 \pm 5.68$ ($2.90$) & $2.47 \pm 4.11$ ($1.93$) & --                     & --                    & $1.3$ & $3.4$\\
077 & $34.527$ & $-3.656$ & $0.202$ & 2 & $1.53^{+0.28}_{-0.21}$ & $5.1 \pm 4.8$ ($4.1$) & $0.36 \pm 0.45$ ($0.29$) & $0.25 \pm 0.31$ ($0.20$) & $0.73^{+0.59}_{-0.34}$ & $0.49^{+0.42}_{-0.24}$ & $1.4$ & $3.5$\\
078 & $33.948$ & $-4.842$ & $0.953$ & 2 & $2.63^{+0.41}_{-0.45}$ & $5.5 \pm 5.1$ ($4.3$) & $0.21 \pm 0.40$ ($0.41$) & $0.14 \pm 0.27$ ($0.28$) & $1.11^{+0.97}_{-0.56}$ & $0.74^{+0.70}_{-0.38}$ & $-1.4$ & $1.8$\\
079 & $34.494$ & $-4.868$ & $0.194$ & 2 & --                     & $5.5 \pm 5.0$ ($4.4$) & $0.44 \pm 0.54$ ($0.35$) & $0.30 \pm 0.37$ ($0.24$) & --                     & --                    & $1.1$ & $2.6$\\
080 & $34.597$ & $-5.413$ & $0.646$ & 2 & $1.65^{+0.32}_{-0.25}$ & $4.7 \pm 4.7$ ($3.7$) & $0.14 \pm 0.22$ ($0.18$) & $0.09 \pm 0.15$ ($0.12$) & $0.71^{+0.63}_{-0.35}$ & $0.47^{+0.45}_{-0.24}$ & $0.5$ & $3.4$\\
082 & $32.714$ & $-6.173$ & $0.427$ & 1 & $3.58^{+0.61}_{-0.50}$ & $4.6 \pm 4.7$ ($3.7$) & $0.39 \pm 0.69$ ($0.40$) & $0.27 \pm 0.47$ ($0.26$) & $1.74^{+1.54}_{-0.87}$ & $1.17^{+1.10}_{-0.60}$ & $1.7$ & $3.1$
\enddata
\tablecomments{We list for each cluster posterior summary statistics ($\CBI\pm \SBI$) of the NFW halo parameters, $c_{200}$, $M_{200}$, and $M_{500}$. Numbers in parentheses represent the median of the posterior probability distribution of each parameter. These parameters have been constrained by fitting a spherical NFW model to the weak-lensing $\Delta\Sigma(R)$ profile of each individual cluster over the comoving radial range $R\in [0.3,3]\,\Mpch$. These cluster mass and concentration measurements are not corrected for mass modeling bias, statistical bias, or selection effects. We provide bias-corrected, weak-lensing-calibrated mass estimates $\Mmttw$ and $\Mmtfv$ (median and 68\percent confidence intervals of the posterior distribution) for each individual cluster based on the measured X-ray temperature, $\Tx$, where available (Section \ref{subsec:forecast}; see Figure \ref{fig:MTR}). We recommend using these statistically corrected $M_{\Delta, \mathrm{MT}}$ as a weak-lensing mass estimate for a given individual cluster. All these mass estimates are subject to a systematic uncertainty of $\pm 5\percent$. Our concentration estimates have a systematic uncertainty of $\pm 16\percent$.}
\tablenotetext{a}{XLSSC cluster identifier (between 1 and 499 or between 500 and 999, for XXL-N or XXL-S, respectively).}
\tablenotetext{b}{X-ray cluster coordinates in R.A. and decl. (J2000.0).}\end{deluxetable*}

\begin{deluxetable*}{ccccccccccccc}
\addtocounter{table}{-1}
\tablecolumns{13}
\tablewidth{0pt}
\tabletypesize{\scriptsize}
\tablecaption{Continued.}
\tablehead{
\multicolumn{1}{c}{ID\tablenotemark{a}} &
\multicolumn{1}{c}{R.A.\tablenotemark{b}} &
\multicolumn{1}{c}{Decl.\tablenotemark{b}} &
\multicolumn{1}{c}{$z$} & 
\multicolumn{1}{c}{Class} &
\multicolumn{1}{c}{$\Tx$} &
\multicolumn{1}{c}{$c_{200}$} &
\multicolumn{1}{c}{$M_\mathrm{200}$} &
\multicolumn{1}{c}{$M_\mathrm{500}$} &
\multicolumn{1}{c}{$\Mmttw$} &
\multicolumn{1}{c}{$\Mmtfv$} &
\multicolumn{1}{c}{SNR} & 
\multicolumn{1}{c}{$($SNR$)_\mathrm{q}$}\\
\colhead{} & \multicolumn{1}{c}{(deg)} & \multicolumn{1}{c}{(deg)} & \colhead{} & \colhead{} & \multicolumn{1}{c}{(keV)} & \colhead{} & \multicolumn{1}{c}{($10^{14}\Msun$)} & \multicolumn{1}{c}{($10^{14}\Msun$)} & \multicolumn{1}{c}{($10^{14}\Msun$)} & \multicolumn{1}{c}{($10^{14}\Msun$)}& \colhead{} & \colhead{} }
\startdata
085 & $32.870$ & $-6.196$ & $0.428$ & 1 & $4.09^{+0.76}_{-0.69}$ & $5.5 \pm 5.0$ ($4.4$) & $0.12 \pm 0.19$ ($0.15$) & $0.08 \pm 0.13$ ($0.10$) & $1.51^{+1.38}_{-0.78}$ & $0.99^{+0.95}_{-0.52}$ & $-1.4$ & $2.6$\\
086 & $32.809$ & $-6.162$ & $0.424$ & 1 & $2.81^{+0.56}_{-0.49}$ & $5.2 \pm 4.9$ ($4.1$) & $0.15 \pm 0.23$ ($0.18$) & $0.10 \pm 0.16$ ($0.12$) & $1.15^{+1.02}_{-0.58}$ & $0.76^{+0.72}_{-0.39}$ & $0.4$ & $3.3$\\
087 & $37.720$ & $-4.348$ & $0.141$ & 1 & $1.61^{+0.12}_{-0.11}$ & $4.5 \pm 4.5$ ($3.8$) & $0.45 \pm 0.55$ ($0.37$) & $0.30 \pm 0.37$ ($0.25$) & $0.78^{+0.60}_{-0.36}$ & $0.53^{+0.43}_{-0.25}$ & $1.5$ & $3.7$\\
088 & $37.611$ & $-4.581$ & $0.295$ & 1 & $1.91^{+0.27}_{-0.24}$ & $4.6 \pm 4.5$ ($3.7$) & $0.74 \pm 0.81$ ($0.63$) & $0.50 \pm 0.54$ ($0.42$) & $1.03^{+0.78}_{-0.46}$ & $0.69^{+0.55}_{-0.32}$ & $2.1$ & $3.7$\\
089 & $37.127$ & $-4.733$ & $0.609$ & 1 & $2.11^{+0.40}_{-0.39}$ & $6.2 \pm 5.3$ ($5.0$) & $1.68 \pm 2.06$ ($1.37$) & $1.22 \pm 1.47$ ($0.97$) & $1.35^{+1.10}_{-0.61}$ & $0.95^{+0.81}_{-0.44}$ & $1.1$ & $2.7$\\
090 & $37.121$ & $-4.857$ & $0.141$ & 1 & $1.09^{+0.12}_{-0.07}$ & $5.5 \pm 5.1$ ($4.4$) & $0.12 \pm 0.17$ ($0.12$) & $0.08 \pm 0.11$ ($0.08$) & $0.42^{+0.34}_{-0.20}$ & $0.28^{+0.25}_{-0.14}$ & $-0.1$ & $2.3$\\
091 & $37.926$ & $-4.881$ & $0.186$ & 1 & $5.15^{+0.31}_{-0.31}$ & $8.9 \pm 4.7$ ($8.3$) & $5.77 \pm 1.66$ ($5.72$) & $4.44 \pm 1.11$ ($4.42$) & $5.87^{+1.79}_{-1.37}$ & $4.60^{+1.22}_{-0.97}$ & $6.0$ & $9.1$\\
095 & $31.962$ & $-5.206$ & $0.138$ & 1 & $0.90^{+0.09}_{-0.08}$ & $6.9 \pm 5.4$ ($5.8$) & $0.91 \pm 0.83$ ($0.82$) & $0.66 \pm 0.60$ ($0.59$) & $0.59^{+0.44}_{-0.25}$ & $0.42^{+0.33}_{-0.18}$ & $-0.0$ & $3.9$\\
096 & $30.973$ & $-5.027$ & $0.520$ & 1 & $4.98^{+0.50}_{-0.86}$ & $5.3 \pm 5.0$ ($4.2$) & $0.17 \pm 0.28$ ($0.19$) & $0.11 \pm 0.18$ ($0.13$) & $1.97^{+1.77}_{-1.02}$ & $1.30^{+1.24}_{-0.68}$ & $0.1$ & $1.1$\\
097 & $33.342$ & $-6.098$ & $0.697$ & 1 & $5.04^{+1.14}_{-0.95}$ & $5.7 \pm 5.1$ ($4.4$) & $0.19 \pm 0.34$ ($0.31$) & $0.12 \pm 0.23$ ($0.21$) & $1.96^{+1.83}_{-1.01}$ & $1.31^{+1.30}_{-0.69}$ & $-1.7$ & $3.4$\\
098 & $33.115$ & $-6.076$ & $0.297$ & 1 & $2.96^{+0.57}_{-0.59}$ & $4.4 \pm 4.6$ ($3.6$) & $0.14 \pm 0.21$ ($0.15$) & $0.10 \pm 0.14$ ($0.10$) & $1.15^{+1.03}_{-0.58}$ & $0.76^{+0.71}_{-0.39}$ & $1.2$ & $3.4$\\
099 & $33.220$ & $-6.202$ & $0.391$ & 1 & $3.72^{+0.87}_{-0.54}$ & $5.6 \pm 5.1$ ($4.4$) & $0.30 \pm 0.56$ ($0.39$) & $0.20 \pm 0.37$ ($0.27$) & $1.75^{+1.59}_{-0.89}$ & $1.18^{+1.14}_{-0.61}$ & $0.4$ & $1.9$\\
100 & $31.549$ & $-6.193$ & $0.915$ & 1 & $5.60^{+0.51}_{-0.43}$ & $5.5 \pm 5.0$ ($4.4$) & $0.48 \pm 1.08$ ($1.07$) & $0.32 \pm 0.73$ ($0.72$) & $2.64^{+2.37}_{-1.34}$ & $1.78^{+1.73}_{-0.92}$ & $-0.8$ & $0.8$\\
101 & $32.193$ & $-4.436$ & $0.756$ & 1 & $2.95^{+0.47}_{-0.39}$ & $4.4 \pm 4.4$ ($3.7$) & $5.33 \pm 7.94$ ($4.44$) & $3.90 \pm 5.50$ ($3.02$) & $1.99^{+1.71}_{-0.91}$ & $1.44^{+1.29}_{-0.68}$ & $1.6$ & $3.0$\\
102 & $31.322$ & $-4.652$ & $0.969$ & 1 & $3.87^{+0.81}_{-0.76}$ & $5.4 \pm 5.0$ ($4.3$) & $0.45 \pm 1.01$ ($1.00$) & $0.30 \pm 0.68$ ($0.67$) & $1.73^{+1.59}_{-0.87}$ & $1.17^{+1.16}_{-0.60}$ & $0.1$ & $1.6$\\
103 & $36.886$ & $-5.961$ & $0.233$ & 1 & $2.53^{+0.40}_{-0.34}$ & $6.2 \pm 5.2$ ($5.0$) & $1.96 \pm 1.44$ ($1.92$) & $1.39 \pm 0.99$ ($1.36$) & $1.79^{+1.15}_{-0.70}$ & $1.27^{+0.83}_{-0.50}$ & $2.3$ & $4.6$\\
104 & $37.324$ & $-5.895$ & $0.294$ & 1 & --                     & $6.4 \pm 5.2$ ($5.3$) & $1.03 \pm 1.04$ ($0.89$) & $0.74 \pm 0.75$ ($0.62$) & --                     & --                    & $0.5$ & $3.0$\\
105 & $38.411$ & $-5.506$ & $0.432$ & 1 & $6.01^{+0.79}_{-0.91}$ & $3.2 \pm 3.4$ ($3.0$) & $1.29 \pm 1.55$ ($1.07$) & $0.85 \pm 0.94$ ($0.69$) & $2.99^{+2.50}_{-1.45}$ & $1.94^{+1.66}_{-0.94}$ & $2.5$ & $4.0$\\
106 & $31.351$ & $-5.732$ & $0.300$ & 1 & $2.78^{+0.20}_{-0.17}$ & $2.1 \pm 1.4$ ($2.1$) & $3.70 \pm 2.31$ ($3.60$) & $2.19 \pm 1.24$ ($2.15$) & $2.69^{+1.58}_{-0.97}$ & $1.81^{+1.01}_{-0.64}$ & $3.9$ & $5.4$\\
108 & $31.832$ & $-4.827$ & $0.254$ & 1 & $2.34^{+0.31}_{-0.24}$ & $4.7 \pm 4.7$ ($3.8$) & $0.14 \pm 0.20$ ($0.14$) & $0.10 \pm 0.14$ ($0.09$) & $0.94^{+0.79}_{-0.47}$ & $0.62^{+0.56}_{-0.31}$ & $1.2$ & $2.3$\\
110 & $33.537$ & $-5.585$ & $0.445$ & 1 & $1.74^{+0.28}_{-0.22}$ & $3.2 \pm 3.1$ ($3.0$) & $3.43 \pm 2.55$ ($3.32$) & $2.23 \pm 1.51$ ($2.20$) & $1.81^{+1.26}_{-0.71}$ & $1.33^{+0.88}_{-0.52}$ & $3.1$ & $4.6$\\
111 & $33.111$ & $-5.627$ & $0.300$ & 1 & $3.70^{+0.52}_{-0.50}$ & $2.0 \pm 1.2$ ($1.9$) & $5.89 \pm 3.02$ ($5.73$) & $3.40 \pm 1.48$ ($3.36$) & $4.43^{+2.33}_{-1.49}$ & $2.98^{+1.36}_{-0.92}$ & $5.6$ & $6.5$\\
112 & $32.514$ & $-5.462$ & $0.139$ & 1 & $1.02^{+0.06}_{-0.05}$ & $3.6 \pm 3.9$ ($3.2$) & $0.74 \pm 0.84$ ($0.63$) & $0.49 \pm 0.52$ ($0.41$) & $0.57^{+0.45}_{-0.25}$ & $0.40^{+0.32}_{-0.18}$ & $3.8$ & $4.4$\\
114 & $30.425$ & $-5.031$ & $0.234$ & 2 & --                     & $4.8 \pm 3.8$ ($4.5$) & $3.47 \pm 1.61$ ($3.42$) & $2.42 \pm 1.03$ ($2.41$) & --                     & --                    & $3.2$ & $6.2$\\
116 & $32.664$ & $-5.945$ & $0.534$ & 2 & $6.03^{+0.29}_{-0.48}$ & $5.4 \pm 5.0$ ($4.3$) & $0.23 \pm 0.38$ ($0.24$) & $0.15 \pm 0.25$ ($0.16$) & $2.68^{+2.45}_{-1.41}$ & $1.78^{+1.72}_{-0.94}$ & $-0.0$ & $2.9$\\
117 & $33.121$ & $-5.528$ & $0.298$ & 1 & $3.42^{+0.47}_{-0.57}$ & $3.6 \pm 3.6$ ($3.3$) & $2.37 \pm 1.63$ ($2.32$) & $1.56 \pm 0.97$ ($1.55$) & $2.35^{+1.45}_{-0.90}$ & $1.60^{+0.95}_{-0.60}$ & $3.9$ & $4.9$\\
121 & $37.015$ & $-5.297$ & $0.317$ & 2 & $2.18^{+0.34}_{-0.33}$ & $4.1 \pm 4.2$ ($3.6$) & $1.30 \pm 1.34$ ($1.13$) & $0.87 \pm 0.87$ ($0.76$) & $1.34^{+1.00}_{-0.59}$ & $0.91^{+0.71}_{-0.41}$ & $1.9$ & $2.9$\\
123 & $36.487$ & $-5.643$ & $0.194$ & 1 & --                     & $5.8 \pm 5.2$ ($4.6$) & $0.17 \pm 0.24$ ($0.15$) & $0.11 \pm 0.16$ ($0.10$) & --                     & --                    & $-0.1$ & $3.4$\\
124 & $34.425$ & $-4.863$ & $0.516$ & 1 & $2.13^{+0.41}_{-0.38}$ & $5.3 \pm 4.9$ ($4.2$) & $0.36 \pm 0.59$ ($0.34$) & $0.24 \pm 0.40$ ($0.23$) & $1.05^{+0.92}_{-0.52}$ & $0.71^{+0.66}_{-0.36}$ & $0.9$ & $3.3$\\
127 & $36.850$ & $-3.566$ & $0.315$ & 2 & $0.91^{+0.14}_{-0.15}$ & $6.0 \pm 5.2$ ($4.9$) & $0.35 \pm 0.46$ ($0.28$) & $0.24 \pm 0.32$ ($0.19$) & $0.48^{+0.41}_{-0.23}$ & $0.33^{+0.30}_{-0.16}$ & $-0.4$ & $2.9$\\
130 & $35.176$ & $-5.430$ & $0.546$ & 2 & $1.53^{+0.25}_{-0.30}$ & $4.4 \pm 4.5$ ($3.7$) & $0.72 \pm 1.16$ ($0.63$) & $0.51 \pm 0.80$ ($0.42$) & $0.88^{+0.77}_{-0.43}$ & $0.60^{+0.56}_{-0.30}$ & $1.5$ & $2.8$\\
135 & $33.868$ & $-4.049$ & $0.371$ & 2 & $1.30^{+0.26}_{-0.20}$ & $5.1 \pm 4.9$ ($4.0$) & $0.58 \pm 0.79$ ($0.47$) & $0.39 \pm 0.54$ ($0.31$) & $0.74^{+0.63}_{-0.35}$ & $0.50^{+0.45}_{-0.24}$ & $1.5$ & $3.5$\\
137 & $34.416$ & $-3.807$ & $0.290$ & 2 & $1.66^{+0.22}_{-0.15}$ & $4.8 \pm 4.7$ ($3.9$) & $0.26 \pm 0.37$ ($0.22$) & $0.17 \pm 0.25$ ($0.15$) & $0.76^{+0.62}_{-0.37}$ & $0.51^{+0.45}_{-0.25}$ & $1.1$ & $2.3$\\
138 & $33.750$ & $-3.905$ & $0.140$ & 2 & --                     & $3.1 \pm 3.1$ ($3.0$) & $1.01 \pm 1.00$ ($0.92$) & $0.66 \pm 0.61$ ($0.60$) & --                     & --                    & $3.2$ & $4.9$\\
139 & $34.267$ & $-3.536$ & $0.216$ & 2 & --                     & $4.9 \pm 4.8$ ($3.8$) & $0.17 \pm 0.24$ ($0.16$) & $0.11 \pm 0.17$ ($0.11$) & --                     & --                    & $1.8$ & $3.7$\\
140 & $36.303$ & $-5.524$ & $0.294$ & 2 & $1.44^{+0.21}_{-0.17}$ & $4.8 \pm 4.8$ ($3.8$) & $0.14 \pm 0.20$ ($0.14$) & $0.09 \pm 0.14$ ($0.09$) & $0.60^{+0.51}_{-0.30}$ & $0.40^{+0.36}_{-0.20}$ & $1.8$ & $3.0$\\
141 & $34.357$ & $-4.659$ & $0.196$ & 2 & --                     & $6.0 \pm 5.2$ ($4.8$) & $0.25 \pm 0.33$ ($0.21$) & $0.17 \pm 0.23$ ($0.14$) & --                     & --                    & $-0.4$ & $3.3$\\
142 & $34.729$ & $-5.469$ & $0.451$ & 2 & $2.10^{+0.54}_{-0.37}$ & $6.0 \pm 5.2$ ($4.8$) & $0.11 \pm 0.16$ ($0.12$) & $0.08 \pm 0.11$ ($0.08$) & $0.81^{+0.71}_{-0.41}$ & $0.53^{+0.49}_{-0.27}$ & $-1.7$ & $2.6$\\
144 & $34.152$ & $-4.450$ & $0.447$ & 2 & $1.72^{+0.28}_{-0.23}$ & $2.4 \pm 1.9$ ($2.4$) & $5.26 \pm 3.26$ ($5.09$) & $3.22 \pm 1.75$ ($3.18$) & $2.56^{+1.67}_{-0.96}$ & $1.93^{+1.12}_{-0.69}$ & $3.8$ & $4.3$\\
145 & $37.388$ & $-4.666$ & $0.627$ & 2 & --                     & $4.8 \pm 4.8$ ($3.9$) & $0.13 \pm 0.20$ ($0.15$) & $0.09 \pm 0.13$ ($0.10$) & --                     & --                    & $0.6$ & $1.7$\\
146 & $37.462$ & $-4.150$ & $0.254$ & 1 & $1.84^{+0.26}_{-0.28}$ & $5.2 \pm 4.9$ ($4.0$) & $0.11 \pm 0.15$ ($0.11$) & $0.07 \pm 0.10$ ($0.07$) & $0.69^{+0.59}_{-0.34}$ & $0.46^{+0.41}_{-0.23}$ & $-0.1$ & $2.9$\\
147 & $37.641$ & $-4.625$ & $0.031$ & 2 & --                     & $5.4 \pm 5.0$ ($4.3$) & $0.21 \pm 0.32$ ($0.20$) & $0.14 \pm 0.22$ ($0.14$) & --                     & --                    & $0.2$ & $3.0$\\
148 & $37.719$ & $-4.859$ & $0.294$ & 2 & $1.19^{+0.06}_{-0.08}$ & $5.2 \pm 4.8$ ($4.3$) & $0.80 \pm 0.79$ ($0.73$) & $0.56 \pm 0.54$ ($0.49$) & $0.69^{+0.51}_{-0.30}$ & $0.48^{+0.37}_{-0.21}$ & $0.8$ & $4.3$\\
149 & $37.634$ & $-4.989$ & $0.292$ & 2 & --                     & $3.6 \pm 3.4$ ($3.4$) & $2.92 \pm 1.70$ ($2.85$) & $1.92 \pm 1.00$ ($1.90$) & --                     & --                    & $3.8$ & $5.0$\\
150 & $37.661$ & $-4.992$ & $0.292$ & 1 & $2.02^{+0.40}_{-0.29}$ & $2.7 \pm 2.3$ ($2.7$) & $2.01 \pm 1.49$ ($1.94$) & $1.27 \pm 0.88$ ($1.24$) & $1.59^{+1.04}_{-0.62}$ & $1.09^{+0.70}_{-0.42}$ & $3.0$ & $4.0$\\
151 & $38.122$ & $-4.788$ & $0.189$ & 1 & $1.86^{+0.27}_{-0.30}$ & $3.3 \pm 2.8$ ($3.2$) & $2.92 \pm 1.56$ ($2.87$) & $1.91 \pm 0.93$ ($1.89$) & $2.12^{+1.15}_{-0.72}$ & $1.51^{+0.77}_{-0.50}$ & $3.1$ & $6.7$\\
152 & $38.082$ & $-4.817$ & $0.205$ & 2 & $0.81^{+0.16}_{-0.15}$ & $2.7 \pm 2.5$ ($2.6$) & $0.83 \pm 0.95$ ($0.71$) & $0.53 \pm 0.56$ ($0.46$) & $0.56^{+0.47}_{-0.25}$ & $0.39^{+0.33}_{-0.18}$ & $3.3$ & $3.6$\\
153 & $38.490$ & $-5.139$ & $0.880$ & 2 & --                     & $5.5 \pm 5.0$ ($4.3$) & $0.16 \pm 0.29$ ($0.27$) & $0.11 \pm 0.19$ ($0.18$) & --                     & --                    & $-1.2$ & $2.6$\\
154 & $38.502$ & $-4.826$ & $0.179$ & 1 & $1.17^{+0.06}_{-0.08}$ & $5.9 \pm 5.2$ ($4.7$) & $0.16 \pm 0.23$ ($0.15$) & $0.11 \pm 0.15$ ($0.10$) & $0.49^{+0.39}_{-0.23}$ & $0.33^{+0.28}_{-0.16}$ & $0.4$ & $3.8$\\
158 & $32.793$ & $-4.349$ & $0.442$ & 2 & $1.72^{+0.31}_{-0.27}$ & $5.8 \pm 4.5$ ($5.2$) & $6.81 \pm 3.16$ ($6.70$) & $4.86 \pm 1.98$ ($4.82$) & $3.94^{+2.11}_{-1.35}$ & $3.25^{+1.50}_{-1.01}$ & $3.5$ & $5.3$
\enddata
\end{deluxetable*}

\begin{deluxetable*}{ccccccccccccc}
\addtocounter{table}{-1}
\tablecolumns{13}
\tablewidth{0pt}
\tabletypesize{\scriptsize}
\tablecaption{Continued.}
\tablehead{
\multicolumn{1}{c}{ID\tablenotemark{a}} &
\multicolumn{1}{c}{R.A.\tablenotemark{b}} &
\multicolumn{1}{c}{Decl.\tablenotemark{b}} &
\multicolumn{1}{c}{$z$} & 
\multicolumn{1}{c}{Class} &
\multicolumn{1}{c}{$\Tx$} &
\multicolumn{1}{c}{$c_{200}$} &
\multicolumn{1}{c}{$M_\mathrm{200}$} &
\multicolumn{1}{c}{$M_\mathrm{500}$} &
\multicolumn{1}{c}{$\Mmttw$} &
\multicolumn{1}{c}{$\Mmtfv$} &
\multicolumn{1}{c}{SNR} & 
\multicolumn{1}{c}{$($SNR$)_\mathrm{q}$}\\
\colhead{} & \multicolumn{1}{c}{(deg)} & \multicolumn{1}{c}{(deg)} & \colhead{} & \colhead{} & \multicolumn{1}{c}{(keV)} & \colhead{} & \multicolumn{1}{c}{($10^{14}\Msun$)} & \multicolumn{1}{c}{($10^{14}\Msun$)} & \multicolumn{1}{c}{($10^{14}\Msun$)} & \multicolumn{1}{c}{($10^{14}\Msun$)}& \colhead{} & \colhead{} }
\startdata
159 & $32.268$ & $-5.305$ & $0.614$ & 2 & $2.44^{+0.67}_{-0.48}$ & $6.1 \pm 5.2$ ($4.9$) & $0.22 \pm 0.37$ ($0.25$) & $0.14 \pm 0.25$ ($0.17$) & $1.10^{+0.99}_{-0.55}$ & $0.74^{+0.71}_{-0.38}$ & $-0.4$ & $2.6$\\
160 & $31.521$ & $-5.194$ & $0.817$ & 2 & --                     & $5.3 \pm 4.9$ ($4.2$) & $0.32 \pm 0.67$ ($0.66$) & $0.21 \pm 0.45$ ($0.45$) & --                     & --                    & $0.5$ & $1.9$\\
161 & $33.915$ & $-5.980$ & $0.306$ & 1 & $2.41^{+0.41}_{-0.34}$ & $5.5 \pm 5.0$ ($4.4$) & $0.74 \pm 0.81$ ($0.63$) & $0.52 \pm 0.56$ ($0.42$) & $1.23^{+0.94}_{-0.56}$ & $0.84^{+0.68}_{-0.39}$ & $1.2$ & $3.3$\\
162 & $32.524$ & $-6.093$ & $0.138$ & 2 & --                     & $4.6 \pm 4.6$ ($3.7$) & $0.21 \pm 0.31$ ($0.20$) & $0.14 \pm 0.21$ ($0.13$) & --                     & --                    & $2.1$ & $2.4$\\
163 & $32.463$ & $-6.117$ & $0.283$ & 1 & --                     & $5.0 \pm 4.8$ ($4.0$) & $0.16 \pm 0.24$ ($0.16$) & $0.11 \pm 0.16$ ($0.11$) & --                     & --                    & $0.7$ & $2.1$\\
165 & $33.356$ & $-4.516$ & $0.180$ & 2 & $0.97^{+0.12}_{-0.15}$ & $6.1 \pm 5.1$ ($5.0$) & $0.71 \pm 0.72$ ($0.62$) & $0.50 \pm 0.51$ ($0.43$) & $0.60^{+0.46}_{-0.26}$ & $0.41^{+0.33}_{-0.19}$ & $0.2$ & $3.5$\\
166 & $33.211$ & $-4.600$ & $0.158$ & 1 & $1.54^{+0.14}_{-0.17}$ & $5.9 \pm 5.2$ ($4.7$) & $0.33 \pm 0.42$ ($0.27$) & $0.23 \pm 0.29$ ($0.18$) & $0.71^{+0.56}_{-0.33}$ & $0.48^{+0.41}_{-0.23}$ & $0.7$ & $3.7$\\
167 & $32.479$ & $-4.630$ & $0.298$ & 1 & $1.84^{+0.25}_{-0.23}$ & $4.6 \pm 4.6$ ($3.8$) & $0.36 \pm 0.49$ ($0.29$) & $0.24 \pm 0.33$ ($0.19$) & $0.88^{+0.72}_{-0.42}$ & $0.59^{+0.51}_{-0.29}$ & $0.8$ & $3.0$\\
168 & $37.387$ & $-5.880$ & $0.295$ & 1 & $2.16^{+0.36}_{-0.31}$ & $5.7 \pm 5.1$ ($4.5$) & $0.09 \pm 0.11$ ($0.09$) & $0.06 \pm 0.07$ ($0.06$) & $0.72^{+0.62}_{-0.36}$ & $0.47^{+0.42}_{-0.24}$ & $-0.6$ & $3.0$\\
169 & $37.538$ & $-5.679$ & $0.498$ & 1 & $4.70^{+0.97}_{-1.05}$ & $3.7 \pm 3.6$ ($3.4$) & $3.16 \pm 2.43$ ($3.04$) & $2.10 \pm 1.48$ ($2.06$) & $3.01^{+2.04}_{-1.23}$ & $2.08^{+1.37}_{-0.83}$ & $2.6$ & $3.5$\\
170 & $37.998$ & $-5.737$ & $0.403$ & 2 & $1.74^{+0.30}_{-0.22}$ & $3.6 \pm 3.8$ ($3.2$) & $0.94 \pm 1.14$ ($0.78$) & $0.63 \pm 0.72$ ($0.51$) & $1.01^{+0.80}_{-0.46}$ & $0.69^{+0.57}_{-0.32}$ & $2.1$ & $3.6$\\
171 & $31.986$ & $-5.871$ & $0.044$ & 1 & --                     & $5.7 \pm 5.2$ ($4.6$) & $0.10 \pm 0.14$ ($0.10$) & $0.07 \pm 0.10$ ($0.07$) & --                     & --                    & $-2.1$ & $3.3$\\
172 & $31.571$ & $-5.893$ & $0.426$ & 2 & --                     & $5.3 \pm 4.9$ ($4.1$) & $0.12 \pm 0.17$ ($0.12$) & $0.08 \pm 0.11$ ($0.08$) & --                     & --                    & $0.5$ & $2.0$\\
173 & $31.251$ & $-5.931$ & $0.413$ & 1 & $4.29^{+0.27}_{-0.22}$ & $9.6 \pm 5.6$ ($9.3$) & $2.90 \pm 1.67$ ($2.91$) & $2.24 \pm 1.30$ ($2.24$) & $3.04^{+1.63}_{-1.08}$ & $2.24^{+1.26}_{-0.81}$ & $1.4$ & $4.5$\\
174 & $30.592$ & $-5.899$ & $0.235$ & 1 & $1.50^{+0.09}_{-0.09}$ & $5.9 \pm 5.2$ ($4.8$) & $0.31 \pm 0.43$ ($0.25$) & $0.21 \pm 0.30$ ($0.17$) & $0.70^{+0.56}_{-0.33}$ & $0.47^{+0.41}_{-0.23}$ & $0.3$ & $2.2$\\
176 & $32.490$ & $-4.980$ & $0.141$ & 1 & $1.42^{+0.18}_{-0.15}$ & $4.9 \pm 4.8$ ($3.9$) & $0.14 \pm 0.20$ ($0.13$) & $0.10 \pm 0.13$ ($0.09$) & $0.57^{+0.47}_{-0.28}$ & $0.38^{+0.34}_{-0.19}$ & $0.8$ & $2.3$\\
177 & $31.290$ & $-4.918$ & $0.211$ & 2 & --                     & $5.6 \pm 5.0$ ($4.5$) & $0.55 \pm 0.62$ ($0.46$) & $0.38 \pm 0.43$ ($0.31$) & --                     & --                    & $1.1$ & $2.8$\\
180 & $33.863$ & $-5.556$ & $0.289$ & 1 & $2.74^{+0.18}_{-0.19}$ & $3.3 \pm 3.3$ ($3.0$) & $2.27 \pm 1.86$ ($2.18$) & $1.47 \pm 1.12$ ($1.41$) & $1.97^{+1.31}_{-0.79}$ & $1.35^{+0.90}_{-0.54}$ & $3.1$ & $4.1$\\
181 & $36.376$ & $-3.817$ & $0.371$ & 2 & $1.09^{+0.08}_{-0.08}$ & $3.7 \pm 3.9$ ($3.4$) & $0.59 \pm 0.79$ ($0.49$) & $0.41 \pm 0.52$ ($0.32$) & $0.58^{+0.48}_{-0.27}$ & $0.40^{+0.35}_{-0.19}$ & $1.6$ & $2.6$\\
182 & $36.227$ & $-3.478$ & $0.174$ & 2 & $0.97^{+0.13}_{-0.15}$ & $5.8 \pm 5.2$ ($4.6$) & $0.08 \pm 0.09$ ($0.07$) & $0.05 \pm 0.06$ ($0.05$) & $0.34^{+0.28}_{-0.16}$ & $0.22^{+0.20}_{-0.11}$ & $-1.6$ & $2.7$\\
183 & $35.065$ & $-4.917$ & $0.511$ & 2 & $4.42^{+0.89}_{-0.69}$ & $4.4 \pm 4.4$ ($3.8$) & $1.99 \pm 2.07$ ($1.80$) & $1.36 \pm 1.35$ ($1.21$) & $2.47^{+1.90}_{-1.11}$ & $1.68^{+1.33}_{-0.76}$ & $1.9$ & $3.8$\\
184 & $35.311$ & $-4.204$ & $0.811$ & 2 & --                     & $5.1 \pm 4.9$ ($4.0$) & $0.14 \pm 0.23$ ($0.19$) & $0.09 \pm 0.15$ ($0.13$) & --                     & --                    & $-0.5$ & $2.9$\\
185 & $36.387$ & $-5.539$ & $0.566$ & 2 & --                     & $5.3 \pm 5.0$ ($4.2$) & $0.16 \pm 0.26$ ($0.19$) & $0.11 \pm 0.18$ ($0.13$) & --                     & --                    & $0.0$ & $1.7$\\
186 & $36.003$ & $-5.864$ & $0.515$ & 2 & $1.04^{+0.08}_{-0.06}$ & $5.4 \pm 5.0$ ($4.2$) & $0.19 \pm 0.31$ ($0.21$) & $0.13 \pm 0.21$ ($0.14$) & $0.45^{+0.41}_{-0.22}$ & $0.30^{+0.30}_{-0.15}$ & $0.3$ & $2.4$\\
187 & $34.136$ & $-4.509$ & $0.447$ & 2 & $3.24^{+0.60}_{-0.59}$ & $2.4 \pm 2.1$ ($2.4$) & $1.70 \pm 2.00$ ($1.44$) & $1.05 \pm 1.14$ ($0.90$) & $1.91^{+1.53}_{-0.87}$ & $1.27^{+1.04}_{-0.58}$ & $2.9$ & $3.4$\\
188 & $33.812$ & $-4.223$ & $0.570$ & 2 & --                     & $6.3 \pm 5.3$ ($5.0$) & $0.13 \pm 0.20$ ($0.15$) & $0.09 \pm 0.14$ ($0.10$) & --                     & --                    & $-1.7$ & $3.1$\\
189 & $34.908$ & $-4.007$ & $0.204$ & 1 & $1.28^{+0.18}_{-0.14}$ & $5.5 \pm 5.0$ ($4.3$) & $0.16 \pm 0.22$ ($0.15$) & $0.11 \pm 0.15$ ($0.10$) & $0.54^{+0.44}_{-0.26}$ & $0.36^{+0.32}_{-0.18}$ & $0.2$ & $2.1$\\
190 & $36.748$ & $-4.589$ & $0.070$ & 1 & $1.07^{+0.07}_{-0.07}$ & $5.7 \pm 5.1$ ($4.6$) & $0.17 \pm 0.24$ ($0.15$) & $0.11 \pm 0.16$ ($0.10$) & $0.43^{+0.35}_{-0.21}$ & $0.29^{+0.25}_{-0.14}$ & $-0.3$ & $3.1$\\
191 & $36.574$ & $-5.078$ & $0.054$ & 1 & $0.94^{+0.05}_{-0.06}$ & $3.1 \pm 2.6$ ($3.0$) & $3.58 \pm 2.27$ ($3.45$) & $2.30 \pm 1.28$ ($2.27$) & $1.36^{+0.95}_{-0.54}$ & $1.10^{+0.68}_{-0.41}$ & $4.4$ & $6.6$\\
192 & $34.509$ & $-5.029$ & $0.341$ & 2 & --                     & $5.0 \pm 4.8$ ($4.0$) & $0.14 \pm 0.20$ ($0.13$) & $0.09 \pm 0.13$ ($0.09$) & --                     & --                    & $1.0$ & $3.7$\\
193 & $34.876$ & $-5.058$ & $0.203$ & 2 & --                     & $5.5 \pm 5.1$ ($4.4$) & $0.07 \pm 0.08$ ($0.07$) & $0.05 \pm 0.05$ ($0.05$) & --                     & --                    & $-0.4$ & $3.6$\\
194 & $34.200$ & $-4.555$ & $0.411$ & 2 & --                     & $6.1 \pm 5.3$ ($4.8$) & $0.18 \pm 0.27$ ($0.17$) & $0.12 \pm 0.18$ ($0.12$) & --                     & --                    & $0.3$ & $3.9$\\
195 & $34.266$ & $-4.478$ & $0.661$ & 2 & --                     & $4.3 \pm 4.4$ ($3.6$) & $1.95 \pm 2.82$ ($1.62$) & $1.39 \pm 1.91$ ($1.09$) & --                     & --                    & $1.6$ & $2.5$\\
198 & $33.496$ & $-5.186$ & $0.356$ & 1 & $1.32^{+0.14}_{-0.09}$ & $4.8 \pm 4.7$ ($3.9$) & $0.23 \pm 0.34$ ($0.21$) & $0.15 \pm 0.23$ ($0.14$) & $0.60^{+0.50}_{-0.29}$ & $0.40^{+0.36}_{-0.20}$ & $0.4$ & $3.1$\\
201 & $32.767$ & $-4.893$ & $0.138$ & 1 & $1.60^{+0.24}_{-0.16}$ & $5.3 \pm 4.9$ ($4.2$) & $0.22 \pm 0.31$ ($0.19$) & $0.15 \pm 0.21$ ($0.13$) & $0.69^{+0.57}_{-0.34}$ & $0.47^{+0.41}_{-0.23}$ & $0.1$ & $3.0$\\
202 & $34.160$ & $-4.617$ & $0.292$ & 2 & --                     & $5.6 \pm 5.1$ ($4.4$) & $0.13 \pm 0.19$ ($0.13$) & $0.09 \pm 0.13$ ($0.09$) & --                     & --                    & $0.5$ & $2.7$
\enddata
\end{deluxetable*}

In Table \ref{tab:stack} we summarize basic characteristics of the
C1+C2, C1, and C2 samples selected for our study.
Figure \ref{fig:sample} shows the distribution of the full (C1+C2)
sample of 136 XXL clusters in the HSC-\XMM\ field (see Section \ref{subsec:hsc}).
Figure \ref{fig:XXL} shows the distribution of our 136 XXL clusters in the
 X-ray flux ($f_{60}$) versus redshift ($z$) plane.
We summarize in Table \ref{tab:clusters} the properties of individual
clusters in our sample.

\subsection{Subaru HSC Survey}
\label{subsec:hsc}

We use the HSC first-year shear catalog for our weak-lensing analysis.
Full details of the creation of the catalog are described in
\citet{Mandelbaum2018shear} and \citet{Mandelbaum2018sim}.
We thus refer the reader to those papers and give a basic summary
here. 

The first-year shear catalog was produced using about 90 nights of
HSC-Wide data taken from 2014 March to 2016 April.
This shear catalog consists of six distinct patches of the sky covering
a total of 137\,deg$^2$, which is larger than the area covered by the
public Data Release 1 (DR1). In this study, we use the shear catalog
updated with a star mask called ``Arcturus'' 
\citep{Coupon2018hsc,Miyatake2019actpol}.

HSC-Wide consists of observations made with the $grizy$ filters,
reaching a typical limiting magnitude of $i\sim 26$\,ABmag
\citep[$5\sigma$ for point sources;][]{hsc2018dr1}. 
The $i$-band imaging was performed under exceptional seeing conditions
for weak-lensing shape measurements, resulting in 
a median seeing FWHM of $\simeq 0.6\arcsec$.
The galaxy shapes were measured on the co-added $i$-band images using the
re-Gaussianization method \citep{Hirata+Seljak2003}.
Basic cuts were applied to select galaxies with robust photometry and   
shape measurements \citep{Mandelbaum2018shear}.
The HSC-\XMM\ field covers an effective area of $29.5$\,deg$^2$ once the  
star mask region is removed (Figure \ref{fig:sample}).
The area of the overlap region between the HSC and XXL surveys is
$21.4$\,deg$^2$. 
The weighted number density of source galaxies
in the HSC-\XMM\ field is $\ngal\simeq 22.1$ galaxies\,arcmin$^{-2}$,
and their mean redshift is $0.82$ \citep[see][]{Miyatake2019actpol}.   

We use the HSC multiband photometry to select background source
galaxies for a given cluster in the XXL sample.
Several different codes were used to estimate photometric redshifts
(photo-$z$'s) for individual galaxies from the multiband imaging data
\citep{Tanaka2018photoz}.
In this work, we employ the point-spread function (PSF) matched aperture 
(afterburner) photometry (Ephor\_AB) code
\citep{Tanaka2018photoz,Hikage2019wl}. 
Additional cuts needed to select background source galaxies are
described in Section \ref{subsec:source}.

\section{HSC Weak-lensing Analysis}
\label{sec:wl}

\subsection{Weak-lensing Basics}
\label{subsec:basics}

The effects of weak gravitational lensing are described by the 
convergence $\kappa$ and the complex shear $\gamma$.
The convergence causes an isotropic magnification, while the shear
induces a quadrupole anisotropy that can be estimated
from the ellipticities of background galaxies
\citep[e.g.,][]{Umetsu2010Fermi}.
These effects depend on the projected matter overdensity
field, as well as on the redshifts of the lens,
$z_l$, and the source galaxy, $z_s$, through the critical surface
mass density for lensing, $\Sigma_\mathrm{cr}(z_l,z_s)$, as
defined below.
In general, the observable quantity for weak lensing is not $\gamma$,
but the reduced shear,
\begin{equation}
 \label{eq:g}
 g=\frac{\gamma}{1-\kappa}.
\end{equation}

The complex shear $\gamma$ can be decomposed into the tangential
component $\gamma_+$ and the $45^\circ$-rotated component $\gamma_\times$.
The tangential shear component $\gamma_+$ averaged around a circle of
projected radius $R$ is related to the excess surface mass density
$\Delta\Sigma(R)$ through the following identity \citep{Kaiser1995}:
\begin{equation}
 \gamma_+(R)=\frac{\Sigma(<R)-\Sigma(R)}{\Sigma_\mathrm{cr}(z_l, z_s)}\equiv
  \frac{\Delta\Sigma(R)}{\Sigma_\mathrm{cr}(z_l, z_s)},
\end{equation}
where $\Sigma(R)$ is the azimuthally averaged surface mass density at $R$,
$\Sigma(<R)$ denotes the average surface mass density interior to $R$, 
and
\begin{equation}
 \Sigma_\mathrm{cr}(z_l, z_s)=\frac{c^2D_s}{4\pi G (1+z_l)^2D_l D_{ls}}
\end{equation}
with $c$ the speed of light, $G$ the gravitational constant,
and $D_l$, $D_s$ , and $D_{ls}$ the observer--lens,
observer--source, and lens--source angular diameter distances,
respectively. 
The extra factor of $(1+z_l)^2$ is due to our use of comoving
surface mass densities.
The quantity $\Sigma_\mathrm{cr}^{-1}(z_l, z_s)$
describes the geometric lensing strength,
where we set $\Sigma_\mathrm{cr}^{-1}(z_l, z_s)=0$ for $z_s \le z_l$.

\subsection{Tangential Shear Profile}
\label{subsec:DSigma}

The X-ray-emitting gas provides an excellent tracer of the total 
gravitational potential of the cluster
\citep[e.g.,][]{Donahue2014clash,Umetsu2018clump3d,OkabeT2018}, except
for massive cluster collisions caught in an ongoing phase of
dissociative mergers \citep[e.g.,][]{Clowe2006Bullet,Okabe+Umetsu2008}.
In this study, we measure the weak-lensing signal around the X-ray peak
location of each cluster (Table \ref{tab:clusters}) as a function of
comoving cluster-centric radius, $R$.
We compute $\Delta\Sigma$ in $N=8$ radial bins of equal logarithmic spacing
$\Delta\ln{R}=\ln(R_\mathrm{max}/R_\mathrm{min})/N\simeq 0.29$ 
from $R_\mathrm{min}=0.3\Mpch$ to $R_\mathrm{max}=3\Mpch$
\citep[e.g.,][]{Medezinski2018planck,Miyatake2019actpol}. 
The chosen inner limit $R_\mathrm{min}$ is sufficiently large so
that our photo-$z$ and shape measurements are not expected to be
affected significantly by masking or imperfect deblending by bright
cluster galaxies
\citep[see discussion in][]{Medezinski2018src}.
Moreover, $R_\mathrm{min}$ is much larger than  the typical offsets
between the brightest cluster galaxy (BCG) and the X-ray peak for 
XXL clusters
\citep[hereafter \citetalias{2016MNRAS.462.4141L}]{2016MNRAS.462.4141L}.
Hence, smoothing of the weak-lensing signal due to miscentering effects
\citep[e.g.,][]{Johnston+2007b,Umetsu+2011stack} is expected to be not
important for our analysis based on X-ray centering information.
However, it should be noted that there is a possibility that a merger has 
boosted the luminosity and made the X-ray peak off-centered during the
compression phase. 
Although the timescale on which this happens is expected to be
short \citep[$\sim 1$\,Gyr; see][]{2001ApJ...561..621R,Zhang+2016merger},
it could possibly induce a selection effect and contribute to the
scatter in scaling relations.

We estimate $\Delta\Sigma$ in each radial bin for either an individual
cluster or a stacked ensemble of multiple clusters
using the following estimator
\citep[][]{Mandelbaum2018shear}:
\begin{equation}
 \label{eq:DSigma}
  \Delta\Sigma_+(R_i) = \frac{1}{2{\cal
  R}(R_i)}\frac{\sum_{l,s\in i} w_{ls}e_{+,ls}[\langle\Sigma_{\mathrm{cr},ls}^{-1}\rangle]^{-1}}
 {[1+K(R_i)]\sum_{l,s\in i}w_{ls}},
\end{equation}
where the double summation is taken over all clusters of interest ($l$)
and over all source galaxies ($s$)
that lie within the cluster-centric radial bin ($i$), and
\begin{equation}
 e_+ = -\cos(2\phi) e_1 - \sin(2\phi)e_2
\end{equation}
is the tangential ellipticity of the source galaxy,
$\phi$ is the angle measured in sky coordinates from the R.A.
direction to the line connecting the lens and the source galaxy,
and ($e_1,e_2$) are the ellipticity components in sky coordinates
obtained from the HSC data analysis pipeline
\citep{Mandelbaum2018shear,Bosch2018pipe}. 
The critical surface mass density for each lens--source pair,
$\langle\Sigma_{\mathrm{cr},ls}^{-1}\rangle^{-1}$,
is averaged with the photo-$z$ probability distribution function (PDF)
of the source galaxy (see Section \ref{subsec:source}), $P_s(z)$, as
\begin{equation}
\label{eq:Sigma_cr}
\langle\Sigma_{\mathrm{cr},ls}^{-1}\rangle
 =\frac{\int_0^\infty\,P_s(z)\Sigma_\mathrm{cr}^{-1}(z_l,z)dz}
 {\int_0^\infty\,P_s(z)dz}.
\end{equation}
The statistical weight factor
$w_{ls}$ in Equation (\ref{eq:DSigma}) is given by
\begin{equation}
 w_{ls} = \left(\langle\Sigma_{\mathrm{cr},ls}^{-1}\rangle\right)^2\frac{1}{\sigma_{e,s}^2+e^2_{\mathrm{rms},s}},
\end{equation}
where $\sigma_{e,s}$ is the shape measurement uncertainty per
ellipticity component (i.e., $\sigma_{e_1,s}=\sigma_{e_2,s}\equiv \sigma_{e,s}$)
and $e_{\mathrm{rms},s}$ is the rms ellipticity estimate per component. 
The $[1+K(R_i)]$ factor statistically corrects for multiplicative
residual shear bias as determined from simulations
\citep{Mandelbaum2018shear,Mandelbaum2018sim},
\begin{equation}
 1+K(R_i) = \frac{\sum_{l,s\in i} w_{ls} (1+m_s)}{\sum_{l,s\in i} w_{ls}},
\end{equation}
where $m_s$ denotes the multiplicative bias factor of individual source
galaxies.
In our ensemble analysis of the XXL sample, we will include a $1\percent$
systematic uncertainty on the residual multiplicative
bias \citep[see Section \ref{subsec:syst};][]{Mandelbaum2018sim,Hikage2019wl}. 
We also conservatively correct for additive residual shear bias by
subtracting off the weighted mean offset from Equation (\ref{eq:DSigma})
\citep[see][]{Mandelbaum2018shear,Miyaoka2018,Okabe2019hsc}.  
The shear responsivity ${\cal R}(R_i)$ is calculated as
\citep[see also][]{Mandelbaum2005ggl}
\begin{equation}
 {\cal R}(R_i) = 1-\frac{\sum_{l,s\in
  i}w_{ls}e^2_{\mathrm{rms},s}}{\sum_{l,s\in i}w_{ls}}.
\end{equation}
The typical value of ${\cal R}$ is $\approx 0.84$
\citep[$e_\mathrm{rms}\approx 0.4$;][]{Medezinski2018src}.
A full description and clarification of the procedure are
given in \citet{Mandelbaum2018shear}.

Similarly, we define the $\times$-component surface mass density,
$\Delta\Sigma_\times$, by replacing $e_+$ in Equation (\ref{eq:DSigma})
with the 45$^\circ$-rotated ellipticity component $e_\times$, defined by
\begin{equation}
 e_\times = -e_2 \cos{2\phi} + e_1\sin{2\phi}.
\end{equation}
The azimuthally averaged $\times$ component, or the B-mode signal, 
is expected to be statistically consistent with zero if the signal is
due to weak lensing.

When interpreting the binned tangential shear profile
$\bd\equiv\{\Delta\Sigma_+(R_i)\}_{i=1}^N$,
it is important to define and determine the corresponding
bin radii $\{R_i\}_{i=1}^N$ accurately so as to minimize systematic
bias in cluster mass measurements. 
Following \citet{Okabe+Smith2016},
we define the effective bin radius $R_i$ using the weighted
harmonic mean of lens--source transverse separations $R_{ls}$ as 
\begin{equation}
\label{eq:binradius}
 R_i \equiv \frac{\sum_{l,s\in i}w_{ls}}{\sum_{l,s\in i} w_{ls} R_{ls}^{-1}},
\end{equation}
which allows for an unbiased determination of the underlying cluster
lensing profile \citep{Okabe+Smith2016,Sereno2017psz2lens}.
Similarly,  when stacking multiple clusters together, we assume that all
the clusters are at a single effective redshift, which is defined as a
weighted average over the lens--source pairs used in the stacked analysis,
\begin{equation}
\label{eq:zeff}
 \langle z\rangle_\mathrm{wl} = \frac{\sum_i\sum_{l,s\in i}w_{ls}z_{l}}{\sum_i\sum_{l,s\in i} w_{ls}}.
\end{equation}


\begin{figure}[!htb] 
  \begin{center}
   \includegraphics[scale=0.45, angle=0, clip]{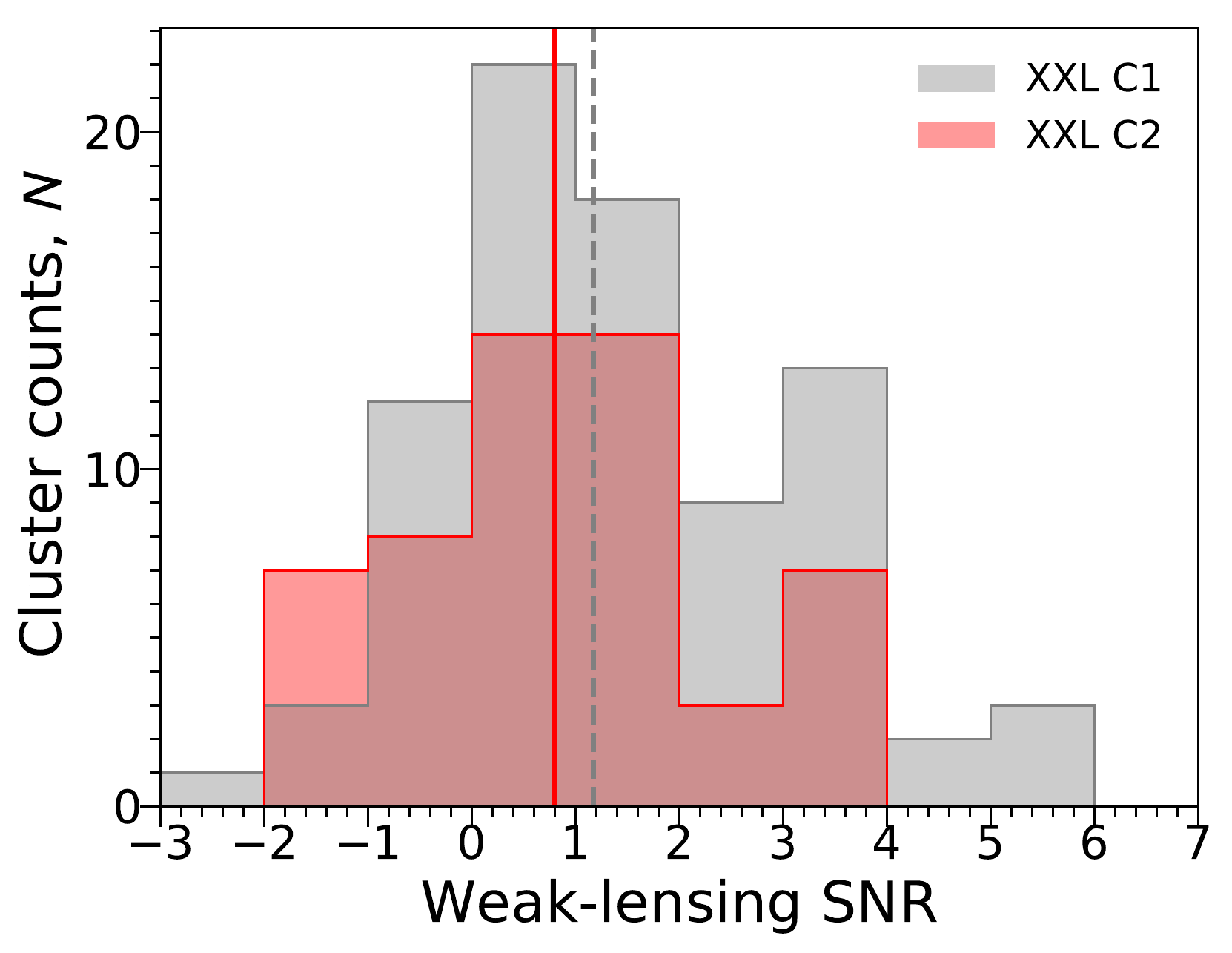} 
  \end{center}
\caption{
 \label{fig:SNR}
Histogram distribution of the weak-lensing SNR, shown separately for the
 C1 (gray) and C2 (red) subsamples. 
The median SNR values of the C1 and C2 subsamples are marked by a
 gray-dashed and a red solid line, respectively.
For the full C1+C2 sample, the observed values of weak-lensing SNR span
the range from $-2.1$ to $6.0$, with a standard deviation of 1.6.
}
\end{figure}

Finally,
to quantify the significance of the shear profile measurement
$\bd = \{\Delta\Sigma_+(R_i)\}_{i=1}^N$
around each individual or stacked cluster,
we define a linear signal-to-noise ratio (SNR) estimator \citep{Sereno2017psz2lens}
by $\mathrm{SNR}=\langle d\rangle/\sigma_{\langle d\rangle}$ with
\begin{equation}
\label{eq:SNR}
  \begin{aligned}
   \langle d\rangle =& \frac{\sum_{i=1}^N
   \Delta\Sigma_+(R_i)/\sigma^2_\mathrm{shape}(R_i)}{\sum_{i=1}^N
   1/\sigma^2_\mathrm{shape}(R_i)},\\
   \sigma_{\langle d\rangle} =& \frac{1}{\sqrt{\sum_{i=1}^N 1/\sigma^2_\mathrm{shape}(R_i)}}, 
  \end{aligned}
\end{equation}
and $\sigma_\mathrm{shape}(R_i)$ the statistical uncertainty in
Equation (\ref{eq:DSigma}) due to the shape noise \citep[e.g.,][]{Miyaoka2018}, 
\begin{equation}
 \label{eq:sigma_stat}
 \sigma_\mathrm{shape}^2(R_i) = \frac{1}{4{\cal R}^2(R_i)[1+K(R_i)]^2
  \sum_{l,s\in i} w_{ls}}.
\end{equation}
This estimator gives a weak-lensing SNR integrated in the
fixed comoving radial range $R\in [0.3,3]\,\Mpch$.
We note that we use the full covariance matrix for our cluster mass
measurements (Section \ref{subsec:cmat}).

This SNR estimator is different from the conventional quadratic
estimator, 
\begin{equation}
 \label{eq:SNRq}
(\mathrm{SNR})_\mathrm{q}\equiv\left[\sum_{i=1}^N (\Delta\Sigma_{+,i})^2/\sigma_{\mathrm{shape},i}^2\right]^{1/2}> 0
\end{equation}
\citep[e.g.,][hereafter \citetalias{2016AA...592A...4L}]{UB2008,Okabe+Smith2016,2016AA...592A...4L}.
As noted by \citet{Umetsu2016clash}, this quadratic definition breaks
down and leads to overestimation of significance in the noise-dominated
regime, in which the actual per-bin SNR is less than unity (see Table 
\ref{tab:clusters}). 

To ensure a statistical ensemble analysis based on weak-lensing
measurements of individual clusters, we require the per-cluster SNR to
be of the order of unity.
Figure \ref{fig:SNR} shows the histogram distributions of the
weak-lensing SNR for the C1 and C2 subsamples.  The median
per-cluster SNR values for the C1 and C2 subsamples are 1.2 and 0.8,
respectively. The median per-cluster SNR of the full (C1+C2) sample is
1.1, so that the above requirement is satisfied.

\subsection{Error Covariance Matrix}
\label{subsec:cmat}

To obtain robust constraints on the mass scaling relation and its
intrinsic scatter,
we need to ensure that the mass likelihood from a weak-lensing
analysis includes all sources of uncertainty 
\citep{Gruen2015}. 
Following \citet{Umetsu2016clash}, we decompose the error covariance
matrix for the binned tangential shear  profile $\bd$ as
\begin{equation}
\label{eq:cmat}
C = C^\mathrm{shape} + C^\mathrm{lss} + C^\mathrm{int},
\end{equation}
where $C^\mathrm{shape}_{ij} = \sigma^2_\mathrm{shape}(R_i)\delta_{ij}$ is
the diagonal statistical uncertainty due to the shape noise
(see Equation (\ref{eq:sigma_stat})), with $\delta_{ij}$ Kronecker's delta;
$C^\mathrm{lss}_{ij}$ is the cosmic noise covariance matrix due to 
uncorrelated large-scale structures projected along the line of sight 
\citep{2003MNRAS.339.1155H};
and $C^\mathrm{int}_{ij}$ accounts for the intrinsic variations of the
projected cluster lensing signal at fixed mass due to variations in halo  
concentration, cluster asphericity, and the presence of correlated halos
\citep{Gruen2015}.\footnote{Strictly speaking, when simultaneously
determining the mass and concentration for a given individual cluster,
the contribution from the intrinsic scatter in the $c$--$M$ relation
should be excluded from $C^\mathrm{int}$. However, for our cluster
sample,
the contribution from  the intrinsic $c$--$M$ variance becomes
important only at $R\simlt 0.3\Mpch$ \citep{Gruen2015}, which is below
the radial range used for our analysis.}
 
We compute the elements of the $C^\mathrm{lss}$ matrix by closely
following the procedure outlined in \citet{Miyaoka2018}
\citep[see also][]{Medezinski2018planck,Miyatake2019actpol}.
To this end, we employ the nonlinear matter power spectrum
of \citet{Smith+2003halofit} for the {\em Wilkinson Microwave Anisotropy  
Probe} (\WMAP) 9\,yr cosmology \citep{Hinshaw+2013WMAP9},
with a source plane at $z_s=1.2$, which closely matches the mean
redshift of the selected background galaxies
\citep{Medezinski2018src}. When stacking 
multiple clusters together, we simply scale the $C^\mathrm{lss}$ matrix
according to the number of independent clusters $\Ncl$ as
$C^\mathrm{lss}\to C^\mathrm{lss}/\Ncl$
\citep[e.g.,][]{Medezinski2018planck}.

We estimate the $C^\mathrm{int}$ matrix for the tangential shear
profile by following \citet[][see their Appendix]{Miyatake2019actpol},
who developed a useful procedure to translate the intrinsic covariance
matrix for the convergence (or $\Sigma$) profile
\citep{Gruen2015,Umetsu2016clash}
to that for the tangential shear (or $\Delta\Sigma$) profile.
In the stacked analysis of multiple independent clusters, we scale the
$C^\mathrm{int}$ matrix as $C^\mathrm{int}\to C^\mathrm{int}/\Ncl$.

As found by \citet{Miyatake2019actpol},
the total uncertainty per cluster is dominated by the shape noise
($C^\mathrm{shape}$) at  $R\simlt 3\Mpch$ (see their Figure 4), beyond  
which the contribution from the cosmic noise ($C^\mathrm{lss}$)
becomes important.
The relative contribution from intrinsic
variance ($C^\mathrm{int}$) increases toward the cluster center but
remains subdominant at all radii for our weak-lensing measurements.

\subsection{Source Galaxy Selection}
\label{subsec:source}

A secure selection of background galaxies is key for obtaining accurate
cluster mass measurements from weak lensing
\citep[e.g.,][]{BTU+05,UB2008,Medezinski+2010,Gruen2014,Okabe+Smith2016,Medezinski2018src}. 
We follow the methodology outlined in
\citet{Medezinski2018src} to select background galaxies for our cluster
weak-lensing analysis.
Two source-selection methods have been tested
and established
in \citet{Medezinski2018src} using the CAMIRA catalog of optically
selected clusters from the HSC survey \citep{Oguri2018camira}: one based
on selection in color-color space (the CC-cut),
and another that employs constraints on the cumulative photo-$z$ PDF
(the $P$-cut). 
Both methods are optimized to
minimize dilution of the lensing signal and
perform comparatively well in removing most of the contamination from
foreground and cluster galaxies  \citep[][]{Medezinski2018src}.
The level of contamination by cluster members depends on and increases
with the cluster mass or richness \citep{Medezinski2018src}.
For our sample that is dominated by low-mass clusters and groups,
we thus expect a less significant degree of dilution of the weak-lensing signal
compared to previous HSC cluster weak-lensing studies
\citep[e.g.,][]{Medezinski2018src,Medezinski2018planck,Miyaoka2018,Miyatake2019actpol,Okabe2019hsc}.

In the present work, we use the $P$-cut method for our fiducial analysis 
because it gives higher SNR values (i.e., higher number densities of
background galaxies) than the CC-cut method.
We use full $P(z)$ data obtained with the Ephor\_AB code
\citep{Tanaka2018photoz,Hikage2019wl} to define the $P$-cut and to
compute the lensing signal (Section \ref{subsec:DSigma}).   
With this method, for each cluster ($l$),
we define a sample of background galaxies ($s=1,2,...$) that satisfy the
following conditions \citep{Oguri2014,Medezinski2018src}:
\begin{equation}
 p_\mathrm{cut} < \int_{z_{\mathrm{min},l}}^\infty\,P_s(z)dz \ \ \
  \mathrm{and} \ \ \ z_{\mathrm{p},s} < z_\mathrm{max},
\end{equation}
where $p_\mathrm{cut}$ is a constant probability set to 0.98,
$z_{\mathrm{min},l}=z_l+\Delta z$ with a constant offset $\Delta z$,
$z_{\mathrm{p},s}$ is a photo-$z$ point estimate for the source galaxy, and
 $z_\mathrm{max}$ is the maximum redshift parameter \citep[see][]{Medezinski2018src}.
 Following \citet{Medezinski2018src}, we set $z_\mathrm{max}=2.5$ and
 adopt $\Delta z=0.2$ for a stringent rejection of cluster and
 foreground galaxies, and we use as $z_\mathrm{p}$ a randomly sampled
 point estimate that is drawn from $P(z)$
 \citep[\texttt{photoz\_mc}; see][]{Tanaka2018photoz,Miyatake2019actpol}.


\begin{figure}[!htb] 
  \begin{center}
   \includegraphics[scale=0.45, angle=0, clip]{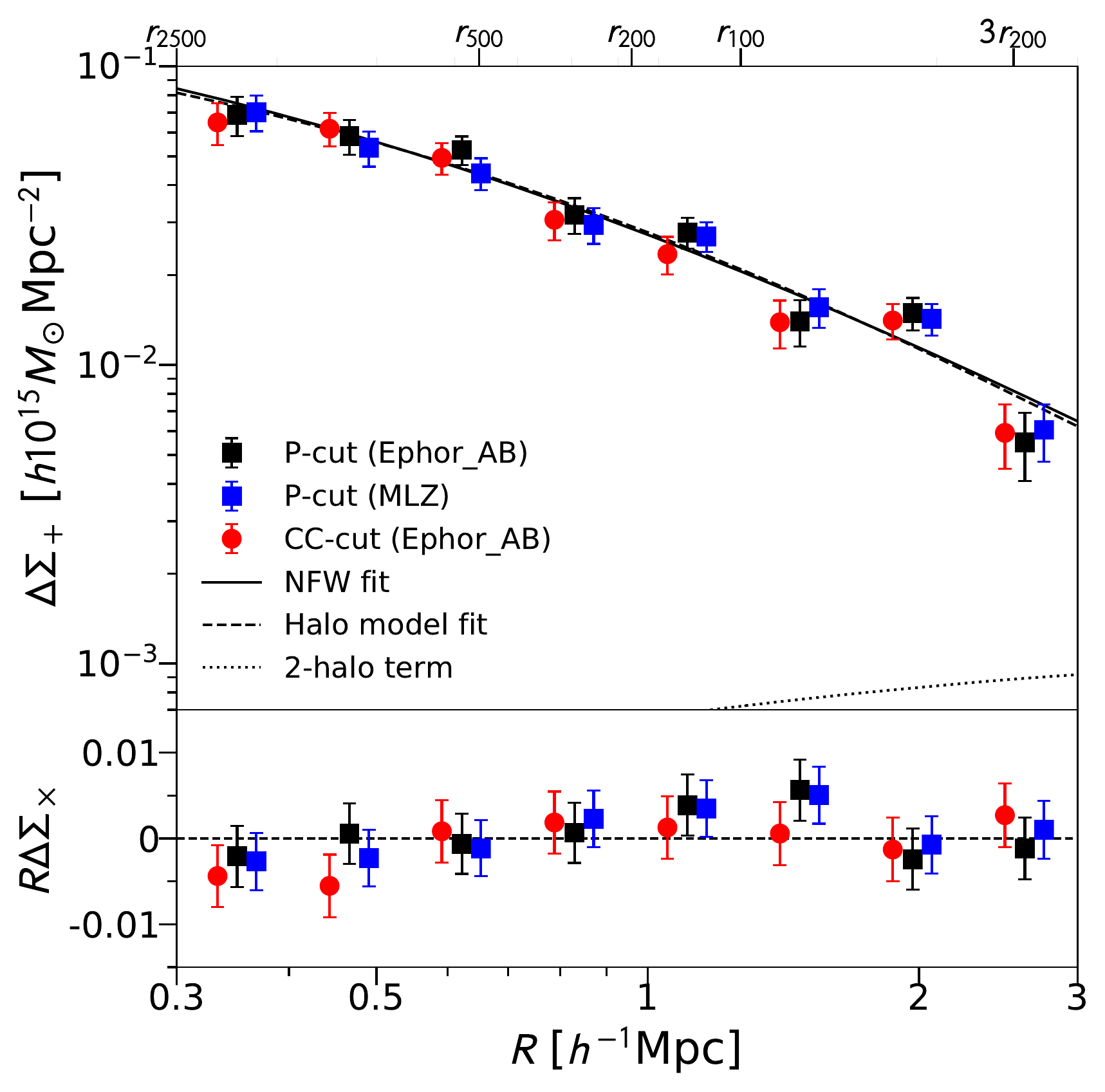} 
  \end{center}
\caption{
Stacked surface mass density of the full C1+C2 sample 
 (upper panel) as a function of cluster-centric comoving radius $R$. The
 results are shown for three different source selection methods.
 The black squares with error bars show our fiducial results obtained
 using the 
 $P$-cut method with the Ephor\_AB photo-$z$ code.
The results obtained using the $P$-cut method with the MLZ code (blue
 squares) and those using the CC-cut method with the Ephor\_AB
 code (red circles) are shown for comparison.
The data points with different selection methods are horizontally
 shifted with each other for visual clarity.
 The solid line and the dashed line represent the best-fit NFW model and the
 halo model (BMO + 2-halo term) derived from the 
 fiducial $P$-cut measurements. The dotted line shows the 2-halo term
 contribution of the best-fit halo model.
The lower panel shows the $45^\circ$-rotated shear component
 $\Delta\Sigma_\times$, expected to be consistent with zero.
 \label{fig:DSigma}
 }
\end{figure}

The top panel of Figure \ref{fig:DSigma} shows the stacked
tangential shear profiles $\Delta\Sigma_+(R)$ obtained for the full sample
using the $P$-cut and CC-cut methods, both with the Ephor\_AB code. 
For comparison, we also show the $P$-cut results obtained with MLZ, an
unsupervised machine-learning method based on self-organizing maps (SOMs)
\citep{Tanaka2018photoz}. 
The comparison shows no significant difference between these profiles
within errors in all bins.

In the bottom panel of Figure \ref{fig:DSigma}, we show the corresponding
stacked B-model profiles $\Delta\Sigma_\times(R)$ (Section
\ref{subsec:DSigma}) obtained with these three selection methods.
Here we use a $\chi^2$ test to assess the
significance of the measured B-model signal against the null
hypothesis. For our fiducial measurement ($P$-cut with Ephor\_AB),
we find $\chi^2=4.73$ per 8 degrees of freedom (dof).
Similarly, we find $\chi^2/\mathrm{dof}=5.30/8$ and
 $\chi^2/\mathrm{dof}=4.88/8$ using the $P$-cut method with MLZ and the
 CC-cut method with Ephor\_AB, respectively.
In all cases, the B-mode signal is statistically consistent with zero.

In what follows, we focus on the
results obtained with the Ephor\_AB code.
In terms of the best-fit NFW mass model (see Section \ref{sec:mass}), we
find a logarithmic mass offset between the $P$-cut and CC-cut methods of
$b_{\mathrm{cont}}\equiv \ln{(M_\mathrm{500,Pcut}/M_{500,\mathrm{CC}})}=(+3.1\pm 5.1)\percent$,
where the error accounts for the covariance between the overlapping
source samples. 
This is consistent with the level of foreground
contamination found by \citet{Medezinski2018src}.
Although we do not find statistical evidence that our $P$-cut method gives
a diluted signal compared to the CC-cut method,
we conservatively assume a systematic mass uncertainty
of $3.1\percent$ 
associated with residual contamination by foreground and cluster galaxies.

\subsection{Photometric Redshift Bias}
\label{subsec:bz}


An accurate estimation of photometric redshifts for source galaxies is
crucial for weak lensing because biased photo-$z$ estimates can lead to
a systematic bias in mass estimates through the calculation of the
critical surface density (see Equation (\ref{eq:Sigma_cr})).  Here we follow
the procedure of \citet{Miyatake2019actpol} to quantify the level of
this bias. For details of the procedure, we refer to
\citet[][see their Section 3.4]{Miyatake2019actpol}.

The photo-$z$ bias in the tangential shear signal of each cluster at
redshift $z_l$ can be estimated as
\citep{Mandelbaum2008photoz,Nakajima2012,Miyatake2019actpol}
\begin{equation}
 \label{eq:photoz_bias}
 \frac{\Delta\Sigma}{\Delta\Sigma^\mathrm{true}}(z_l)=1+b_z(z_l)
  =\frac{\sum_s w_{ls} \langle\Sigma^{-1}_{\mathrm{cr},ls}\rangle^{-1}[\Sigma^\mathrm{true}_{\mathrm{cr},ls}]^{-1}}{\sum_s w_{ls}},
\end{equation}
where the quantities with the superscript ``true'' denote those that
would be measured with an unbiased spectroscopic sample, and the sum
over $s$ runs over all source galaxies.
Ideally, such a photo-$z$ bias should be examined using a
spectroscopic-redshift (spec-$z$) sample that is independent from those
used to calibrate the photo-$z$'s and that matches the population
properties (i.e., magnitude and color distribution) of our source galaxy
sample. In practice, however, it is difficult to obtain such a
representative spec-$z$ sample matching the depth of our source
sample, $i<24.5$\,ABmag \citep{Miyatake2019actpol}.

Following \citet{Miyatake2019actpol},
we use the 2016 version of the 30-band photo-$z$ catalog of the
$2$\,deg$^2$ COSMOS field 
\citep{Ilbert+2009COSMOS,Laigle2016cosmos} as a representative redshift
sample and compute the photo-$z$ bias $b_z$ for a given cluster
redshift, $z_l$. 
As discussed in \citet[][see their Section 5.2]{Hikage2019wl},
there are some caveats associated with this assumption.
We thus use a reweighting method to match the populations between COSMOS
galaxies and our background source galaxies
\citep[for details, see][]{Hikage2019wl,Miyatake2019actpol}. 
The procedure is summarized as follows.
For a given cluster redshift $z_l$,
we define a sample of background source galaxies from the entire shear 
catalog using the $P$-cut method described in Section
\ref{subsec:source}. 
We then decompose source galaxies in the weak-lensing sample using their  
$i$-band magnitude and four colors into cells 
of an SOM \citep[S. More et al. 2019, in preparation; see][]{Masters2015}.
We use a subsample of COSMOS galaxies \citep{Hikage2019wl}\footnote{This
subsample composes $20\percent$ of galaxies in the COSMOS 30-band
catalog, which were not used for training the HSC photo-$z$ codes. We
use this subsample for our testing purposes.}
and classify them into SOM cells
defined by the weak-lensing sample and compute their new weights,
$w_\mathrm{SOM}$,  
such that the weighted distributions of the photometric observables 
match those of the corresponding distributions of the weak-lensing
sample. 
We compute the photo-$z$ bias
(see Equation (\ref{eq:photoz_bias}))
by including $w_\mathrm{SOM}$ in the definition of $w_{ls}$.

For our full sample of 136 XXL clusters,
we find a weighted average of $\langle b_z\rangle \simeq 0.68\percent$.
We find that
our estimate for the average level of photo-$z$ bias is insensitive to the
chosen weighting scheme (e.g., a sample median of $\simeq 0.87\percent$).
The photo-$z$ bias of $\langle b_z\rangle\simeq 0.68\percent$
is translated into the cluster mass uncertainty as
$\langle b_z\rangle/\Gamma_{200}\simeq 0.9\percent$ with
$\Gamma_{200}\equiv d\ln{\Delta\Sigma}/d\ln{M_{200}}\simeq 0.75$,
the typical value of the logarithmic derivative of the weak-lensing signal
with respect to cluster mass for our cluster weak-lensing analysis
\citep{Melchior2017des,Sereno2017psz2lens}. 
Hence, the mass calibration uncertainty due to photo-$z$ calibration
errors is estimated to be $0.9\percent$ (Section \ref{sec:scaling}).
\citet{Miyatake2019actpol} found a similar level of photo-$z$
bias ($2\percent$) for a sample of eight ACTPol-selected SZE clusters
with a median redshift of $z\sim 0.5$.

\section{Weighing XXL Clusters}
\label{sec:mass}

In this section, we use the HSC weak-lensing data to infer the mass and 
concentration parameters for our XXL cluster sample.
In Section \ref{subsec:modeling}, our procedure for weak-lensing mass
modeling is outlined, and the systematic effects in ensemble mass
calibration are discussed on the basis of simulations (Appendix
\ref{appendix:test}). 
In Section \ref{subsec:syst}, we discuss and summarize systematic errors
in ensemble modeling of the XXL sample with weak lensing. Section
\ref{subsec:single} presents our 
weak-lensing mass estimates of individual clusters in 
the XXL sample. Section \ref{subsec:stack} presents the results
of stacked weak-lensing measurements.

\subsection{Mass Modeling}
\label{subsec:modeling}

We model the radial mass distribution of galaxy 
clusters with a spherical NFW profile,
which has been motivated by cosmological $N$-body simulations
\citep[e.g.,][]{1996ApJ...462..563N,1997ApJ...490..493N,Oguri+Hamana2011},
as well as by direct lensing measurements
\citep[e.g.,][]{Umetsu+2012,Umetsu2014clash,Umetsu2016clash,Oguri+2012SGAS,Newman+2013a,Niikura2015,Okabe+Smith2016,Umetsu+Diemer2017}.
The radial dependence of the NFW density profile is given by
\citep{1996ApJ...462..563N}:
\begin{equation}
 \label{eq:NFW}
 \rho(r)=\frac{\rho_\mathrm{s}}{(r/r_\mathrm{s})(1+r/r_\mathrm{s})^2}
\end{equation}
with $\rho_\mathrm{s}$ the characteristic density parameter and
$r_\mathrm{s}$ the characteristic scale radius at which the logarithmic
density slope equals $-2$.
The overdensity mass $M_\Delta$ is given by integrating Equation
(\ref{eq:NFW}) out to the corresponding overdensity radius $r_\Delta$ at
which the mean interior density is $\Delta\times \rho_\mathrm{c}(z_l)$ 
(Section \ref{sec:intro}), and given as $M_\Delta=(4\pi\Delta/3)\rho_{\mathrm{c}}(z_l)r_\Delta^3$.
We specify the NFW model by the mass, $M_{200}$, and the concentration
parameter, $c_{200}=r_{200}/r_\mathrm{s}$.
The characteristic density $\rho_\mathrm{s}$ is then given by
\begin{equation}
 \rho_\mathrm{s}=
\frac{\Delta}{3}
 \frac{c_{\Delta}^3}{\ln(1+c_{\Delta})-c_{\Delta}/(1+c_{\Delta})}\rho_\mathrm{c}(z).
\end{equation}
%

We use a Markov Chain Monte Carlo (MCMC) method to obtain
well-characterized inference of the mass and concentration parameters
from our weak-lensing data \citep{Umetsu2014clash,Umetsu2016clash}.  
We adopt
log-uniform priors for $M_{200}$ and $c_{200}$
(or uniform priors for $\log{M_{200}}$ and $\log{c_{200}}$)
in the range
$10^{12}\le M_{200}/(\Msunh)\le 10^{16}$
and
$1\le c_{200} \le 20$.

We note that it is appropriate to assume a log-uniform prior, instead of
a uniform prior, for a positive-definite quantity, especially when the
quantity spans a wide dynamic range  
\citep[e.g.,][]{Sereno+Covone2013,Umetsu2014clash,Umetsu2016clash,Umetsu2018clump3d,Okabe2019hsc}.
Such a treatment is also self-consistent with  our scaling relation 
analysis, where we work with logarithmic quantities,
$\log{M_{\Delta}}$ and $\log{c_{200}}$
 (Section \ref{sec:scaling}).  
Since the corresponding prior distributions in $M_{200}$ and $c_{200}$
scale as $1/M_{200}$ and $1/c_{200}$,  the choice of their lower bounds
is relatively important.
The chosen priors allow for a sufficiently wide range of mass and
concentration relevant for group--cluster-scale halos
with $10^{13}\Msunh \simlt M_{200}\simlt 10^{15}\Msunh$.
If the lower prior boundary of $M_{200}$ is increased toward the mass
limit of the sample ($M_{200}\sim 10^{13}\Msunh$),
this will lead us to overestimate $M_{200}$ for low-mass groups
and to underestimate the uncertainty of their mass estimates, owing
to the edge effect.

The log-likelihood function for our observations
$\bd=\{\Delta\Sigma_+(R_i)\}_{i=1}^N$ is written as
\begin{equation}
  \begin{aligned}
   -\ln{\cal L}(\bp)
   =& \frac{1}{2}\sum_{i,j=1}^{N}\left[\Delta\Sigma_+(R_i)-f_\mathrm{mod}(R_i|\bp)\right]\\
   &\times (C^{-1})_{ij}\left[\Delta\Sigma_+(R_j)-f_\mathrm{mod}(R_j|\bp)\right] + \mathrm{const.},
  \end{aligned}
\end{equation}
where $C^{-1}$ is the inverse covariance matrix and
$f_\mathrm{mod}(R_i|\bp)$ denotes the theoretical prediction
of the model given a set of parameters $\bp=(M_{200},c_{200})$.
We use analytic expressions given by
\citet{2000ApJ...534...34W}
for the radial dependence of the projected
NFW profiles
$\Sigma_\mathrm{NFW}(R|\bp)$ and
$\Delta\Sigma_\mathrm{NFW}(R|\bp)$,
which provide a good approximation for the projected matter distribution 
around clusters
 \citep{Oguri+Hamana2011}.
The contribution from the 2-halo term to $\Delta\Sigma$
becomes significant at about several virial radii
\citep{Oguri+Hamana2011}, which is larger than the outer
radial limit, $R_\mathrm{max}=3\Mpch$ (see also Section
\ref{subsec:stack}). 
We thus fit the tangential shear profile
$\bd=\{\Delta\Sigma_+(R_i)\}_{i=1}^N$ over the full radial range 
$R\in [0.3,3]\,\Mpch$ in comoving length units. 

Since the relation between the observable image distortion and the
lensing fields is nonlinear (see Equation (\ref{eq:g})), the observed
$\Delta\Sigma$ profile
 is nonlinearly related to the averaged lensing fields.
 Here we use the following approximation
 to include next-to-leading-order corrections \citep{Umetsu2014clash}:
\begin{equation}
 \begin{aligned}
 f_\mathrm{mod}(R_i|\bp) &=
  \frac{\Delta\Sigma_\mathrm{NFW}(R_i|\bp)}{1-\llangle\Sigma_{\mathrm{cr},i}^{-1}\rrangle\times\Sigma_\mathrm{NFW}(R_i|\bp)},
 \end{aligned}
\end{equation}
where $\llangle\Sigma_{\mathrm{cr},i}^{-1}\rrangle$ is the sensitivity-weighted,
inverse critical surface mass density evaluated in each radial bin,
defined by
\begin{equation}
  \llangle\Sigma_{\mathrm{cr},i}^{-1}\rrangle = \frac{\sum_{l,s\in i}
   w_{ls}\langle\Sigma_{\mathrm{cr},ls}^{-1}\rangle}{\sum_{l,s\in i}
   w_{ls}}. 
\end{equation}

As summary statistics, we employ the biweight estimator of
\citet{1990AJ....100...32B} to represent the center location ($\CBI$)
and the scale or spread ($\SBI$) of marginalized one-dimensional
posterior distributions 
\citep[e.g.,][]{Stanford1998,Sereno+Umetsu2011,Biviano2013,Umetsu2014clash,Umetsu2016clash,Umetsu2018clump3d}.
Biweight statistics are insensitive to and stable (robust) against noisy
outliers because they assign higher weights to data points that are
closer to the center of the distribution   
\citep{1990AJ....100...32B}.
For a lognormally distributed quantity,
$\CBI$ approximates the median of the distribution. 
From the posterior samples, we derive marginalized constraints on
the total mass $M_\Delta$
and the concentration $c_\Delta$
at several characteristic interior overdensities $\Delta$.

Our modeling procedure and assumptions
have been tested and validated with simulations.
In Appendix \ref{appendix:test}, we describe the details of tests of our 
``shear-to-mass'' procedure and pipeline.
There are two possible main sources of systematics in an ensemble
weak-lensing analysis of the XXL sample that includes low-mass groups: 
modeling of those groups/clusters detected with low values of  
weak-lensing SNR (Figure \ref{fig:SNR}),  and the modeling
uncertainty due to systematic deviations from the assumed NFW form in
projection. 
To this end, we use two different sets of simulations to assess the
impact of these systematic effects.
To examine the first possibility (Appendix
\ref{appendix:test_lognormal}), we analyze synthetic  
weak-lensing data based on simulations of analytical NFW lenses.
These simulations closely match our weak-lensing observations in terms
of the noise level and the SNR distribution. 
To address the second possibility (Appendix
\ref{appendix:test_BAHAMAS}), we analyze a set of synthetic data 
created from a DM-only realization of BAHAMAS simulations
\citep{McCarthy2017}.

Our simulations show that the overall mass scale of a sample of  
XXL-like clusters can be recovered within $3.3\percent$ accuracy
from individual cluster weak-lensing measurements
(Appendix \ref{appendix:test}).
Specifically, we find the level of mass bias
(see Equation (\ref{eq:bsim})) to be
$b_{\mathrm{sim}, M_{200}}=(2.1\pm 1.5)\percent$ 
and
$b_{\mathrm{sim}, M_{500}}=(0.9\pm 1.3)\percent$
in $M_{200}$ and $M_{500}$, respectively,
with the BAHAMAS simulation (Appendix \ref{appendix:test_BAHAMAS}). 
With synthetic data from simulations of NFW lenses (Appendix   
\ref{appendix:test_lognormal}),
we find
$b_{\mathrm{sim}, M_{200}}=(0.1\pm 2.4)\percent$ 
and
$b_{\mathrm{sim}, M_{500}}=(3.3\pm 2.3)\percent$, with no
systematic dependence on cluster mass over the full range in true
cluster mass (Figure \ref{fig:cM_lognormal}). 

However, the results from the BAHAMAS simulation suggest a significant
level of mass bias of $\sim -20\percent$ for low-mass group 
systems with $\Mtrue\simlt 4\times 10^{13}\Msunh$
(Appendix \ref{appendix:test_BAHAMAS}; see Table \ref{tab:BAHAMAS}).  
Since we do not find any mass-dependent behavior when using the true
density profile assumed in our simulations of NFW lenses, it is likely
that this negative bias is caused by systematic deviations of
``projected'' halos from the NFW profile shape.
In fact, we find such a systematic trend in the outskirts
($1\simlt R/r_{200}\simlt 3$) of
projected $\Delta\Sigma(R)$ profiles around low-mass group-scale halos
selected from DM-only BAHAMAS simulations, whereas their spherically
averaged density profiles $\rho(r)$ in three dimensions are well
described by the NFW form (M. Lieu et al.\ 2020, in preparation).
However, we note that the typical mass measurement uncertainty for
such low-mass groups is $\sigma(M)/M\sim 140\percent$ per cluster
(see Appendix \ref{appendix:test_lognormal}),
and that even when averaging over all such clusters,
the statistical uncertainty on the mean mass is of the order of
$\simgt 20\percent$ (Section \ref{subsec:single}). This level of
systematic bias ($\simlt 1\sigma$) is not expected to significantly
affect our ensemble weak-lensing analysis of the XXL sample.

On the other hand, we find a significant systematic offset in the mean
concentration recovered from weak lensing:  
$b_{\mathrm{sim}, c_{200}}=(-18\pm 2)\percent$ from the BAHAMAS
simulation and $b_{\mathrm{sim}, c_{200}}=(13\pm 3)\percent$
from our simulations of NFW lenses.
This is because the typical scale radius for our sample,
$r_\mathrm{s}\sim 0.25\Mpch$, lies slightly below the radial range for  
fitting, $R\in [0.3,3]\,\Mpch$ (comoving),
and the characteristic profile curvature around $r_\mathrm{s}$ is poorly
constrained by our data.

\subsection{Systematic Uncertainties in Ensemble Modeling}
\label{subsec:syst}

We have accounted for various sources of statistical errors associated
with cluster weak-lensing measurements (Section \ref{subsec:cmat}).
All of these errors are encoded in the total covariance matrix
$C=C^\mathrm{shape}+C^\mathrm{lss}+C^\mathrm{int}$ (see Equation
(\ref{eq:cmat})) of the binned tangential shear profile,
$\bd=\{\Delta\Sigma_+(R_i)\}_{i=1}^N$ (Section \ref{subsec:DSigma}).
We have statistically corrected our tangential shear
measurements for multiplicative and additive residual shear bias
estimated from the dedicated image simulations
\citep[Section \ref{subsec:DSigma};
see][]{Mandelbaum2018sim,Mandelbaum2018shear}.

We have also quantified unaccounted-for sources of systematic errors in 
cluster mass calibration by considering the following effects:
(i) the residual systematic uncertainty in the overall shear
calibration (Section \ref{subsec:DSigma}), $1\percent$;
(ii) dilution of the weak-lensing signal by residual contamination from  
foreground and cluster members (Section \ref{subsec:source}), 
$b_\mathrm{cont}\simeq 3.1\percent$;
(iii) photo-$z$ bias in the $\langle\Sigma_\mathrm{cr}^{-1}\rangle$
estimates (Section \ref{subsec:bz}),
$\langle b_z\rangle/\Gamma\simeq 0.9\percent$;
and
(iv) the systematic uncertainty in the overall mass modeling (Section
\ref{subsec:modeling}),
$b_\mathrm{sim}\simeq 3.3\percent$. 
These systematic errors add up in quadrature to a total systematic
uncertainty of $\simeq \dfmcal$ in the ensemble mass calibration of the 
XXL sample. This level of systematic uncertainty is
below the statistical precision of the current full sample,
$\simeq 9\percent$ at
$M_{200}\sim 9\times 10^{13}\Msunh$ (Table \ref{tab:stack}).
We account for these systematics and marginalize over the mass
calibration uncertainty of $\pm\dfmcal$ in our scaling relation
analyses (Section \ref{sec:scaling}).

Regarding the concentration parameter, since we find systematic errors
of opposite signs from the two sets of simulations (Section
\ref{subsec:modeling}), 
we include a systematic uncertainty of
$\pm\sqrt{(0.18^2+0.13^2)/2} \simeq \pm 16\percent$
(Appendix \ref{appendix:test}) on the normalization of the $c$--$M$ 
relation (Section \ref{subsec:cMR}).
Possible sources of bias are the mass-dependent deviations from the
NFW form, the effect of correlated structure, and noise.
All these are functions of the projected cluster-centric radius, and the
net effect is sensitive to the radial fitting range. 
We note that the level of systematic errors in $c_{200}$ is
below the statistical uncertainty, even for our ensemble measurements of
the XXL sample (see Equation (\ref{eq:cMbest}) and Table
\ref{tab:stack}).

\subsection{Individual Cluster Weak-lensing Analysis}
\label{subsec:single}

In Table \ref{tab:clusters} we list posterior summary statistics
($\CBI\pm \SBI$ and median values) of the mass and concentration
parameters $(c_{200}, M_{200}, M_{500})$ for all individual clusters in
the full C1+C2 sample.

There are 31 clusters whose weak-lensing SNR values are negative as 
dominated by statistical noise fluctuations
\citep[Table \ref{tab:clusters}; see also][]{Sereno2017psz2lens}.
These clusters span a wide range of redshift ($0.044\le z \le 0.953$)
with a median of $0.324$. 
The typical mass uncertainty for these clusters
is $\SBI/\CBI\sim 140\percent$,
so that their mass estimates are consistent with zero.
According to our simulations based on analytical NFW lenses, such low
SNR clusters are distributed over a fairly representative range in true
mass (Appendix \ref{appendix:test_lognormal}; 
see Figures \ref{fig:Mtrue_lognormal} and \ref{fig:SNR_lognormal}).
At a given true mass, it is expected that there is a statistical
counterpart of up-scattered clusters with apparently boosted SNR values
and thus overestimated weak-lensing masses.
In fact, the simulations show that the inclusion of low-SNR clusters does
not significantly bias our ensemble mass measurements at particular mass
scales  (see Figure \ref{fig:cM_lognormal}).
It must be stressed that if one selects a subsample of clusters
according to their weak-lensing SNR values, they are no more
representative of the parent population, and 
such a selection will bias high the weak-lensing mass
estimates at a given X-ray cut, an effect known as the Malmquist bias
\citep[e.g.,][see also Appendix \ref{appendix:test_lognormal}]{CoMaLit5}.

As a robust estimator for the average $M_\Delta$ over a given cluster
sample ($n=1,2,...,\Ncl$), we use geometric means, instead of 
arithmetic means.
An advantage of using this geometric estimator is that
error-weighted geometric means of cluster properties, such as $M_{200}$
and $c_{200}$, are relevant to our scaling relation analysis, where
we work with logarithmic quantities (Section \ref{sec:scaling}).
Specifically, we employ an error-weighted, geometric mean estimator for
the sample average
\citep{Umetsu2014clash,Umetsu2016clash,Okabe+Smith2016}, defined by 
\begin{equation}
\label{eq:geom}
 \langle M_\Delta\rangle_\mathrm{g} :=
 e^{\langle \ln M_\Delta\rangle}
= \exp{\left(
       \frac{\sum_{n=1}^{\Ncl} u_n \ln{M_{\Delta,n}}}{\sum_n u_n}
      \right)}
\end{equation}
and its uncertainty,
\begin{equation}
 \begin{aligned}
 \label{eq:dgeom}
\sigma_{\langle M_\Delta\rangle_\mathrm{g}}
&=
  \frac{1}{2} \langle M_\Delta\rangle_\mathrm{g}\\
\times& \left[
\exp\left(\frac{1}{\sqrt{\sum_{n=1}^{\Ncl} u_n}}\right)
-
\exp\left(-\frac{1}{\sqrt{\sum_{n=1}^{\Ncl} u_n}}\right)
\right],
\end{aligned}
\end{equation}
where $u_n$ is the inverse variance weight for the $n$th cluster,
$u_n^{-1}=\sigma^2(M_{\Delta,n})/M_{\Delta,n}^2$,
with $M_{\Delta,n}$ and $\sigma(M_{\Delta,n})$
being $\CBI$ and $\SBI$ (Section \ref{subsec:modeling}), respectively,
of the marginalized posterior distribution of
$M_\Delta$
for the $n$th cluster. 
The geometric means are symmetric with respect to an exchange of the
numerator and denominator
(i.e., $\langle A/B\rangle_\mathrm{g}=\langle B/A\rangle_\mathrm{g}^{-1}$),
so that this weighted geometric estimator is also suitable for use in
estimating mean mass ratios between two cluster samples
\citep{Donahue2014clash, Umetsu2014clash, Umetsu2016clash}.

Using this estimator, we find weighted geometric means of
$\langle M_{200}\rangle_\mathrm{g}=(9.8\pm 0.8)\times 10^{13}\Msunh$,
$\langle M_{200}\rangle_\mathrm{g}=(11.6\pm 1.2)\times 10^{13}\Msunh$,
and
$\langle M_{200}\rangle_\mathrm{g}=(6.5\pm 1.0)\times 10^{13}\Msunh$
for the C1+C2, C1, and C2 samples, respectively (Table \ref{tab:stack}).

\subsection{Stacked Weak-lensing Analysis}
\label{subsec:stack}

Stacking an ensemble of clusters helps average out large statistical
fluctuations inherent in noisy weak-lensing measurements of individual
clusters (Section \ref{subsec:cmat}).
The statistical precision can be greatly improved by stacking together a
large number of clusters, allowing for tighter and more robust
constraints on the cluster mass distribution.
A stacked analysis is complementary to our primary approach based
on individual weak-lensing mass measurements. A comparison of the two
approaches thus provides a useful consistency check in different SNR
regimes. 
It is noteworthy, however, that interpreting the effective mass from
stacked lensing requires caution because the amplitude of the lensing
signal is weighted by the redshift-dependent sensitivity
\citep{Umetsu2016clash} and is not linearly proportional to the cluster
mass \citep{Mandelbaum2005halomodel,Melchior2017des,Sereno2017psz2lens,Miyatake2019actpol}.

\begin{deluxetable*}{cccccccccccc}
\tablecolumns{12}
\tablewidth{0pt}
\tabletypesize{\scriptsize}
\tablecaption{\label{tab:tstack}
Characteristics of the $T_\mathrm{X}$-binned Subsamples}
\tablehead{
\multicolumn{1}{c}{Bin} & 
\multicolumn{1}{c}{$N_\mathrm{cl}$} & 
\multicolumn{1}{c}{$T_\mathrm{X}$} & 
\multicolumn{1}{c}{$\langle T_\mathrm{X}\rangle_\mathrm{wl}$} & 
\multicolumn{1}{c}{$\zmed$} & 
\multicolumn{1}{c}{$\langle z\rangle_\mathrm{wl}$} & 
\multicolumn{1}{c}{$c_{200}$} & 
\multicolumn{1}{c}{$M_{200}$} & 
\multicolumn{1}{c}{$\langle M_{200}\rangle_\mathrm{wl}$} & 
\multicolumn{1}{c}{$\langle M_{200}\rangle_\mathrm{g}$} & 
\multicolumn{1}{c}{SNR} & 
\multicolumn{1}{c}{$($SNR$)_\mathrm{q}$} \\
\colhead{} & 
\colhead{} & 
\multicolumn{1}{c}{(keV)} & 
\multicolumn{1}{c}{(keV)} & 
\colhead{} & 
\colhead{} & 
\colhead{} & 
\multicolumn{1}{c}{($10^{13}h^{-1}M_\odot$)} & 
\multicolumn{1}{c}{($10^{13}h^{-1}M_\odot$)} & 
\multicolumn{1}{c}{($10^{13}h^{-1}M_\odot$)} & 
\colhead{} & 
\colhead{} 
}
\startdata
T1 & $22$ & $1.1$ & $1.0$ & $0.18$ & $0.18$ & $5.7\pm 4.6$ & $4.5\pm 1.2$ & $4.8\pm 1.4$ & $4.1\pm 1.0$ & $5.3$ & $7.9$\\
T2 & $21$ & $1.6$ & $1.5$ & $0.29$ & $0.22$ & $3.2\pm 1.9$ & $8.3\pm 1.9$ & $6.5\pm 1.7$ & $7.9\pm 1.8$ & $6.9$ & $9.1$\\
T3 & $17$ & $1.9$ & $1.9$ & $0.30$ & $0.29$ & $3.0\pm 1.4$ & $12.6\pm 2.7$ & $9.5\pm 2.2$ & $13.6\pm 3.0$ & $6.6$ & $9.5$\\
T4 & $19$ & $2.4$ & $2.4$ & $0.33$ & $0.31$ & $2.0\pm 1.0$ & $11.3\pm 3.0$ & $8.0\pm 2.4$ & $6.7\pm 1.8$ & $6.2$ & $7.9$\\
T5 & $17$ & $3.5$ & $3.4$ & $0.43$ & $0.33$ & $4.3\pm 2.4$ & $19.8\pm 4.1$ & $19.6\pm 4.4$ & $20.2\pm 4.1$ & $8.7$ & $10.4$\\
T6 & $9$ & $5.0$ & $5.1$ & $0.51$ & $0.31$ & $5.9\pm 3.0$ & $25.8\pm 5.3$ & $22.6\pm 5.2$ & $25.8\pm 5.7$ & $7.3$ & $10.2$
\enddata
\tablecomments{The definitions of the columns are the same as in Table \ref{tab:stack}.}
\end{deluxetable*}



\begin{figure}[!htb] 
  \begin{center}
   \includegraphics[scale=0.45, angle=0, clip]{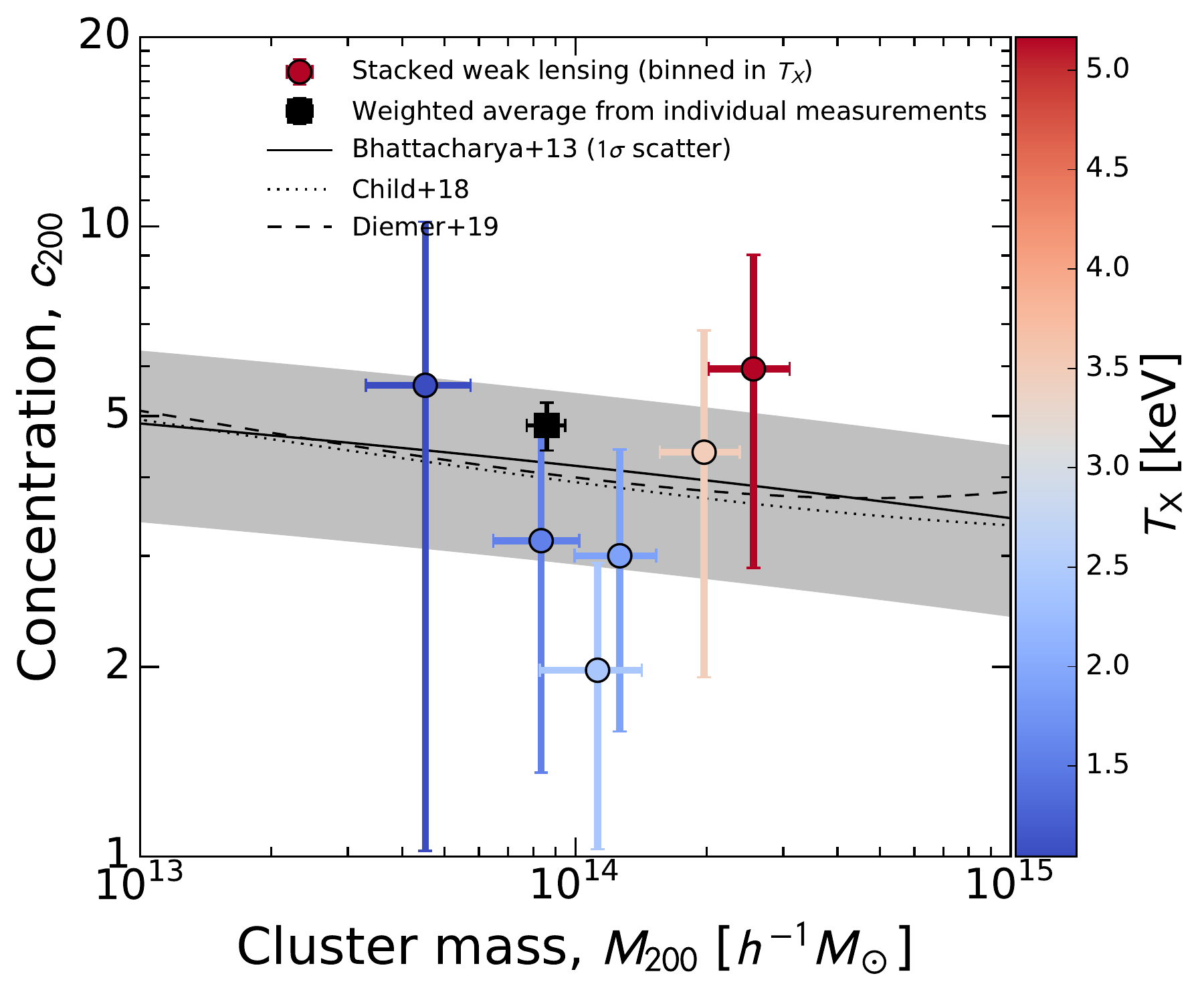} 
  \end{center}
\caption{
Stacked weak-lensing constraints on the NFW concentration and mass
 parameters (circles with error bars)
 for six subsamples of our XXL clusters (see Table \ref{tab:tstack})
 binned in X-ray temperature. This analysis is limited to 105 C1+C2
 clusters with measured X-ray temperatures $\Tx$ from the XXL survey.
The X-ray temperature of the data points is color-coded according to the
 color bar on the right side.
The black square with error bars shows the weighted average of individual
 weak-lensing measurements over the sample of 105 XXL clusters. 
The results are compared to theoretical $c$--$M$
 relations evaluated at $z=0.3$ for the full population of DM halos from 
 numerical simulations of $\Lambda$CDM cosmologies
 \citep{Bhatt+2013,Child2018cm,Diemer+Joyce2019}.
The gray shaded region represents the lognormal intrinsic dispersion
 $\sigma(\ln{c_{200}})=1/3$ around the $c$--$M$ relation of   
 \citet{Bhatt+2013}. 
 \label{fig:cM_Tbins}
 }
\end{figure}

First, we examine the effective mass and concentration parameters of the
full C1+C2 sample of 136 XXL clusters from the stacked
$\Delta\Sigma$ profile shown in Figure \ref{fig:DSigma} (fiducial).
The lensing-weighted mean redshift of the full sample is
$\langle z\rangle_\mathrm{wl}\simeq 0.25$, which is smaller than
the sample median redshift, $\zmed=0.30$.
From a single-mass-bin NFW fit to the stacked $\Delta\Sigma$ profile 
(see Section \ref{sec:mass}), we obtain
$M_{200}=(8.7\pm 0.8)\times 10^{13}\Msunh$ and
$c_{200}=3.5\pm 0.9$ for the C1+C2 sample.
This is in agreement with the degree
of concentration expected for DM halos in the standard $\Lambda$CDM
cosmology, $c_{200}\simeq 4.1$ at $M_{200}=8.7\times 10^{13}\Msunh$ and
$z=0.25$ \citep{Diemer+Kravtsov2015,Diemer+Joyce2019}.
The effective mass and concentration parameters for the C1+C2, C1, and
C2 samples are summarized in Table \ref{tab:stack}.

In Figure \ref{fig:DSigma}, we also show the best-fit two-parameter halo
model including the effects of surrounding large-scale structure as a
2-halo term. Here we follow the standard halo model prescription
of \citet{Oguri+Hamana2011} using the linear halo bias
$b_\mathrm{h}(M_{200};z)$ of \citet{Tinker+2010} in a \WMAP\ 9\,yr
based $\Lambda$CDM cosmology \citep{Hinshaw+2013WMAP9}.  
The 2-halo term contribution to the $\Delta\Sigma(R)$ profile in
comoving length units is expressed as
\begin{equation}
 \label{eq:2ht}
 \Delta\Sigma_\mathrm{2h}(R) =
  \frac{\rho_\mathrm{m}(z)
  b_\mathrm{h}(M_{200};z)}{(1+z)^3 d_\mathrm{A}^2(z)}
  \int\!\frac{ldl}{2\pi}\,J_2(l\theta)P(k_l; z),
\end{equation}
where $\rho_\mathrm{m}(z)$ is the mean matter density of the universe at   
the cluster redshift $z$,
$d_\mathrm{A}(z)$ is the comoving angular diameter distance,
$P(k;z)$ is the linear matter power spectrum,
$k_l\equiv l/d_\mathrm{A}(z)$,
$\theta\equiv R/d_\mathrm{A}(z)$, and
$J_n$ is the Bessel function of the first kind and $n$th order.
The 2-halo term is proportional to the product
$b_\mathrm{h}\sigma_8^2$.

As demonstrated in Figure \ref{fig:DSigma}, the 2-halo term 
$\Delta\Sigma_\mathrm{2h}(R)$ in the radial range $R\in [0.3,3]\,\Mpch$
is negligibly small, even in low-mass groups  
\citep[see][]{Leauthaud2010cosmos,Covone2014,Sereno2015s8,Sereno2017psz2lens}.
This is because the tangential shear, or the excess surface mass density
$\Delta\Sigma(R)=\Sigma(<R)-\Sigma(R)$, is insensitive to
flattened sheet-like structures \citep{Schneider+Seitz1995}. 
When the 2-halo term is neglected, the standard halo model 
reduces to the Baltz--Marshall--Oguri \citep[][BMO]{BMO} model that
describes a smoothly truncated NFW profile \citep[][see their Section
5.2.2]{Umetsu2016clash}.  
Using synthetic weak-lensing data based on the DM-only BAHAMAS
simulation (Appendix \ref{appendix:test_BAHAMAS}), we find that the
standard halo modeling does not significantly improve the accuracy of
weak-lensing mass estimates for a sample of XXL-like objects (see Table 
\ref{tab:BAHAMAS}).

As a consistency check of our ensemble weak-lensing analysis, we compare
the stacked lensing constraints on $M_{200}$ with those from individual 
cluster measurements (see Section \ref{subsec:single}).
It is reassuring that the effective $M_{200}$ masses extracted from the
stacked $\Delta\Sigma$ profiles are in good agreement with the
respective weighted geometric means
$\langle M_{200}\rangle_\mathrm{g}$ obtained from individual cluster 
mass estimates (see Table \ref{tab:stack}).  
Alternatively, we can estimate the average mass by using the lensing
weight to be consistent with the stacked weak-lensing analysis
\citep[see Equation
\ref{eq:DSigma};][]{Umetsu2016clash,Medezinski2018planck,Miyatake2019actpol}
as
\begin{equation}
\label{eq:wlmean}
\langle M_{\Delta}\rangle_\mathrm{wl}=(\sum_{l,s}w_{ls})^{-1}\sum_l M_{\Delta,l}\sum_s w_{ls}.
\end{equation}
Incorporating the lens weighting, we find
$\langle M_{200}\rangle_\mathrm{wl}=(8.0\pm 0.8)\times 10^{13}\Msunh$,
$(9.0\pm 1.0)\times 10^{13}\Msunh$,
and
$(6.1\pm 1.1)\times 10^{13}\Msunh$
for the C1+C2, C1, and C2 samples, respectively, all consistent with the 
results from the stacked analysis within the errors (see Table \ref{tab:stack}).
This agreement suggests that those clusters detected with low values of
weak-lensing SNR are not biasing the ensemble-averaged mass with respect
to the stacked weak-lensing analysis.


Next, we perform a stacked analysis by dividing the full sample into 6 
subsamples with roughly equal numbers (except for the highest
temperature bin) according to the X-ray temperature, $\Tx$. 
This analysis is limited to 105 clusters with measured X-ray
temperatures $\Tx$ from the XXL survey
(Section \ref{subsec:xxl}).
This subsample has a weighted average mass of 
$\langle M_{200}\rangle_\mathrm{g}=(8.6\pm 0.9)\times 10^{13}\Msunh$
and a weighted average concentration of
$\langle c_{200}\rangle_\mathrm{g}=4.8\pm 0.4 \,(\mathrm{stat.})\pm 0.8\,(\mathrm{syst.})$
(Figure \ref{fig:cM_Tbins}).

The results of stacked weak-lensing measurements are summarized in Table  
\ref{tab:tstack}.
These subsamples have similar SNR values, ranging from 5.3 to 8.7,
with a median of 6.7.
For each subsample, we derive $(M_{200},c_{200})$ 
from a single-mass-bin fit to the stacked $\Delta\Sigma$
profile.
The mass extracted from the stacked lensing signal ranges from
$M_{200}=(4.5\pm 0.9)\times 10^{13}\Msunh$ 
at $\Tx\simeq 1.0$\,keV
to
$M_{200}=(2.6\pm 0.3)\times 10^{14}\Msunh$ 
at $\Tx\simeq 5.2$\,keV.
The effective $M_{200}$ mass extracted from the stacked analysis and the
corresponding lensing-weighted mass $\langle M_{200}\rangle_\mathrm{wl}$
from individual cluster measurements
are consistent within the errors in all $\Tx$ bins
(Table \ref{tab:tstack}).
Overall, these lensing-weighted mass estimates are in agreement with the
error-weighted geometric means $\langle M_{200}\rangle_\mathrm{g}$ from
individual cluster mass estimates (Tables \ref{tab:stack} and
\ref{tab:tstack}).

In Figure \ref{fig:cM_Tbins} we show the distribution of
$(M_{200},c_{200})$ for the six subsamples along with with theoretical
predictions for the full population of $\Lambda$CDM halos
\citep{Bhatt+2013,Child2018cm,Diemer+Joyce2019}. All these models are
evaluated at a reference redshift of $\zref=0.3$ and designed  
for a qualitative comparison and a consistency check only (see Table
\ref{tab:tstack}). 
The average X-ray temperature of each subsample is color-coded according
to the color bar on the right side. 
Figure \ref{fig:cM_Tbins} shows that $M_{200}$ correlates well with
$\Tx$, and that $c_{200}$ is scattered
around the theoretical $c$--$M$ relations, with no hint of
significant overconcentration for the XXL sample. A complete regression
analysis of the $c$--$M$ relation, accounting for various statistical
effects, 
is given in Section \ref{subsec:cMR}.

\section{XXL Mass Scaling Relations}
\label{sec:scaling}

In this section we examine and characterize the concentration--mass
($c_{200}$--$M_{200}$) and temperature--mass ($\Tx$--$M_{500}$) scaling
relations separately for the XXL sample using our HSC and XXL data
products presented in the previous sections.

\subsection{Bayesian Regression Scheme}
\label{subsec:regression}

Here we outline the Bayesian regression scheme of \citet{Sereno2016lira}
used in our scaling relation analysis.
Our regression approach allows for a self-consistent treatment of
redshift evolution, intrinsic scatter, and selection effects through
Bayesian population modeling of the cluster sample.
For full details of the formalism, we refer the reader to
\citet{Sereno2016lira} and the companion paper by
\citet{Sereno2020xxl}. 

In this analysis, we use the publicly available \texttt{LIRA} package
\citep{Sereno2016lira,Sereno2016lirapackage}.
We have tested and validated our analysis procedure and its
\texttt{LIRA} implementation by performing a regression analysis of the 
$c_{200}$--$M_{200}$ relation using realistic synthetic data based on the
DM-only BAHAMAS simulation (see Appendix \ref{appendix:test_BAHAMAS}).
We find that we can accurately recover the true (input) parameters of
the $c_{200}$--$M_{200}$ relation except for the normalization, which is 
subject to a systematic offset (see Section \ref{subsec:syst} and
Appendix \ref{appendix:test}).

\subsubsection{Mass Scaling Relations}

We consider a power-law function of the following form that describes
the average mass scaling relation of a given cluster observable ${\cal O}$:
\begin{equation}
{\cal O} \propto 10^\alpha M_\Delta^\beta F_z(z)^\gamma,
\end{equation}
where
$\alpha$, $\beta$, and $\gamma$ denote the normalization, mass trend,
and redshift trend, respectively;
$F_z(z)$ describes the redshift evolution of the scaling
relation and is normalized to unity at a
reference redshift, $\zref$.
In this work, we consider
$F_z(z)= (1+z)/(1+\zref)$ for the $c_{200}$--$M_{200}$ relation
\citep[e.g.,][]{Duffy+2008,Dutton+Maccio2014}
and
$F_z(z)= E(z)/E(\zref)$ for the $\Tx$--$M_{500}$ relation
\citep[e.g.,][]{Vikhlinin2009cccp2,Ettori2015,Mantz2016}. 
In what follows, we set $\zref=0.3$.

We focus on the logarithms of quantities that describe global cluster
properties of interest.  These logarithmic quantities are then linearly
related to each other. 
We consider the cluster mass $M_\Delta$ as the most fundamental property
of galaxy clusters and define the corresponding logarithmic quantity as
\begin{equation}
 Z=\log{\left(\frac{M_\Delta}{M_{\Delta,\mathrm{pivot}}}\right)}
\end{equation}
with
$M_{\Delta,\mathrm{pivot}}$ the pivot in the $M_\Delta$ mass.
We use the weak-lensing mass $M_{\Delta,\mathrm{wl}}$ as a mass proxy
and introduce the logarithmic weak-lensing mass,
\begin{equation}
 X=\log{\left(\frac{M_{\Delta,\mathrm{WL}}}{M_{\Delta,\mathrm{pivot}}}\right)}.
\end{equation}

For a regression analysis of the $c_{200}$--$M_{200}$ relation, we
choose the pivot in $M_{200}$ to be $M_{200,\mathrm{pivot}}=10^{14}\Msunh$ and
define the logarithmic observable,
\begin{equation}
 Y = \log{c_{200}}.
\end{equation}
For the $\Tx$--$M_{500}$ relation, we set
$M_{500,\mathrm{pivot}}=7\times 10^{13}\Msunh=10^{14}\Msun$ and define
\begin{equation}
 Y = \log{\left(\frac{\Tx}{1\,\mathrm{keV}}\right)}.
\end{equation}

For any observable cluster property,
we distinguish the following three quantities:
(i) $Y_Z$, the quantity that is exactly linked to $Z$ through a
deterministic functional relation $Y_Z(Z)$ \citep{Maughan2014};
(ii) $Y$, a scattered version of $Y_Z$;
and
(iii) $y$, a measured realization of $Y$ that includes observational
noise. 
As defined, $Y$ is intrinsically scattered with respect to $Y_Z$, which
we may express as $Y=Y_Z(Z) \pm \sigma_{Y|Z}$,
with $\sigma_{Y|Z}$ the intrinsic dispersion of $Y$ at fixed cluster
mass or $Z$. 

To proceed, we assume that the weak-lensing mass ($X$) is an unbiased
but scattered proxy of the true cluster mass ($Z$). 
The mass scaling relations $Y_Z(Z)$ and $X_Z(Z)$ are then expressed as 
\begin{eqnarray}
 \label{eq:YZ}
  Y_Z &=& \alpha_{Y|Z} + \beta_{Y|Z} Z + \gamma_{Y|Z}\log{F_z(z)},\\
 \label{eq:XZ}
  X_Z &=& Z,
\end{eqnarray}
where $\alpha_{Y|Z}$, $\beta_{Y|Z}$, and $\gamma_{Y|Z}$ are the
intercept, mass-trend, and redshift-trend parameters, respectively. 
We may rewrite Equation (\ref{eq:XZ}) as $X=Z\pm \sigma_{X|Z}$ with $\sigma_{X|Z}$ the
intrinsic dispersion of $X$ at fixed $Z$.

\subsubsection{Mass Calibration Uncertainty}
\label{subsubsec:mcal}

Any mass calibration bias (i.e.,
$Z_X=\alpha_{Z|X}+X$ with $\alpha_{X|Z}\ne 0$) 
can lead to a biased estimate of the normalization of the scaling
relation, $\alpha_{Y|Z}$. 
We assume a zero-centered Gaussian prior on $\alpha_{X|Z}$ of
$\alpha_{X|Z} = \pm \dfmcal/\ln{10}$ to 
marginalize over the remaining mass calibration uncertainty of
$\pm\dfmcal$ (see Section \ref{subsec:syst}).

\subsubsection{Measurement Errors}

The measured quantities $x$ and $y$ are noisy realizations of the latent
variables $X$ and $Y$, respectively.
We assume that the measurement errors
for the two cluster observables ($X,Y$) follow a bivariate Gaussian
distribution \citep{Sereno2016lira}. 

In the XXL survey, the X-ray temperature $\Tx$ was measured in a
fixed aperture of 300\,kpc
\citepalias{2016AA...592A...2P}.
The errors in the X-ray temperature $\Tx$ and the weak-lensing mass
$M_{\Delta,\mathrm{WL}}$ are thus independent of each other.

On the other hand, for a given
cluster, the measurement errors between the NFW parameters are
correlated (Section \ref{subsec:modeling}). For the regression of
the $c_{200}$--$M_{200}$ relation, we thus compute the error covariance
matrix of the $(\log{\Mwl},\log{c_{200}})$ parameters using the
MCMC posterior samples (see Section \ref{subsec:modeling}) and account
for the covariance between the two parameters \citep[e.g.,][]{Umetsu2014clash,Umetsu2016clash,Okabe+Smith2016}.
We note that the correlation coefficient
$r_\mathrm{CC}(\log{\Mwl},\log{c_{200}})$
between the two NFW parameters is close to zero on average
$(\langle r_\mathrm{CC}\rangle \sim -0.03)$
for clusters with noisy weak-lensing measurements ($\mathrm{SNR}<1$), so
that the two parameters are nearly independent in the low-SNR
regime. The correlation coefficient becomes more negative with
increasing weak-lensing SNR  
(and with increasing $\Mwl$; see the right panel of Figure
\ref{fig:SNR_lognormal}), reaching $r_\mathrm{CC}\sim -0.6$ at
$\mathrm{SNR} \sim 5$.

\subsubsection{Intrinsic Scatter}

The true cluster properties ($X,Y$), which one would
 measure in a hypothetical noiseless experiment, are intrinsically
 scattered with respect to ($X_Z,Y_Z$) \citep{Sereno2016lira}. 
 We assume that the intrinsic scatter of the true quantity ($X$ or $Y$)
 around its model prediction ($X_Z$ or $Y_Z$) at fixed $Z$
follows a Gaussian distribution. For a given observable--mass relation
(i.e., $c_{200}$--$M_{200}$ or $\Tx$--$M_{500}$),
we have two intrinsic dispersion parameters, $\sigma_{Y|Z}$ and
$\sigma_{X|Z}$, which are assumed to be constant with mass and redshift.

\subsubsection{Intrinsic Distribution and Selection Effects}

A proper modeling of the mass probability distribution $P(Z)$
  is crucial. Cluster samples are usually biased
  with respect to the underlying parent population (i.e., the mass
  function) because clusters are selected according to their observable 
  properties. Moreover, even in absence of selection effects, the parent 
  population is not uniformly distributed in logarithmic 
  mass $Z$, which can cause tail effects 
  \citep[e.g.,][]{Kelly2007}. 

The intrinsic distribution of the selected clusters
is mainly shaped by the following two effects: first, as predicted  
by the mass function, more massive objects are rarer;
second, less massive objects are typically fainter and more
difficult to detect. Accordingly, the resulting mass probability
distribution tends to be unimodal, and it evolves with redshift 
\citep{CoMaLit4}.

The combined evolution of the completeness and the mass function can
be modeled through the evolution of the mean and dispersion of the
effective mass probability distribution.
In general, the intrinsic mass probability distribution $P(Z)$ of
the selected clusters can be approximated with a mixture of
time-evolving Gaussian functions \citep{Kelly2007,CoMaLit2,CoMaLit4}.

We properly account for these effects and Eddington bias
in Bayesian regression.
In this work, we model the intrinsic probability distribution $P(Z)$ of
the selected sample with a time-evolving single Gaussian function
characterized by the mean $\mu_Z(z)$ and the dispersion $\sigma_Z(z)$.
In general, this treatment provides a good approximation for a regular
unimodal distribution \citep{Kelly2007,Andreon+Berge2012,CoMaLit4,Sereno2016lira}.
It should be stressed that modeling of $P(Z)$ as a Gaussian is to
account for the effect of the XXL selection that depends primarily on
the flux and the extent of the X-ray emission. Such a statistical
treatment is needed even though the parameters involved in the
regression, ($c_{200}, \Tx, M_{\Delta}$), are not directly influencing 
the XXL selection.

We parameterize the time-evolving mean and dispersion of $P(Z)$ as
\citep{Sereno2016lira}  
\begin{equation}
   \begin{aligned}
    \mu_Z(z) =&  \mu_{Z,0} 
    + \gamma_{\mu_Z,{\cal D}}\log {\cal D}(z),\\
    \sigma_Z(z) =& \sigma_{Z,0} {\cal D}(z)^{\gamma_{\sigma_Z, {\cal D}}},
   \end{aligned}   
\end{equation}
where 
${\cal D}(z)=D_\mathrm{L}(z)/D_\mathrm{L}(\zref)$, with
$D_\mathrm{L}$ the luminosity distance at redshift $z$;
$\mu_{Z,0}$ is the local mean at the reference redshift $\zref$;
$\gamma_{\mu_Z,{\cal D}}$ describes the redshift trend of the mean
function;
$\sigma_{Z,0}$ is the local dispersion at the reference redshift
$\zref$; and
$\gamma_{\sigma_Z, {\cal D}}$ describes the redshift trend of the
dispersion function.
In this modeling, we might expect $\mu_z(z)$ to exhibit some positive 
evolution ($\gamma_{\mu_Z,{\cal D}}>0$), 
reflecting the fact that the characteristic cluster mass will increase
as the X-ray selection excludes less massive clusters at higher
redshifts.

\subsubsection{Priors}
\label{subsubsec:priors}

Bayesian statistical inference requires an explicit declaration
of the chosen prior distributions.
In our regression analysis, we have a total of nine regression
parameters,
\begin{equation}
(\alpha_{Y|Z},
\beta_{Y|Z}, \gamma_{Y|Z}, \sigma_{Y|Z}, \sigma_{X|Z}, \mu_{Z,0},
\gamma_{\mu_Z,{\cal D}}, \sigma_{Z,0}, \gamma_{\sigma_Z,{\cal D}}),
\end{equation}
and one calibration nuisance parameter, $\alpha_{X|Z}$, for which we
assume a zero-centered Gaussian prior (Section
\ref{subsubsec:mcal}). 
In the \texttt{LIRA} approach, we choose to assume sufficiently
noninformative priors for all regression parameters
\citep[for details, see][]{CoMaLit1,Sereno2016lira}.

First, the priors on the intercepts $\alpha_{Y|Z}$ and on the mean
$\mu_{Z,0}$ are uniform,
\begin{equation}
 \alpha_{Y|Z},\, \mu_{Z,0} \sim {\cal U}(-1/\epsilon,+1/\epsilon),
\end{equation}
where $\epsilon$ is a small number, which is set to $\epsilon=10^{-4}$.

Next, for the mass-trend and redshift-trend parameters ($\beta,\gamma$),
we consider uniformly distributed direction angles, $\arctan{\beta}$ and
$\arctan{\gamma}$ and model the prior probabilities as a Student's $t_1$
distribution  with one degree of freedom,
\begin{equation}
 \beta_{Y|Z},\, \gamma_{Y|Z},\, \gamma_{\mu_Z, {\cal D}},\,
  \gamma_{\sigma_Z,{\cal D}} \sim t_1.
\end{equation}

Finally, a noninformative prior on the dispersion $\sigma (>0)$ should
have a very long tail to large values. This can be achieved with the nearly
scale-invariant Gamma distribution $\Gamma$ for the inverse of the variance,
\begin{equation}
\label{eq:Pr_Gamma} 
 1/\sigma^2_{Z,0} \sim \Gamma(\epsilon,\epsilon).
\end{equation}

For the analysis of the $c_{200}$--$M_{200}$ relation, we choose to fix
the value of  $\gamma_{\sigma_Z,D}$ to zero (i.e.,
$\sigma_Z(z)=\mathrm{const.}$) because it is poorly constrained by the
weak-lensing data alone and is highly degenerate with other regression parameters. We checked that this
simplification does not significantly affect our regression results.

\subsection{Concentration--Mass Relation}
\label{subsec:cMR}

\subsubsection{Regression Results}
\label{subsubsec:cM_results} 

\begin{deluxetable*}{ccccccccccc}
\tablecolumns{11}
\tablewidth{0pt}
\tabletypesize{\scriptsize}
\tablecaption{\label{tab:cM}
Summary Statistics of Regression Parameters for the XXL Concentration--Mass Relation}
\tablehead{
\multicolumn{1}{c}{Sample} & 
\multicolumn{1}{c}{$N_\mathrm{cl}$} & 
\multicolumn{1}{c}{$\alpha_{Y|Z}$} &
\multicolumn{1}{c}{$\beta_{Y|Z}$} &
\multicolumn{1}{c}{$\gamma_{Y|Z}$} &
\multicolumn{1}{c}{$\sigma_{Y|Z}$} &
\multicolumn{1}{c}{$\alpha_{X|Z}$} &
\multicolumn{1}{c}{$\sigma_{X|Z}$} &
\multicolumn{1}{c}{$\mu_{Z,0}$} &
\multicolumn{1}{c}{$\gamma_{\mu_{Z},D}$} &
\multicolumn{1}{c}{$\sigma_{Z,0}$} }
\startdata
C1+C2 & $136$ & $0.68 \pm 0.10$ & $-0.07 \pm 0.28$ & $-0.03 \pm 0.47$ & $0.023 \pm 0.015$ & $0.00 \pm 0.02$ & $0.37 \pm 0.14$ & $-0.29 \pm 0.07$ & $0.20 \pm 0.29$ & $0.09 \pm 0.17$\\
C1 & $83$ & $0.69 \pm 0.08$ & $-0.06 \pm 0.33$ & $-0.05 \pm 0.60$ & $0.027 \pm 0.019$ & $0.00 \pm 0.02$ & $0.33 \pm 0.14$ & $-0.18 \pm 0.08$ & $0.23 \pm 0.29$ & $0.09 \pm 0.14$
\enddata
\tablecomments{The $\gamma_{\sigma_{Z},D}$ parameter is set to zero in the regression. The intercept $\alpha_{X|Z}$ is a nuisance parameter to marginalize over the residual mass calibration uncertainty of $\pm\dfmcal$.}
\end{deluxetable*}


\begin{figure*}[!htb] 
  \begin{center}
   \includegraphics[scale=0.8, angle=0, clip]{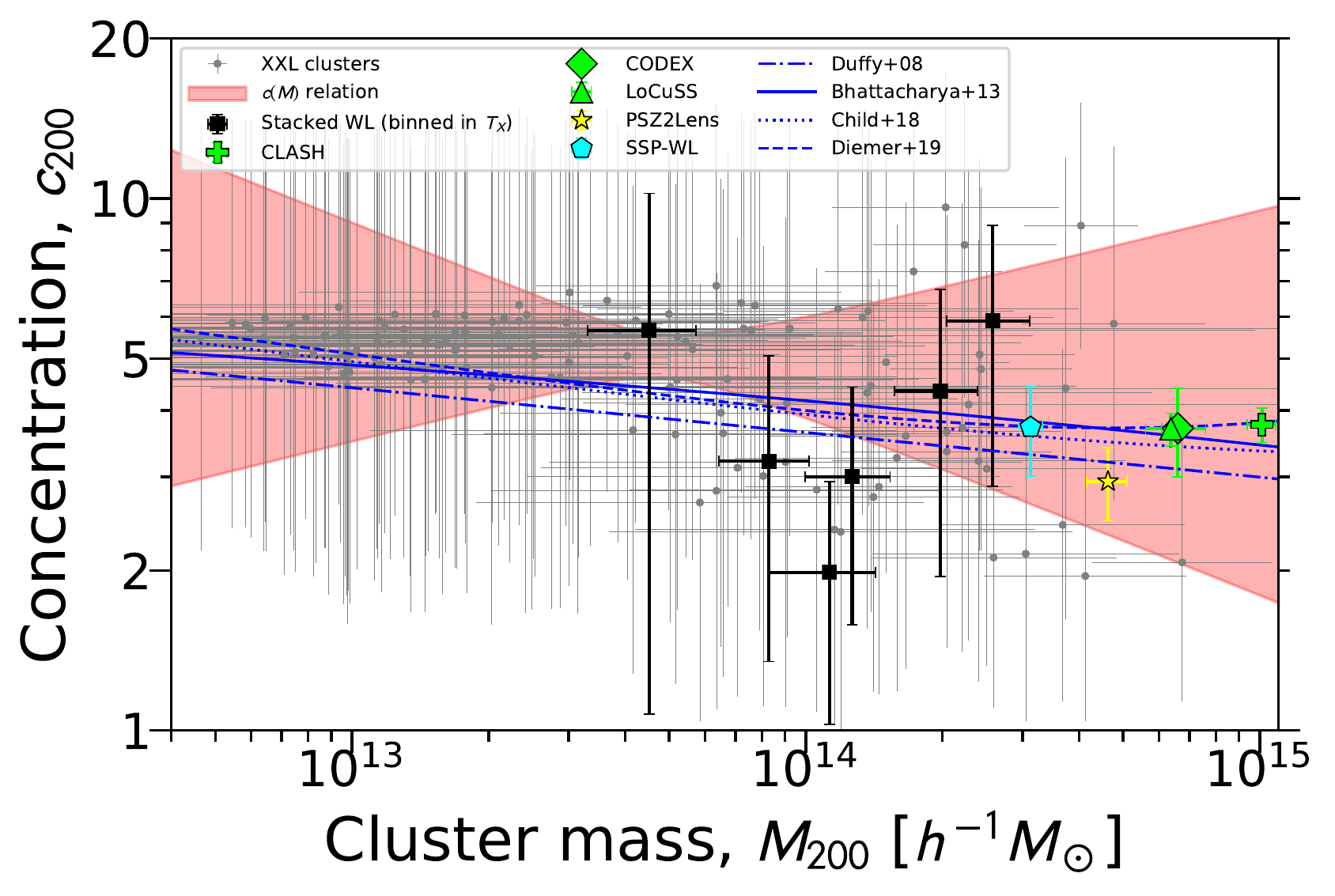} 
  \end{center}
\caption{
 \label{fig:cMR}
The $c$--$M$ relation for the XXL sample of 136 spectroscopically
 confirmed X-ray-selected systems obtained from our weak-lensing
 analysis of the HSC-SSP data. 
The gray circles with error bars represent the measured parameters
 ($\CBI$) and their $1\sigma$ uncertainties ($\SBI$) for individual XXL
 clusters. The red shaded region shows the $1\sigma$ confidence range 
 of the mean $c$--$M$ relation at a reference redshift of $\zref=0.3$ 
 obtained from our Bayesian regression using the \texttt{LIRA} package.
The black squares with error bars show our stacked weak-lensing
 constraints on $M_{200}$ and $c_{200}$
obtained for six subsamples of C1+C2 XXL clusters binned in X-ray
 temperature  (see Table \ref{tab:tstack} and Figure \ref{fig:cM_Tbins}).
 The stacked weak-lensing results of high-mass X-ray-selected clusters
 (CLASH, \citet{Umetsu2016clash};
  LoCuSS, \citet{Okabe+Smith2016};
  CODEX, \citet{Cibirka2017}),
 SZE-selected clusters (PSZ2Lens, \citet{Sereno2017psz2lens}),
 and
 weak-lensing-selected clusters (SSP-WL, \citet{Miyazaki2018wl})
 are also shown for comparison.
These weak-lensing observations are compared to theoretical $c$--$M$
 relations evaluated at $\zref=0.3$ for the full population of DM halos in
 $\Lambda$CDM cosmologies \citep{Duffy+2008,Bhatt+2013,Child2018cm,Diemer+Joyce2019}. 
 }
\end{figure*}

The main results of Bayesian inference for the $c_{200}$--$M_{200}$
relation are summarized in Table \ref{tab:cM} and Figure \ref{fig:cMfit}.
In addition to the regression of the C1+C2 sample, we have also
analyzed the C1 subsample separately.
Posterior summary statistics ($\CBI\pm\SBI$; see Section
\ref{subsec:modeling}) of for all regression parameters (see Section 
\ref{subsubsec:priors}) are listed in Table \ref{tab:cM}. 
Figure \ref{fig:cMfit} shows the marginalized one- and
two-dimensional posterior PDFs for the C1+C2 sample.

In Figure \ref{fig:cMR}, we show the resulting $c_{200}$--$M_{200}$
relation at a reference redshift of $\zref=0.3$ for the C1+C2 sample,
along with the theoretical $c_{200}$--$M_{200}$ relations for the full
population of DM halos predicted by
\citet{Duffy+2008},
\citet{Bhatt+2013},
\citet{Child2018cm},
and
\citet{Diemer+Joyce2019} \citep[see also][]{Diemer+Kravtsov2015}
(see Section \ref{subsubsec:cM_literature}). 
In Figure \ref{fig:cMR}, we overplot the measured values of
($M_{200},c_{200}$) and their $1\sigma$ uncertainties for individual
clusters.

Our inference of the $c_{200}$--$M_{200}$ relation for the C1+C2 sample is
summarized as follows (Table \ref{tab:cM}):  
\begin{equation}
\label{eq:cMbest}
 \begin{aligned}
 c_{200}&=\left[\Cpivot\right]\\
  &\times\left(\frac{M_{200}}{10^{14}\Msunh}\right)^{\betaCM}
  \left(\frac{1+z}{1+\zref}\right)^{\gammaCM},
 \end{aligned}
\end{equation}
with a lognormal intrinsic dispersion at fixed $M_{200}$ of
$\sigma(\ln{c_{200}}) = \ln{10}\,\sigma_{Y|Z} = \sigmalnC$,
and an upper limit of $< \sigmalnCup$ at the $99.7\percent$ CL. 
Here we have included a systematic uncertainty of $16\percent$ in the
normalization of the concentration parameter (Section \ref{subsec:syst}
and Appendix \ref{appendix:test}). 
We find no statistical evidence for redshift evolution of the
$c_{200}$--$M_{200}$ relation for the XXL sample:
$\gamma_{Y|Z}=\gammaCM$.

The $c_{200}$--$M_{200}$ relation inferred for the C1 subsample is
highly consistent with that obtained for the full C1+C2 sample (Table
\ref{tab:cM}), indicating that the underlying mass distribution of the
XXL cluster population is not sensitive to the details of the X-ray
selection function.

The $\Mwl$--$M_{200}$ relation is found to be poorly
constrained given the large statistical uncertainties 
in our weak-lensing mass estimates. 
The posterior distribution of $\sigma_{X|Z}$ is bimodal
(Figure \ref{fig:cMfit}), and there is a distinct lower-scatter
solution of $\sigma_{X|Z}\simlt 0.1$ with a tail extending toward the  
higher-scatter solution.
The lower-scatter solution is associated with $\sigma_{Z,0}\sim 0.4$,
which is reasonable for the XXL sample
\citepalias{2016AA...592A...2P, 2018AA...620A...5A}.
On the other hand, the higher-scatter solution is considerably larger
than the theoretically expected level of intrinsic scatter in the
weak-lensing mass, 
$\sim 20\percent$ \citep{Becker+Kravtsov2011,Gruen2015}.
The higher-scatter solution associated with $\ln{10} \sigma_{Z,0}\ll 1$
(see Table \ref{tab:cM}) is unlikely for the XXL sample
\citepalias{2016AA...592A...2P,2018AA...620A...5A}.

To assess the impact of this higher-scatter solution on our results,
we repeated our regression analysis assuming an informed prior of
$\sigma_{X|Z}\sim {\cal U}(0.05, 0.15)$
centered at $\sigma_{X|Z}=0.1$, which is approximately the theoretically
expected level of intrinsic scatter 
\citep{Becker+Kravtsov2011}. 
With this informed prior, we find
$c_{200}=4.9\pm 0.5$
at $M_{200}=10^{14}\Msunh$ and $z=0.3$, with
$\beta_{Y|Z}=-0.05\pm 0.08$,
$\gamma_{Y|Z}=-0.03\pm 0.43$,
and
$\sigma(\ln{c_{200}}) = (5.4 \pm 3.5)\percent$,
which is fully consistent with our baseline results
(see Equation (\ref{eq:cMbest})). 
This comparison shows that the higher-scatter solution
has negligible impact on the central values of the regression parameters,
whereas the size of errors for the
normalization and mass slope has been largely decreased, as the parameter
space is reduced substantially. On the other hand, we find that the
higher-scatter solution has little influence on the central value and
uncertainty of $\sigma(\ln{c_{200}})$.

Another source of systematic errors is the choice of the concentration
prior in the NFW profile fitting (Section \ref{subsec:modeling}). 
Posterior constraints on the NFW parameters for noisy objects, especially
on $c_{200}$, are poor and dominated by the priors.  We thus tested the
sensitivity of our results to the prior chosen for $c_{200}$. We have
repeated our NFW fits and regression using an even less informative
prior uniform  in the range $\log{c_{200}}\in [0,\log{30}]$,
obtaining a slightly higher normalization of
$c_{200}(\zref)=5.6\pm 1.6$ at $M_{200}=10^{14}\Msunh$,
$\beta_{Y|Z}= -0.05\pm 0.33$,
$\gamma_{Y|Z}= 0.03\pm 0.52$, and
$\sigma(\ln{c_{200}}) = (5.8\pm 3.9)\percent$.
The changes in the slopes and the intrinsic dispersion are
thus negligibly small compared to their respective uncertainties.
Hence, the choice of the concentration prior does influence the
normalization to some degree, but it does not significantly alter
our main results and conclusions even with priors that include
unrealistically large concentrations.

\subsubsection{Comparison with the Literature}
\label{subsubsec:cM_literature}

Overall, our regression results are in good agreement with the
theoretical predictions from DM-only numerical simulations calibrated
for recent $\Lambda$CDM cosmologies
\citep[see Figure \ref{fig:cMR};][]{Bhatt+2013,Diemer+Kravtsov2015,Klypin2016,Child2018cm,Diemer+Joyce2019}.
In particular, the inferred normalization and mass slope are in good
agreement with these DM-only $\Lambda$CDM predictions,
which yield mean concentrations in the range
$c_{200}(z=0.3) \simeq 3.9$--$4.2$ at 
$M_{200}=10^{14}\Msunh$, with a shallow negative
slope of $\beta \simeq -0.09$ \citep[e.g.,][]{Child2018cm}.
The inferred intrinsic dispersion $\sigma(\ln{c_{200}})$,
however, is significantly smaller than predicted for the full population
of $\Lambda$CDM halos, $\sigma(\ln{c_{200}})\simeq 33\percent$
\citep{Bhatt+2013,Child2018cm}.
We note that our test using simulated weak-lensing observations shows
that we can accurately recover the true value of $\sigma(\ln{c_{200}})$
(Figure \ref{fig:cM_BAHAMAS}; see Appendix \ref{appendix:cM_BAHAMAS}).\footnote{The intrinsic scatter is defined
at fixed $\Mtrue$, not at fixed $\Mwl$. Since $\Mtrue$ is a latent variable
that cannot be directly observed, we statistically constrain the
intrinsic scatter by forward-modeling the weak-lensing data.} 
This discrepancy could be due to the X-ray selection bias in
terms of the cool-core or relaxation state as found by previous studies 
\citep[e.g.,][]{Buote+2007,Ettori+2010,Eckert2011cc,Rasia+2013,Meneghetti2014clash,Rosetti2017cc}. 
Another possibility is that the statistical errors on $c_{200}$ are
overestimated as a consequence of the conservative prior,
so that the intrinsic dispersion $\sigma(\ln{c_{200}})$
of the $c_{200}$--$M_{200}$ relation is underestimated.

Although no evidence of redshift evolution for the XXL
$c_{200}$--$M_{200}$ relation is found, the average level of
concentration for $\Lambda$CDM halos is predicted to decrease with
increasing redshift, where the predicted values of the redshift slope
range from 
$\simeq -0.47$ \citep{Duffy+2008},
to $-0.42$ \citep{Child2018cm},
to $-0.29$ \citep{Meneghetti2014clash},
to $-0.16$ \citep{Ragagnin+2019}.
Our results are broadly consistent with these predictions within the
large statistical uncertainty.
We note that the redshift evolution of the concentration parameter is
sensitive to the relaxation state of clusters
\citep[e.g.,][]{DeBoni+2013,Meneghetti2014clash}. 

Numerical simulations suggest that relaxed subsamples have
concentrations that are on average $\sim 10$\% higher than for the full  
population of halos
\citep{Duffy+2008,Bhatt+2013,Meneghetti2014clash,Child2018cm,Ragagnin+2019}.
This indicates that mean concentrations for relaxed halos are
$c_{200}(z=0.3)\simeq 4.3$--$4.6$ at
$M_{200}=10^{14}\Msunh$, which are consistent with the observational
constraint (see Equation (\ref{eq:cMbest})).
At face value, the
$c_{200}$--$M_{200}$ relation obtained for the XXL sample is in better
agreement with those predicted for relaxed systems.
Another important effect of the relaxation state is that relaxed halos
are predicted to have a smaller intrinsic dispersion in the
$c_{200}$--$M_{200}$ relation, $\sigma(\ln{c_{200}})\sim 25\percent$
\citep[e.g.,][]{2007MNRAS.381.1450N,Duffy+2008,Bhatt+2013}, which is
again in better agreement with our observational constraint on the XXL
sample. 

\citet{Meneghetti2014clash} characterized a sample of halos that 
closely matches the selection function of the CLASH X-ray-selected
subsample with $M_{200}\sim 10^{15}\Msunh$
\citep{Donahue2014clash,Umetsu2014clash,Umetsu2016clash,Umetsu2018clump3d,Merten2015clash}.
These clusters were selected to have a high degree of regularity
in their X-ray morphology 
\citep{Postman+2012CLASH}. 
Cosmological hydrodynamical simulations suggest that this subsample is
prevalently composed of relaxed clusters ($\sim 70\percent$) and largely 
free of orientation bias \citep{Meneghetti2014clash}. Another important
effect of the selection function based on X-ray regularity is to reduce
the scatter in concentration down to $\sigma(\ln{c_{200}})\sim 16\percent$
\citep[see also][]{Rasia+2013}.
Although the XXL sample was not selected
explicitly according to their X-ray morphology, the X-ray selection in
favor of relaxed systems is likely to considerably affect the level of
scatter in the $c_{200}$--$M_{200}$ relation \citep{Rasia+2013}.

In Figure \ref{fig:cMR}, we also compare our results with previously
published weak-lensing constraints on
X-ray-selected high-mass clusters
from the
CLASH \citep[][$z=0.34$]{Umetsu2016clash},
LoCuSS \citep[][$z=0.23$]{Okabe+Smith2016}, and
CODEX \citep[][$z=0.50$]{Cibirka2017} surveys;
PSZ2 clusters detected by the {\em Planck} mission
\citep[][$z=0.20$]{Sereno2017psz2lens};
and weak-lensing-selected clusters from the HSC survey 
\citep[][$z=0.27$]{Miyazaki2018wl}.
Their stacked weak-lensing constraints are in excellent agreement with the
DM-only predictions calibrated for recent $\Lambda$CDM cosmologies
\citep[e.g.,][]{Bhatt+2013,Diemer+Kravtsov2015,Child2018cm,Diemer+Joyce2019}
and agree with our results.
We note that the effect of the redshift evolution is not accounted for
in the comparison given in Figure \ref{fig:cMR}.

\citet{Biviano+2017} performed a Jeans dynamical analysis of 49
nearby clusters ($0.04\simlt z\simlt 0.07$)
with the projected phase-space distribution of cluster
members available from the WINGS and OmegaWINGS survey
\citep{Fasano+2006,Gullieuszik+2015}. From their dynamical analysis,
\citet{Biviano+2017} determined total mass density profiles for
individual clusters in their sample and derived the $c_{200}$--$M_{200}$
relation over a wide range of cluster mass
($10^{14}\simlt M_{200}/M_\odot\simlt 2\times 10^{15}$).
They found a flat $c_{200}$--$M_{200}$ relation,
$c_{200}\propto M_{200}^{-0.03\pm 0.09}$, normalized to
$c_{200}\simeq 3.8$ at $M_{200}=10^{14}M_\odot$, which is in excellent
agreement with our results.

\subsection{Temperature--Mass Relation}
\label{subsec:TMR}

\begin{deluxetable*}{cccccccccc}
\tablecolumns{10}
\tablewidth{0pt}
\tabletypesize{\scriptsize}
\tablecaption{\label{tab:TM}
Summary Statistics of Regression Parameters for the XXL Temperature--Mass Relation}
\tablehead{
\multicolumn{1}{c}{$\alpha_{Y|Z}$} &
\multicolumn{1}{c}{$\beta_{Y|Z}$} &
\multicolumn{1}{c}{$\gamma_{Y|Z}$} &
\multicolumn{1}{c}{$\sigma_{Y|Z}$} &
\multicolumn{1}{c}{$\alpha_{X|Z}$} &
\multicolumn{1}{c}{$\sigma_{X|Z}$} &
\multicolumn{1}{c}{$\mu_{Z,0}$} &
\multicolumn{1}{c}{$\gamma_{\mu_{Z},D}$} &
\multicolumn{1}{c}{$\sigma_{Z,0}$} &
\multicolumn{1}{c}{$\gamma_{\sigma_{Z},D}$} }
\startdata
$0.44 \pm 0.09$ & $0.85 \pm 0.31$ & $0.18 \pm 0.66$ & $0.061 \pm 0.049$ & $0.00 \pm 0.02$ & $0.31 \pm 0.08$ & $-0.17 \pm 0.07$ & $0.34 \pm 0.14$ & $0.20 \pm 0.07$ & $-0.05 \pm 0.14$\\
$0.42 \pm 0.07$ & $0.75 \pm 0.27$ & $2/3$ & $0.070 \pm 0.050$ & $-0.00 \pm 0.02$ & $0.29 \pm 0.08$ & $-0.17 \pm 0.07$ & $0.29 \pm 0.11$ & $0.22 \pm 0.06$ & $-0.05 \pm 0.15$\\
$0.41 \pm 0.05$ & $2/3$ & $2/3$ & $0.070 \pm 0.043$ & $-0.00 \pm 0.02$ & $0.29 \pm 0.06$ & $-0.18 \pm 0.07$ & $0.34 \pm 0.09$ & $0.25 \pm 0.04$ & $-0.04 \pm 0.14$
\enddata
\tablecomments{The $T$--$M$ relation is derived for a subset of 105 clusters that have both measured HSC $M_{500}$ masses and X-ray temperatures $\Tx$. The intercept $\alpha_{X|Z}$ is a nuisance parameter to marginalize over the residual mass calibration uncertainty of $\pm\dfmcal$.}
\end{deluxetable*}


\begin{figure*}[!htb] 
  \begin{center}
   \includegraphics[scale=0.8, angle=0, clip]{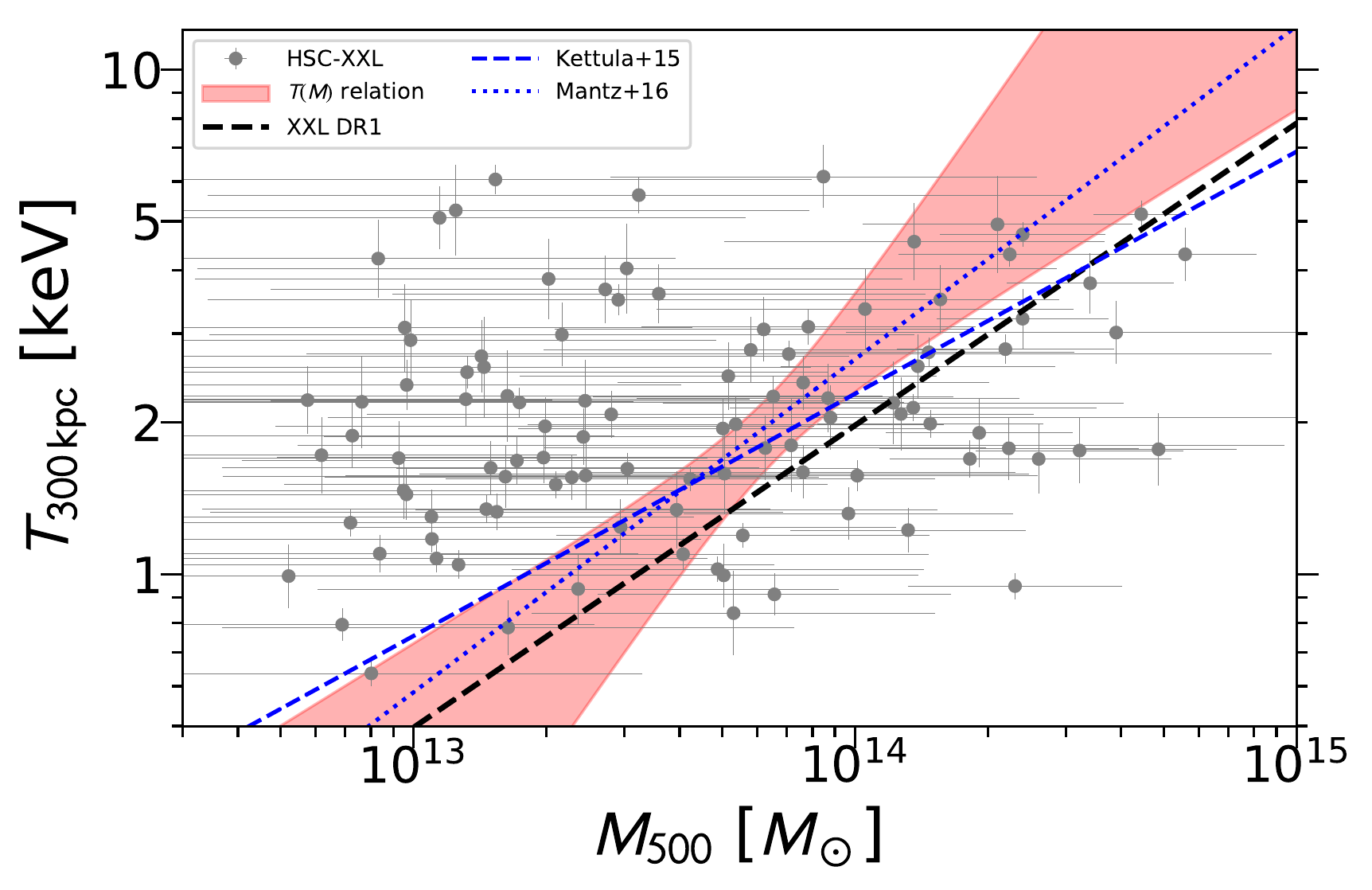} 
  \end{center}
 \caption{The $T_\mathrm{X}$--$M_{500}$ relation for the XXL sample
 obtained using a subsample of 105 clusters having both XXL temperature
 and HSC weak-lensing measurements.
 The gray circles with error bars represent the measured parameters and
 their $1\sigma$ uncertainties for individual XXL clusters. The
 red shaded region shows the $1\sigma$ confidence region of the mean
 $T_\mathrm{X}$--$M_{500}$ relation at a reference redshift of
 $\zref=0.3$ obtained from our
 Bayesian regression using the \texttt{LIRA} package.
The thick black dashed line shows the XXL DR1 results of
\citetalias{2016AA...592A...4L}. 
 Our results are also compared with previously published results for
 massive clusters obtained by \citet{Kettula2015} and \citet{Mantz2016}.  
\label{fig:TMR}
 }
\end{figure*}


\begin{figure*}[!htb] 
  \begin{center}
   \includegraphics[scale=0.8, angle=0, clip]{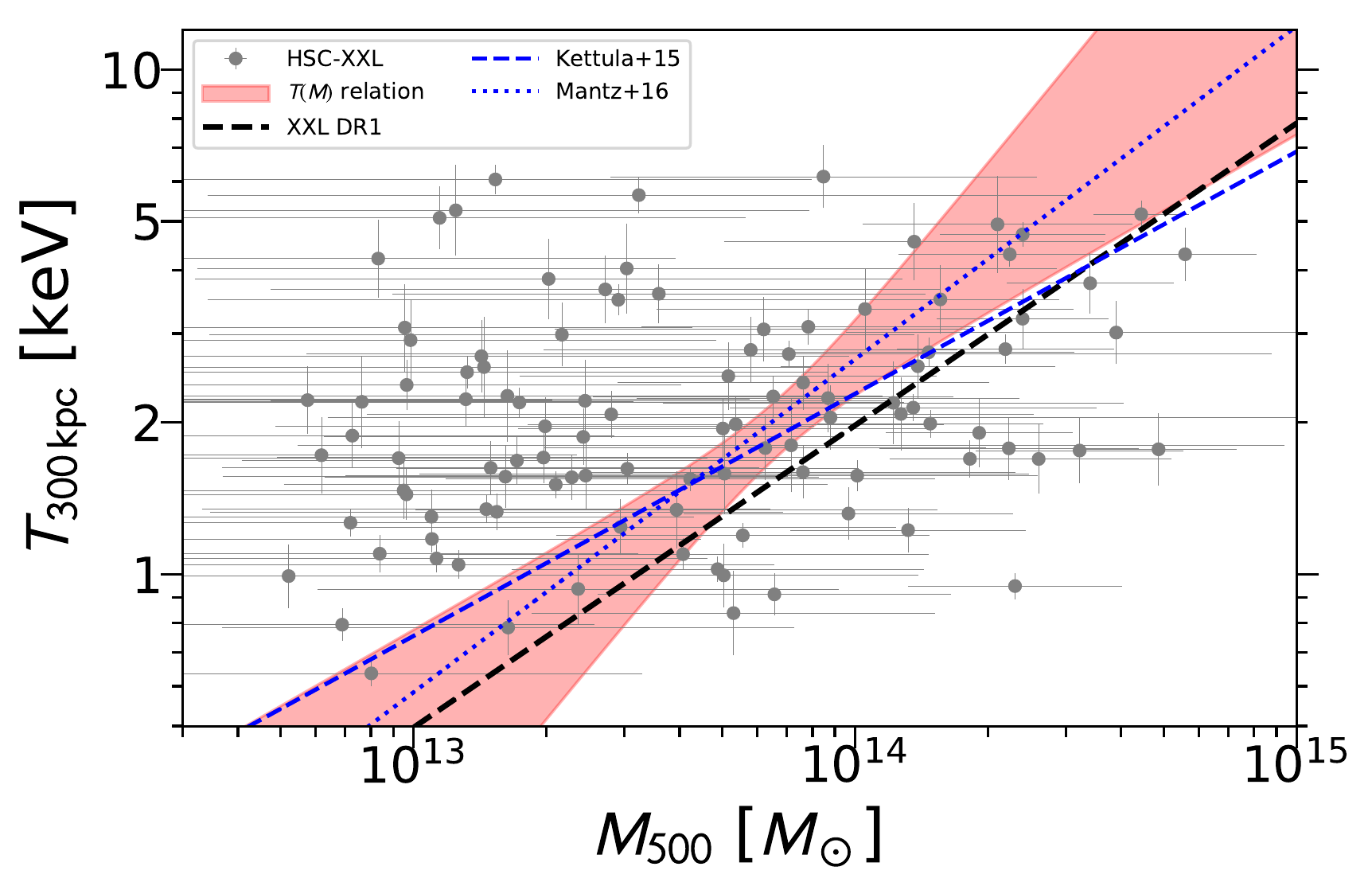} 
  \end{center}
 \caption{
Same as Figure \ref{fig:TMR}, but with the
 $E(z)$-trend parameter fixed to the self-similar model expectation of
 $\gamma_{Y|Z}=2/3$.
  \label{fig:TMR_gamma066}
 }
\end{figure*}

Models of self-similar gravitational collapse in an expanding universe
predict scale-free, power-law relations between cluster properties
\citep{Kaiser1986,Ettori2015}.
Deviations from self-similar behavior are often interpreted as
evidence of feedback into the intracluster gas associated with star
formation and AGN activities, as well as with radiative cooling in the
cluster cores \citep[e.g.,][]{Czakon2015}.
The self-similar prediction for the $T_\mathrm{X}$--$M$ relation is 
$T_\mathrm{X}\propto E^{2/3}(z)M_\Delta^{2/3}$,
in which the virial condition
$GM_{\Delta}/r_{\Delta}\sim T_\mathrm{X}$ with
$r_{\Delta}\propto[M_{\Delta}/\rho_\mathrm{c}(z)]^{1/3}$ is assumed.  

On the other hand, secondary infall and continuous accretion from
the surrounding large-scale structure can lead to a departure
from virial equilibrium \citep{Bertschinger1985}, while scaling
relations of clusters preserve the power-law structure
\citep{Fujita+2018a,Fujita+2018b}. 
The large scatter in growth histories of clusters translates into a
significant diversity in their density profiles
\citep{Diemer+Kravtsov2014}, 
thus contributing to the scatter of the $T_\mathrm{X}$--$M$ relation
\citep{Fujita+2018b}.
The mass dependence of the $c$--$M$ relation and the halo
fundamental-plane (FP) relation \citep{Fujita+2018a} make the mass trend
of the $T_\mathrm{X}$--$M$ relation on cluster scales steeper than the 
self-similar prediction 
\citep[$T_\mathrm{X}\propto
E^{0.75}(z)M_{500}^{0.75}$;][]{Fujita+2018b}. 
However, the mass trend of the $T_\mathrm{X}$--$M$ relation is predicted
to become shallower and closer to the self-similar expectation toward
group scales
\citep[$T_\mathrm{X}\propto E^{0.65}(z)M_{500}^{0.65}$;][]{Fujita+2018b}.

Now we turn to results of Bayesian inference for the $\Tx$--$M_{500}$ relation.
Posterior summary statistics ($\CBI\pm\SBI$; see Section
\ref{subsec:modeling}) for all regression parameters (Section
\ref{subsubsec:priors}) are listed in Table \ref{tab:TM}.   
Figure \ref{fig:TMfit} shows the marginalized one- and two-dimensional
posterior PDFs for the regression parameters of the $\Tx$--$M_{500}$
relation.

Figure \ref{fig:TMR} shows the resulting $\Tx$--$M_{500}$ relation for
the XXL sample at a reference redshift of $\zref=0.3$.
Our inference of the $\Tx$--$M_{500}$ relation is summarized as follows:
\begin{equation}
 \begin{aligned}
  \Tx&=\left(\Tpivot\right)\,\mathrm{keV}\\
  &\times\left(\frac{M_{500}}{10^{14}\Msun}\right)^{\betaTM}
  \left[\frac{E(z)}{E(\zref)}\right]^{\gammaTM},
 \end{aligned}
\end{equation}
with a lognormal intrinsic dispersion of
$\sigma(\ln{\Tx})=\sigmalnT$ at fixed $M_{500}$.
A tighter statistical constraint on the normalization can be obtained   
around the log-mean mass of clusters inferred for the sample, 
$\mu_{Z,0}=\muZTM$ at $\zref=0.3$. 
The inferred mean mass of the population is
$M_{500}=(7.6\pm 1.4)\times 10^{13}\Msun$ at $\zref=0.3$.
For $M_{500}=8\times 10^{13}\Msun$ and $z=0.3$, we find
$\Tx = \Tbest\,\mathrm{keV}$.

We find no statistical evidence for redshift evolution of the
$\Tx$--$M_{500}$ relation for the XXL sample:
$\gamma_{Y|Z}=\gammaTM$, which is also consistent with the
self-similar expectation, $\gamma_{Y|Z}=2/3$.
A slightly shallower mass slope of $\beta_{Y|Z}=\betaTMalt$
is found when performing the regression by
setting the $E(z)$-trend parameter to the self-similar expectation,
$\gamma_{Y|Z}=2/3$ (see Table \ref{tab:TM}).  The resulting constraints
on the $\Tx$--$M_{500}$ relation are shown in Figure \ref{fig:TMR_gamma066}.
When we fix the slope parameters to $\beta_{Y|Z}=\gamma_{Y|Z}=2/3$
expected from the self-similar model, we find
$\Tx=\left(\TpivotFS\right)\,\mathrm{keV}\times(M_{500}/10^{14}\Msun)^{2/3}[E(z)/E(\zref)]^{2/3}$,
with a lognormal intrinsic dispersion of $\sigma(\ln{\Tx})=\sigmalnTFS$.

Overall, our regression results are in agreement within the errors with
the theoretical predictions (Table \ref{tab:TM}).
We find the mass slope parameter $\beta_{Y|Z}$ to be slightly steeper
but consistent with the self-similar expectation,  
$\beta_{Y|Z}=2/3$, as well as with the range 
$\beta_{Y|Z}\simeq 0.65$--$0.75$ predicted by the halo FP relation of  
\citet{Fujita+2018b}. 
The $E(z)$-trend parameter $\gamma_{Y|Z}$ is still consistent with the
self-similar expectation $\gamma_{Y|Z}=2/3$ within the large
uncertainty.
It should be stressed that we measure the X-ray temperatures
 $T_\mathrm{X}=\Tx$ in a
core-included aperture of 300\,kpc (physical), whereas the $r_{500}$
aperture for the XXL sample is typically $\sim 500$--$600$\,kpc (physical).
Hence, a quantitative interpretation of the observed
$\Tx$--$M_{500}$ relation is not straightforward.
For the $M_{500,\mathrm{WL}}$--$M_{500}$ relation, we observe a similar
trend of the intrinsic dispersion
$\sigma_{X|Z}=\sigma(\ln{M_{500,\mathrm{WL}}})/\ln{10}$ to that in the 
$c_{200}$--$M_{200}$ relation (Section \ref{subsec:cMR}).

Recently, \citet{Bulbul+2019} studied mass scaling relations of X-ray  
observables for a sample of 59 SZE-selected high-mass clusters
($3\times 10^{14}M_\odot\le M_{500}\le 1.8\times 10^{15}M_\odot$,
$0.20<z<1.5$) from the South Pole Telescope (SPT) survey. They used
SPT SZE-based cluster mass estimates.
Since \citet{Bulbul+2019} examined the scaling relations with both
core-included and core-excised quantities measured from \XMMNewton\
data (albeit in the high-mass regime),
their results are of critical relevance to our study (see Figures
\ref{fig:TMR} and \ref{fig:TMR_gamma066}).
Overall, they found that the mass trends of the X-ray observables
are steeper than self-similar behavior in all cases
(e.g., $T_\mathrm{X}\propto M_{500}^{0.80\pm 0.10}$ including the core
region), while the redshift trends are consistent with the self-similar 
expectation.
Their mass and $E(z)$ trends of the $T_\mathrm{X}$--$M_{500}$ relation
with and without the core region are both consistent with our results
\citep[see Table 4 of][their fitting results of Form I]{Bulbul+2019}.
According to the findings of \citet{Bulbul+2019}, the mass and redshift
trends, as well as the normalization of the core-included
$T_\mathrm{X}$--$M_{500}$ relation, are consistent within the errors with
those for their core-excised case. The most noticeable difference
between the two cases comes from the intrinsic scatter.
They found a lognormal intrinsic dispersion of
$\sigma(\ln{T_\mathrm{X}})= (13\pm 5)\percent$ for the
core-excised case and $\sigma(\ln{T_\mathrm{X}})= (18\pm 4)\percent$ for
the core-included case. When the core region is included, the intrinsic
lognormal dispersion in the $T_\mathrm{X}$--$M_{500}$ relation is
increased by $\simeq 40\%$, although the difference is not statistically 
significant.

Our $T_\mathrm{X}$--$M_{500}$ relation is in good agreement with that of 
\citet{Mantz2016} obtained for a sample of 40 dynamically relaxed, X-ray
hot ($\simgt 5$\,keV) clusters based on \Chandra\ X-ray observations (see
Figures \ref{fig:TMR} and \ref{fig:TMR_gamma066}). We note that
\citet{Mantz2016} used cluster mass estimates obtained from X-ray data 
assuming hydrostatic equilibrium. They found no significant bias in
their X-ray hydrostatic mass estimates relative to weak lensing.

At group scales of $M_{500}\simlt 5\times 10^{13}\Msun$,
our regression results agree with the XXL DR1 results of
\citetalias{2016AA...592A...4L}
based on weak-lensing mass estimates for a subsample of 38 XXL-N
clusters at $z<0.6$.  
Their analysis used the weak-lensing shear catalog from the
 Canada--France--Hawaii Telescope Lensing Survey 
 \citep[CFHTLenS;][]{Heymans+2012CFHTLenS,Erben2013cfhtlens} to obtain
 the mass--temperature relation for the XXL sample.
Our $\Tx$--$M_{500}$ relation has a slightly steeper mass
 trend than the XXL DR1 results, implying a smaller mass scale in the
 cluster regime. 
The overall offset from the XXL DR1 relation of
\citetalias{2016AA...592A...4L}
is at the $\sim 1.5\sigma$ level (Figure \ref{fig:TMR}).
When the $E(z)$-trend parameter is fixed to $2/3$, our results are in
closer agreement with the XXL DR1 results (Figure
\ref{fig:TMR_gamma066}).  In Section \ref{subsec:dr1dr2}, we provide a
detailed comparison of weak-lensing mass estimates between the XXL DR1
and XXL DR2 (this work) results. 

\citet{Kettula2015} presented a weak-lensing and X-ray analysis of 12
low-mass clusters selected from the CFHTLenS and \XMM-CFHTLS surveys, in
combination with high-mass systems from the Canadian Cluster Comparison 
Project and low-mass systems from the COSMOS survey. Their 
combined sample comprises 70 systems, spanning 
more than two orders of magnitude in mass. After correcting for
Malmquist and Eddington bias, they found a mass slope of
$\beta=0.48\pm 0.06$
in the $T_\mathrm{X}$--$M_{500}$ relation with a lognormal intrinsic
dispersion of $\sigma(\ln{T_\mathrm{X}})=(14\pm 5)\percent$.
The $T_\mathrm{X}$--$M_{500}$ relation of \citet{Kettula2015} is in
agreement with our results (see Figures \ref{fig:TMR} and
\ref{fig:TMR_gamma066}).

\subsection{Comparison with the XXL DR1 Mass Calibration}
\label{subsec:dr1dr2}


\begin{figure*}[!htb] 
  \begin{center}
   \includegraphics[scale=0.48, angle=0, clip]{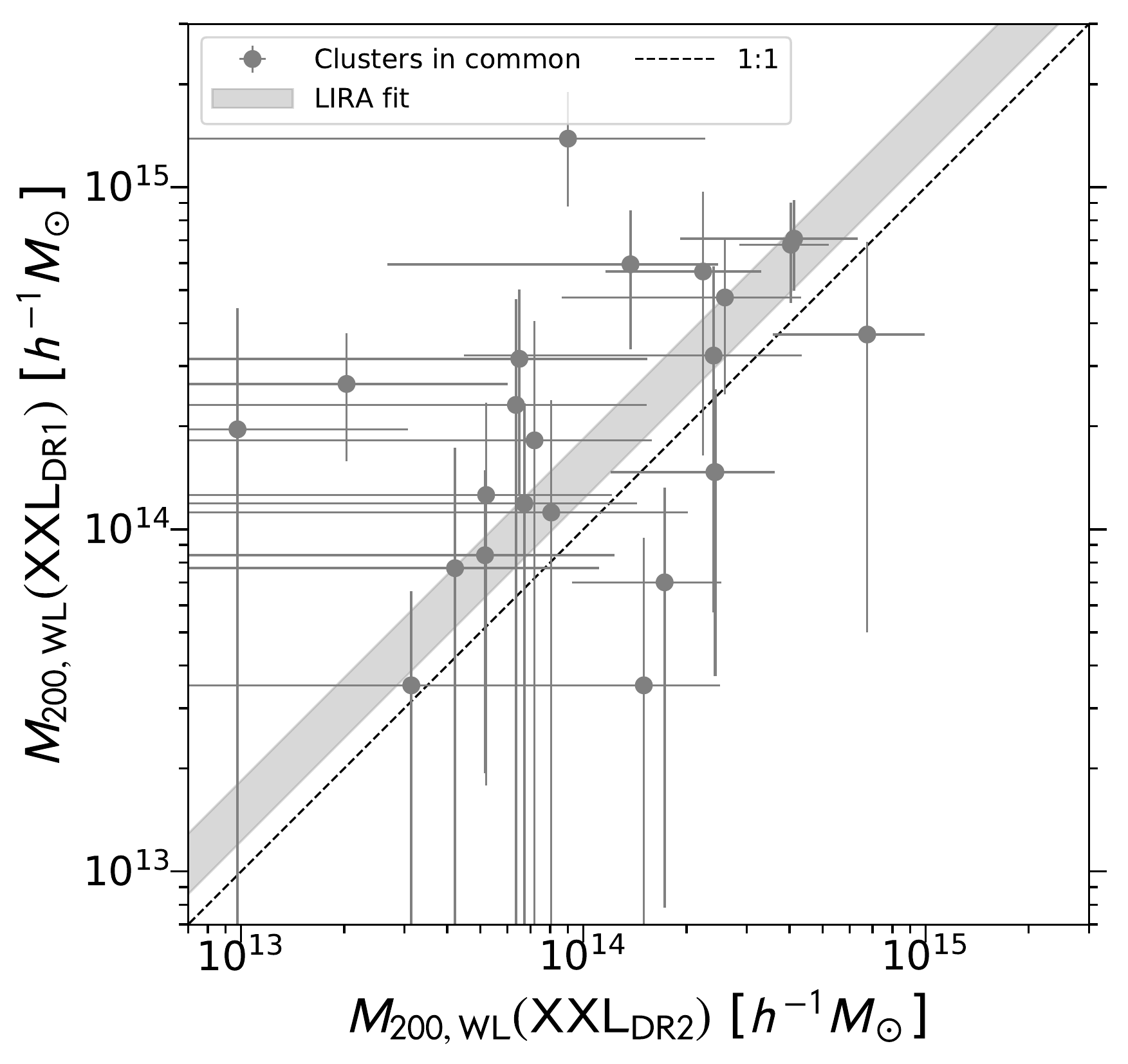} 
   \includegraphics[scale=0.48, angle=0, clip]{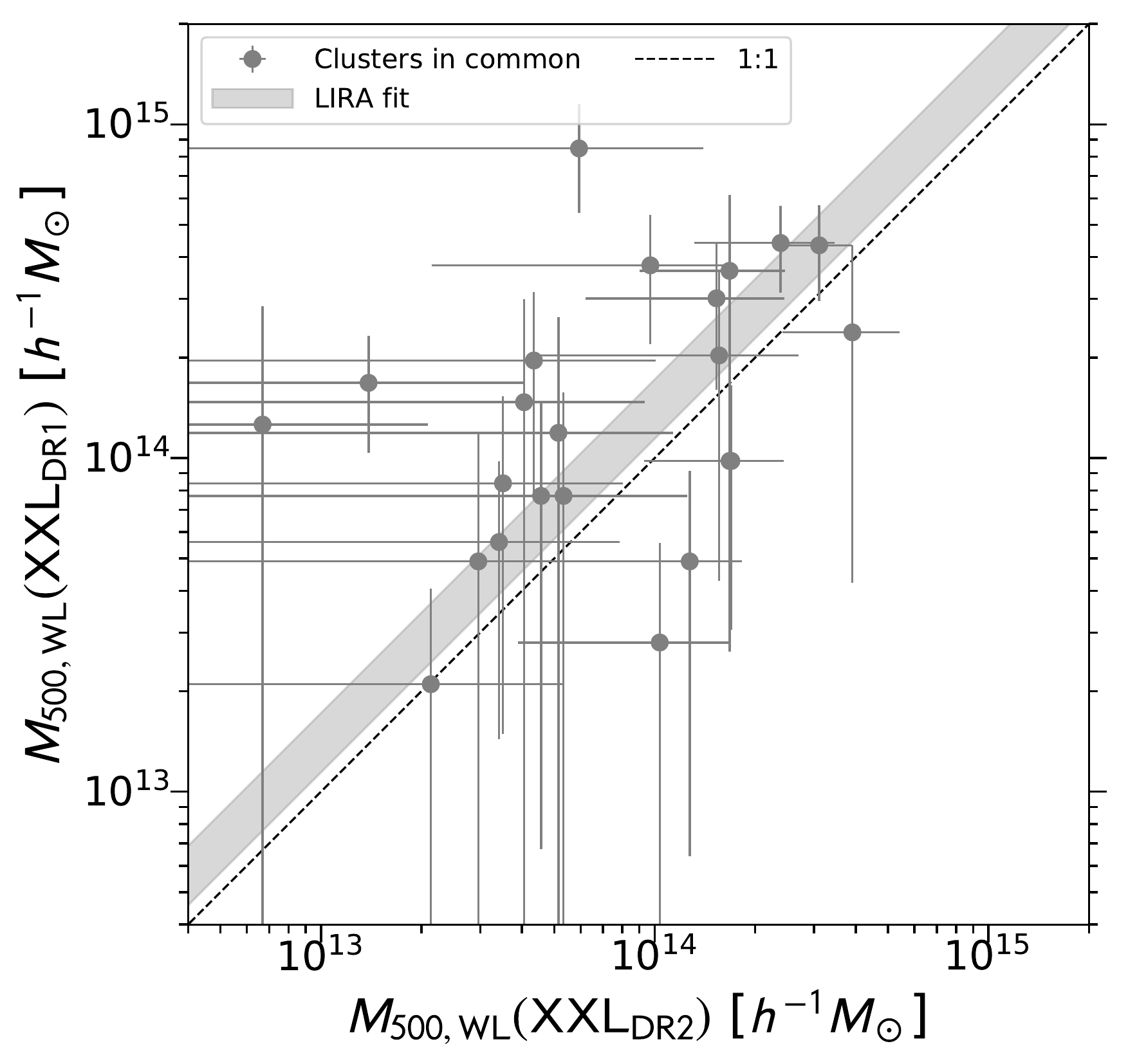}
  \end{center}
 \caption{
Comparison of weak-lensing mass estimates for a subset of 23 XXL
 clusters in common between this work (XXL DR2) and the XXL DR1 results 
\citepalias{2016AA...592A...4L}.
We characterize the discrepancy between these two sets of weak-lensing 
 mass estimates in the \texttt{LIRA} framework, finding a mean mass
 offset of 
 $\bXXL$ in $M_\mathrm{500,WL}$
 and
 $\bXXLtwo$ in $\Mwl$.
 \label{fig:dr1dr2}
 }
\end{figure*}

\citetalias{2016AA...592A...4L}
derived the mass--temperature ($M_{500}$--$\Tx$)
relation for 38 XXL-N clusters at $z<0.6$ selected from the 100
brightest galaxy cluster (XXL-100-GC) sample 
\citepalias{2016AA...592A...2P}
by using weak-lensing mass estimates based on the CFHTLenS shear catalog 
\citep{Heymans+2012CFHTLenS,Erben2013cfhtlens}.
The CFHTLenS survey covers a total survey area of
$\simeq 154$\,deg$^2$,
which overlaps with the XXL-N field. Their shear
catalog comprises galaxy shape measurements with an unweighted
(weighted)
source density of $\ngal \simeq 17$ ($14$) galaxies\,arcmin$^{-2}$,
compared to $\ngal\simeq 25$ ($22$) galaxies\,arcmin$^{-2}$ for the HSC
survey (see Section \ref{subsec:hsc}).

\citet[hereafter \citetalias{2016AA...592A..12E}]{2016AA...592A..12E}
studied the baryon fractions of XXL-100-GC clusters using X-ray gas mass
measurements and the weak-lensing-calibrated $M_{500}$--$\Tx$ relation
of
\citetalias{2016AA...592A...4L}.
They found a low gas mass fraction
($f_\mathrm{gas,500}\simeq 0.048$ at $M_{500}=5\times 10^{13}\Msun$)
that requires a relative mass bias of 
$b_\mathrm{HE}\equiv 1-M_\mathrm{500,X}/M_\mathrm{500,WL}=0.28^{+0.07}_{-0.08}$
to match the gas fractions obtained with weak-lensing and X-ray
hydrostatic-equilibrium mass estimates,
$M_\mathrm{500,WL}$ and $M_{500,X}$, respectively.


As summarized below, the shear-to-mass procedure implemented by
\citetalias{2016AA...592A...4L}
is somewhat different from ours.  
\citetalias{2016AA...592A...4L}
used the same fitting function as in this study
\citep[i.e., the projected NFW functional of][]{2000ApJ...534...34W},
with a similar mass prior that is uniform in the logarithm of $M_{200}$ 
in the range 
$\log{(M_{200}/\Msun)}\in [13, 16]$.
The concentration parameter was fixed to the mean 
$c_{200}$--$M_{200}$ relation of \citet{Duffy+2008}, which is calibrated
for a \WMAP\ 5\,yr cosmology \citep{Komatsu+2009WMAP5}.
At $M_{200}=10^{14}\Msunh$ and $z=0.3$, this model predicts
$c_{200}\simeq 3.6$, which is $\simeq 14\percent$ lower than predicted 
by the \citet{Bhatt+2013} relation, $c_{200}\simeq 4.2$
(see Figure \ref{fig:cMR}).
Because of the $c_{200}$--$M_{200}$ degeneracy \citep[see Section 5.4.1
of][]{Umetsu2014clash}, assuming a lower concentration will result in an
overestimation of the total mass, $M_{200}$.
By repeating the analysis assuming the fixed $c_{200}$--$M_{200}$
relation of \citet{Duffy+2008}, we find that the geometric mean 
mass scale $\langle M_{200}\rangle_\mathrm{g}$ of the C1+C2 sample
is overestimated by $22\percent$ with respect to our fiducial analysis
(Table \ref{tab:stack}).

The fitting radial range chosen by
\citetalias{2016AA...592A...4L}
is $R\in[0.15,3]$\,Mpc (physical), corresponding to
$R\in[0.1365,2.73]\,\Mpch$ (comoving) at $z=0.3$.
Their fitting range is comparable to our choice $R\in [0.3,3]\,\Mpch$
(comoving), but their fits are more sensitive to the inner region.
\citetalias{2016AA...592A...4L}
only accounted for the shape noise (see Equation (\ref{eq:cmat})) in
their error analysis.

Another possible cause of the mass discrepancy is the choice of
posterior summary statistics for the mass scale of each individual
cluster \citep{Sereno2017psz2lens}.
\citetalias{2016AA...592A...4L}
employed the mode and asymmetric confidence limits
of $M_{200}$ as posterior summary statistics. In contrast, we use
symmetrized biweight statistics, $\CBI\pm\SBI$. For a lognormally
distributed quantity, the biweight center location $\CBI$ typically
approximates the median of the distribution. 
By analyzing simulations of NFW lenses, \citet{Sereno2017psz2lens}
found that the mode estimator is less stable and noisier than the
biweight estimator for low-SNR objects (see their Appendix C).
For their NFW lenses with $M_{200}>10^{14}\Msunh$,
\citet{Sereno2017psz2lens} found that the mode estimator overestimates
$M_{200}$ by $4\percent$ relative to the biweight estimator, where the
actual level of bias depends on the mass range of 
the sample and the quality of data \citep{Sereno2017psz2lens}.

It should be emphasized again that
\citetalias{2016AA...592A...4L}
adopted the quadratic weak-lensing SNR estimator (Equation
(\ref{eq:SNRq})),  
which is positive by construction (Section \ref{subsec:DSigma}) and can
lead to overestimation of the true significance if the actual SNR per
radial bin is less than unity (see Table \ref{tab:clusters}). It is also
sensitive to the choice of the number of radial bins (or the number of
degrees of freedom).

We have identified 23 XXL clusters in common between the
XXL DR1
\citepalias{2016AA...592A...4L}
and DR2 (this work) mass calibrations, excluding seven clusters for
which only upper bounds were obtained by
\citetalias{2016AA...592A...4L}.
We characterize the discrepancy between the two sets of weak-lensing
mass estimates by accounting for the respective scatters with respect to
the true mass. 
To this end, we solve the following coupled, scattered relations
in the \texttt{LIRA} framework (Section \ref{subsec:regression}): 
\begin{equation}
 \begin{aligned} 
  X_1 &= \alpha + Z \pm \sigma_{X_1|Z},\\
  X_2 &= Z \pm \sigma_{X_2|Z},
 \end{aligned}
\end{equation}
where $Z$ denotes the true logarithmic mass,
$X_1$ and $X_2$ are the logarithmic weak-lensing masses
from the XXL DR1 and XXL DR2 mass calibrations, respectively,
$\sigma_{X_1|Z}$ and $\sigma_{X_2|Z}$ are the respective intrinsic
dispersions at fixed logarithmic mass $Z$,
and
$\alpha$ describes the logarithmic mass offset. We simultaneously model
the underlying $P(Z)$ characterized by the mean $\mu_{Z,0}$ and the
dispersion $\sigma_{Z,0}$ (see Section \ref{subsec:regression}). For
each cluster, we account for correlations between $X_1$ and $X_2$
assuming a cross-correlation coefficient of $0.7$ (approximately the
ratio of the number densities of source galaxies between the CFHTLenS
and HSC shear catalogs).

The results are  shown in Figure \ref{fig:dr1dr2}.
We find a mean mass offset of
$\ln{10}\alpha=\bXXL$ in $M_{500}$
and
$\bXXLtwo$ in $M_{200}$.
If we exclude the most discrepant cluster with
an XXL DR1 estimate of $\Mwl\simgt 10^{15}\Msunh$,
the mass discrepancy is reduced to
$\bXXLexc$ in $M_{500}$ and
$\bXXLexctwo$ in $M_{200}$.
This is consistent with the level of mass bias found by
\citetalias{2016AA...592A..12E}.

This level of mass discrepancy with respect to DR1 is also comparable to
that found by \citet{Lieu2017}, who reanalyzed the same CFHT
weak-lensing data for the DR1 sample of \citetalias{2016AA...592A...4L}. 
By fitting $M_{200}$ and $c_{200}$ together in Bayesian hierarchical
modeling, \citet{Lieu2017} found weak-lensing $M_{200}$ masses that are
on average $\sim 28\percent$ smaller (in terms of the weighted geometric
mean) than those of
\citetalias{2016AA...592A...4L}.
This discrepancy is reduced to $\sim 14\percent$ when
$c_{200}$ is treated as a free parameter in the DR1 analysis of
\citetalias{2016AA...592A...4L} \citep[][see also Section 4.1 of \citetalias{2016AA...592A...4L}]{Lieu2017}. 

We thus conclude that the discrepancy between the XXL DR1 and XXL DR2
mass calibrations is likely due to the combination of the fixed $c$--$M$
relation assumed in \citetalias{2016AA...592A...4L}
and the different fitting procedures for extracting cluster
masses from weak-lensing data.

\subsection{Mass Forecasting}
\label{subsec:forecast}


\begin{figure*}[!htb] 
  \begin{center}
   \includegraphics[scale=0.8, angle=0, clip]{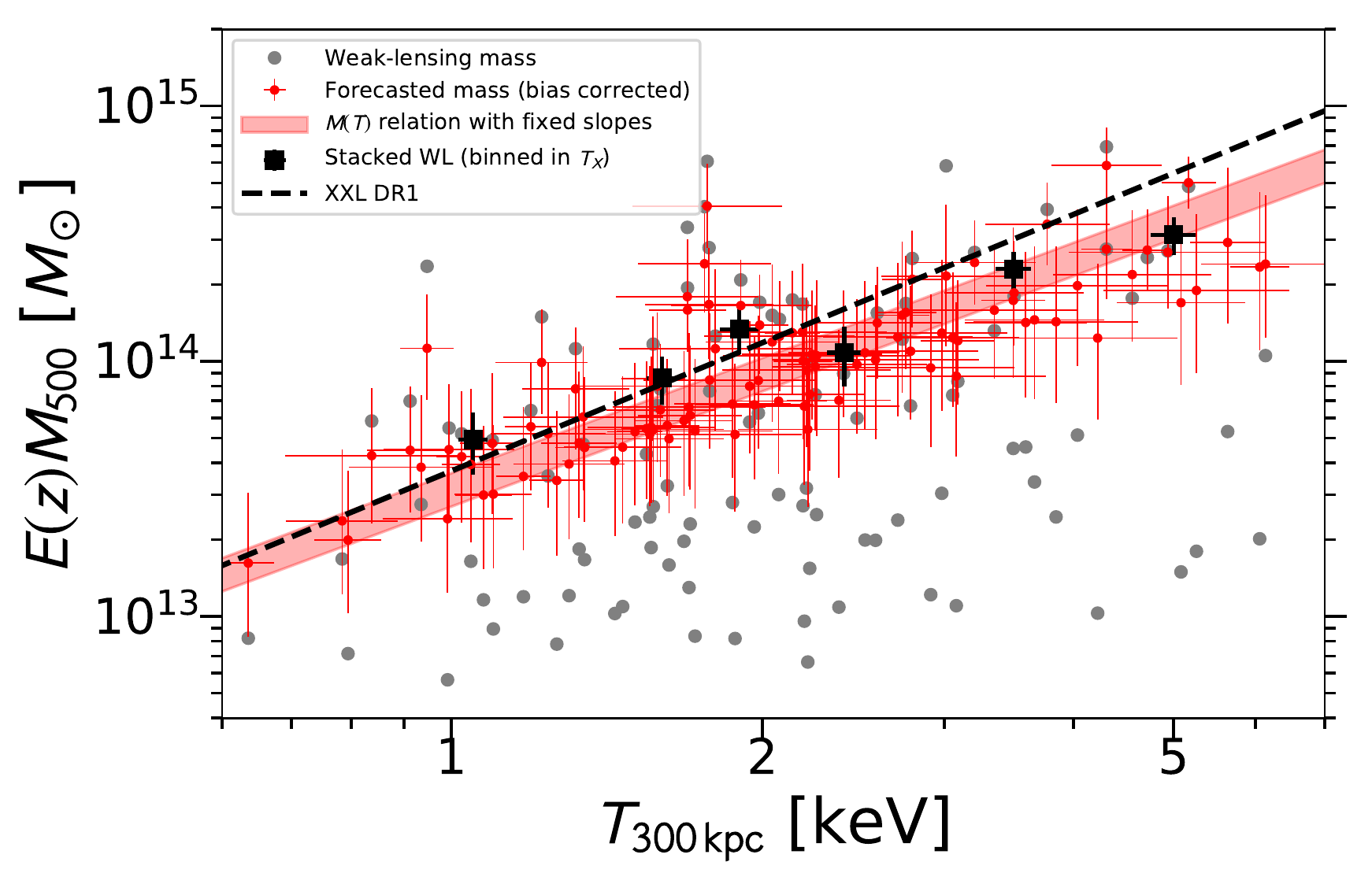} 
  \end{center}
 \caption{Mass forecasting based on X-ray temperature measurements.
 The red shaded region shows the $1\sigma$ confidence range of the
 mean $M_{500}$--$T_\mathrm{X}$ relation calibrated with a subsample of
 105 XXL  clusters having both XXL temperature and HSC weak-lensing
 measurements (gray circles). The $M_{500}$--$T_\mathrm{X}$ relation is
 obtained by fixing the slopes to the self-similar values. 
 The red circles with error bars show bias-corrected,
 weak-lensing-calibrated estimates of $M_{500}$ based on $\Tx$ (Table
 \ref{tab:clusters}).
 The black squares with error bars show the stacked weak-lensing
 constraints obtained for six subsamples of C1+C2 XXL clusters binned in X-ray
 temperature (see Table \ref{tab:tstack} and Figure \ref{fig:cM_Tbins}).
 The $M_{500}$--$T_\mathrm{X}$ relation from the XXL DR1 results  
\citepalias{2016AA...592A...4L}
 is shown with the thick black dashed line.
 \label{fig:MTR}
 }
\end{figure*}

Mass forecasting given a low-scatter mass proxy can be performed in the 
framework of Bayesian hierarchical modeling
\citep[e.g.,][]{Sereno2016lirapackage,CoMaLit5}.
Here we obtain bias-corrected, weak-lensing-calibrated estimates of
$M_{200}$ and $M_{500}$ for individual XXL clusters from their X-ray 
temperatures by using the \texttt{LIRA} package.
To this end, we use the subset of 105 C1+C2 clusters with measured $\Tx$
values as a calibration sample. 
In this backward forecasting analysis, we simultaneously model the 
proxy distribution and determine the $M_{\Delta}$--$\Tx$ scaling
relation \citep{CoMaLit5} for each overdensity $\Delta$.

Figure \ref{fig:MTR} shows the resulting distribution of
weak-lensing-calibrated $M_{500}$ as a function of $\Tx$ for the
calibration sample along with the $M_{500}$--$\Tx$ relation.
Here we considered the scaling relation with the slopes fixed to the
self-similar expectation, i.e.,
$E(z)M_{\Delta}\propto T_\mathrm{X}^{3/2}$.
In Table \ref{tab:clusters}, we provide cluster mass estimates
$\Mmttw$ and $\Mmtfv$, where available, based on the
$M_{\Delta}$--$\Tx$ relation.
These cluster mass estimates are corrected for statistical bias and
selection effects, and the errors of forecasted masses include
uncertainties associated with the X-ray temperature measurements, the
determination of the scaling relation with the calibration sample, and
the intrinsic scatter \citep{CoMaLit5}.
Additionally, we have included a constant bias correction factor of 
$1/(1+b_{M_{\Delta}})\simeq 1.1$ to account for mass
modeling bias as $\Mmt\to \Mmt/(1+b_{M_{\Delta}})$.
Here we adopted
$b_{M_{\Delta}}\simeq -11\percent$
evaluated at $M_\mathrm{500, true}=10^{14}\Msun$, the typical mass scale of the XXL
sample (see Appendix \ref{appendix:NFW_BAHAMAS}).

The bias-corrected $M_{\Delta}$--$\Tx$ relation is summarized as
\begin{equation}
\label{eq:MTR}
 \begin{aligned}
E(z) M_{500} &= (3.15\pm 0.48) \times 10^{13}\Msun \times
  \left(\frac{\Tx}{1\,\mathrm{keV}}\right)^{3/2},\\
E(z) M_{200} &= (4.58\pm 0.70) \times 10^{13}\Msun \times
  \left(\frac{\Tx}{1\,\mathrm{keV}}\right)^{3/2}.
 \end{aligned}
\end{equation} 
It should be noted that these weak-lensing-calibrated mass estimates are
subject to an overall systematic uncertainty of $\pm 5\percent$ (Section 
\ref{subsec:syst}).\footnote{Unlike the analysis of observable--mass
scaling relations, the overall mass calibration uncertainty is not
marginalized over in this backward forecasting analysis.}
Our results are in good agreement with those of
\citet[][hereafter \citetalias{2018AA...620A...8F}]{2018AA...620A...8F},
who constrained the characteristic mass scale of the XXL sample to be
$E(z)M_{200}\simeq 1.3\times 10^{14}\Msun$ at $\Tx=2$\,keV and $z=0.3$
from an ensemble spectroscopic analysis of 132 
spectroscopically confirmed C1 and C2 clusters in the XXL-N field.

\section{Summary and Conclusions}
\label{sec:summary}

In this paper, we have presented an ensemble weak-lensing analysis of 
 X-ray galaxy groups and clusters selected from the XXL DR2 catalog
\citepalias{2018AA...620A...5A}
 using the HSC survey data \citep{hsc2018dr1,Mandelbaum2018shear}.
 Our joint weak-lensing and X-ray analysis focused on 136
 spectroscopically confirmed X-ray-selected systems of class C1 and C2 
 ($0.031 \le z \le 1.033$)
 detected in the 25\,deg$^2$ XXL-N region, which largely
 overlaps with the HSC-\XMM\ field (Figure \ref{fig:sample}).
The area of the overlap region between the two surveys is
 $21.4$\,deg$^2$.

With the HSC weak-lensing data, we have measured the tangential 
shear signal around each individual XXL cluster.
We constrained the mass and concentration parameters individually 
for each cluster by fitting an NFW profile to the
$\Delta\Sigma$ profile over the comoving radial range
$R\in [0.3,3]\,\Mpch$.
In the fitting, we used the covariance
matrix $C=C^\mathrm{shape}+C^\mathrm{lss}+C^\mathrm{int}$ that
accounts for various sources of statistical errors (Section
\ref{subsec:cmat}).
We find an excellent internal consistency between individual and
stacked weak-lensing measurements in terms of the weighted average 
mass of each sample (Table \ref{tab:stack}; 
see Equations (\ref{eq:geom}) and (\ref{eq:wlmean})).
In this consistency check, we find no systematic trend with respect to
the X-ray temperature $\Tx$ (Table \ref{tab:tstack}).

We have characterized the systematic uncertainties in the mass and 
concentration measurements using both empirical approaches and
simulations (Section \ref{subsec:syst}).
There are two possible main sources of systematics in our weak-lensing 
analysis of the XXL sample:
(i) modeling of systems detected with low values of weak-lensing SNR   
(Figure \ref{fig:SNR}) and (ii) the modeling uncertainty due to
systematic deviations from the assumed NFW form in projection.
We used two complementary sets of simulations to assess the
impact of these systematic effects (Appendix \ref{appendix:test}). 

To examine the first possibility, we analyzed synthetic weak-lensing
data based on simulations of analytical NFW lenses (Appendix
\ref{appendix:test_lognormal}), which closely match our observations in
terms of the weak-lensing SNR distribution
(Figures \ref{fig:Mtrue_lognormal} and \ref{fig:SNR_lognormal}).  
Simulations show that the overall mass scale of an XXL-like sample
can be recovered within $3.3\percent$ accuracy from individual cluster 
mass estimates, with no systematic dependence on cluster mass
$M_\mathrm{true}$.
This level of systematic uncertainty is below the statistical precision
of the current full sample, $\simeq 9\percent$ at
$M_{200}\sim 9\times 10^{13}\Msunh$
(Table \ref{tab:stack}). 
Our shear-to-mass procedure is also stable and unbiased against the
presence of low-SNR clusters (Figure \ref{fig:cM_lognormal}).

On the other hand, the results from the DM-only BAHAMAS simulation
suggest a significant level of mass bias of $\sim -20\percent$ for
low-mass group systems with $\Mtrue\simlt 4\times 10^{13}\Msunh$
(Appendix \ref{appendix:test_BAHAMAS}; see Table \ref{tab:BAHAMAS}). 
Since we do not find such a mass-dependent behavior when using the
correct mass profile shape (Appendix \ref{appendix:test_lognormal}), 
this negative bias is likely caused by systematic deviations
from the assumed NFW profile shape in projection (Section
\ref{subsec:modeling}). 
With the present data, the typical mass measurement uncertainty for 
such low-mass groups is $\sigma(M)/M\sim 140\percent$ per cluster.
Even when averaging over all such clusters, the statistical uncertainty
on the mean mass is of the order of $\simgt 20\percent$ (Section
\ref{subsec:single}). Therefore, this level of systematic bias ($\simlt
1\sigma$) is not expected to significantly affect the present
analysis. In principle, one can correct for such mass-dependent
calibration bias using a Bayesian regression approach to
forward-modeling such systematic effects.

We have established the $c_{200}$--$M_{200}$ relation for the full C1+C2 
 sample of 136 XXL clusters,
  by accounting for selection bias and statistical effects and
  marginalizing over the overall mass calibration uncertainty of 
  $\dfmcal$ (Section \ref{subsec:cMR}).  
We find the mass slope of the $c_{200}$--$M_{200}$ relation to 
 be $\beta_{Y|Z}=\betaCM$ and the normalization to be
 $c_{200}=\Cpivot$ at $M_{200}=10^{14}\Msunh$ and $z=0.3$
 (Table \ref{tab:cM} and Figure \ref{fig:cMfit}).

As shown in Figure \ref{fig:cMR}, our weak-lensing results on the
 $c_{200}$--$M_{200}$ relation are in good agreement with those found
 for X-ray, SZE, and weak-lensing-selected high-mass clusters 
 \citep{Umetsu2016clash,Okabe+Smith2016,Cibirka2017,Sereno2017psz2lens,Miyazaki2018wl},  
 as well as with DM-only predictions calibrated
 for recent $\Lambda$CDM cosmologies
  \citep[e.g.,][]{Bhatt+2013,Diemer+Kravtsov2015,Child2018cm,Diemer+Joyce2019}.
Our results are also in excellent agreement with the
 $c_{200}$--$M_{200}$ relation obtained by \citet{Biviano+2017} for a
 sample of 49 nearby clusters from a dynamical analysis of the projected
 phase-space distribution of cluster members.
 
The lognormal intrinsic dispersion in the $c_{200}$--$M_{200}$ relation  
 for the XXL sample is constrained as
$\sigma(\ln{c_{200}}) < \sigmalnCup$  
 ($99.7\percent$ CL), which is smaller than predicted for the full
 population of $\Lambda$CDM halos,
 $\sigma(\ln{c_{200}})\sim 33\percent$ \citep{Bhatt+2013,Child2018cm}.
 This discrepancy is likely caused in part by the X-ray selection bias in
 terms of the cool-core or relaxation state
 \citep[e.g.,][]{Buote+2007,Ettori+2010,Rasia+2013}. 
 Alternatively, the intrinsic dispersion $\sigma(\ln{c_{200}})$ can be
 underestimated if the statistical errors on $c_{200}$ for individual
 clusters are overestimated.

We have also determined the $T_\mathrm{X}$--$M_{500}$ relation
 for a subset of 105 XXL clusters that have both measured HSC
 lensing masses, $M_{500}$, and X-ray temperatures, $\Tx$ (Section
 \ref{subsec:TMR}; see Table \ref{tab:TM} and Figure \ref{fig:TMfit}). 
Again, we have accounted for selection bias and statistical effects,
marginalizing over the mass calibration uncertainty of $\dfmcal$.
We find the mass slope of the $T_\mathrm{X}$--$M_{500}$ relation to 
 be $\beta_{Y|Z}=\betaTM$ and the normalization to be
 $\Tx=\Tpivot$\,keV at $M_{500}=10^{14}\Msun$ and
 $z=0.3$, with a lognormal intrinsic dispersion of
 $\sigma(\ln{\Tx}) = \sigmalnT$.

The resulting $T_\mathrm{X}$--$M_{500}$ relation is consistent within
 the errors with the secondary-infall prediction based on the halo FP
 relation \citep{Fujita+2018a,Fujita+2018b}, as well as with the
 self-similar expectation.
Our $T_\mathrm{X}$--$M_{500}$ relation is also in agreement with those
 obtained by \citet{Kettula2015} and \citet{Mantz2016} (Figures \ref{fig:TMR} and
 \ref{fig:TMR_gamma066}).
At group scales, our results agree with the XXL DR1 results of
\citetalias{2016AA...592A...4L}
 based on the CFHTLenS shear catalog (Figure \ref{fig:TMR}). 
 However, our $T_\mathrm{X}$--$M_{500}$ relation has a slightly steeper
 mass trend, implying a smaller mass scale in the cluster regime. 
The overall offset in the $T_\mathrm{X}$--$M_{500}$ relation is at the
 $\sim 1.5\sigma$ level (Figures \ref{fig:TMR} and
 \ref{fig:TMR_gamma066}), corresponding to a mean mass offset of
 $\bXXL$ (Section \ref{subsec:dr1dr2}; see Figure \ref{fig:dr1dr2}).  
This discrepancy is likely due to the different fitting procedures for
 extracting cluster masses from weak-lensing data 
\citep[Section \ref{subsec:dr1dr2}; see also][]{Lieu2017}.

The change of the mass scale has important implications for cluster
astrophysics probed with the XXL sample. Compared to the XXL DR1
results
\citepalias{2016AA...592A...4L},
our HSC mass calibration leads to a higher gas mass fraction,
$f_\mathrm{gas,500}=0.053\pm 0.015$ at $M_{500}=5\times 10^{13}\Msun$
and $z=0.3$, and a lower level of hydrostatic mass bias,
$b_\mathrm{HE}=(9\pm 17)\percent$ \citep{Sereno2020xxl}.
Our HSC weak-lensing analysis thus alleviates the tension reported by 
\citetalias{2016AA...592A..12E}.
On the other hand, this slight decrease of the mass scale has a direct 
impact on the cosmological interpretation of the abundance
\citep[][hereafter \citetalias{2018AA...620A..10P}]{2018AA...620A..10P}
and clustering properties
\citep[][hereafter \citetalias{2018AA...620A...1M}]{2018AA...620A...1M}
of the XXL sample across cosmic time.

Finally, we have produced bias-corrected, weak-lensing-calibrated mass  
estimates, $\Mmttw$ and $\Mmtfv$, for individual XXL clusters
based on their X-ray temperatures (Section \ref{subsec:forecast}; see
Table \ref{tab:clusters}).
We recommend using these statistically corrected
$M_{\Delta,\mathrm{MT}}$ as a mass estimate for a given individual
cluster.
It is important to note that the weak-lensing-calibrated
$M_{\Delta}$--$T_\mathrm{X}$ relation (Equation (\ref{eq:MTR})) allows
us to estimate $M_{200}$ and $M_{500}$ for all XXL clusters with measured
X-ray temperatures, including those in the XXL-S region.
Such lensing-calibrated mass estimates corrected for statistical and
selection effects will be particularly useful for a statistical
characterization of cluster properties through multiwavelength follow-up
observations.


\acknowledgments

XXL is an international project based on an XMM Very Large Program
surveying two 25\,deg$^2$ extragalactic fields at a depth of  
$\sim 6\times10^{-15}$\,erg\,s$^{-1}$\,cm$^{-2}$ in the
0.5--2\,keV band. 
The XXL website is \href{http://irfu.cea.fr/xxl}{http://irfu.cea.fr/xxl}.
Multiband information and spectroscopic follow-up of the X-ray sources are
obtained through a number of survey programs, summarized at
\href{http://xxlmultiwave.pbworks.com/}{http://xxlmultiwave.pbworks.com/}.

The HSC Collaboration includes the astronomical communities of Japan and
Taiwan, as well as Princeton University. The HSC instrumentation and
software were developed by the National Astronomical Observatory of Japan (NAOJ),
the Kavli Institute for the Physics and Mathematics of the Universe
(Kavli IPMU), the University of Tokyo, the High Energy Accelerator
Research Organization (KEK), the Academia Sinica Institute
for Astronomy and Astrophysics in Taiwan (ASIAA), and
Princeton University. Funding was contributed by the FIRST
program from the Japanese Cabinet Office, the Ministry of
Education, Culture, Sports, Science and Technology (MEXT),
the Japan Society for the Promotion of Science (JSPS), the
Japan Science and Technology Agency (JST), the Toray Science
Foundation, NAOJ, Kavli IPMU, KEK, ASIAA, and Princeton
University.

This paper makes use of software developed for the Large
Synoptic Survey Telescope. We thank the LSST Project for
making their code available as free software at
\href{http://dm.lsst.org}{http://dm.lsst.org}. 

The Pan-STARRS1 Surveys (PS1) have been made possible
through contributions of the Institute for Astronomy, the
University of Hawaii, the Pan-STARRS Project Office, the
Max-Planck Society and its participating institutes, the Max
Planck Institute for Astronomy, Heidelberg and the Max
Planck Institute for Extraterrestrial Physics, Garching, The
Johns Hopkins University, Durham University, the University
of Edinburgh, Queen's University Belfast, the Harvard-
Smithsonian Center for Astrophysics, the Las Cumbres
Observatory Global Telescope Network Incorporated, the
National Central University of Taiwan, the Space Telescope
Science Institute, the National Aeronautics and Space Administration
under grant No. NNX08AR22G issued through the 
Planetary Science Division of the NASA Science Mission
Directorate, the National Science Foundation under grant No.
AST-1238877, the University of Maryland, Eotvos Lorand
University (ELTE), and the Los Alamos National Laboratory.

This work is based on data collected at the Subaru Telescope
and retrieved from the HSC data archive system, which is
operated by the Subaru Telescope and Astronomy Data Center,
National Astronomical Observatory of Japan.

This work is based on observations obtained with \XMMNewton, an ESA
science mission with instruments and contributions directly funded by
ESA Member States and NASA.

We thank the anonymous referee for providing useful suggestions and
comments.
We thank Lucio Chiappetti for his careful examination of the
manuscript.
K.U. acknowledges fruitful discussions with Sandor M. Molnar, Shutaro
Ueda, and August Evrard. 
This work is supported by the Ministry of Science and Technology of
Taiwan (grant MOST 106-2628-M-001-003-MY3) and by the Academia Sinica 
Investigator award (grant AS-IA-107-M01).
This work is supported in part by Hiroshima University's in-house
grant for international conferences, under the MEXT's 
Program for Promoting the Enhancement of Research Universities, Japan. 
The Saclay group acknowledges long-term support from the Centre National
d'Etudes Spatiales (CNES).
M.S. and S.E. acknowledge financial contributions from contract ASI-INAF 
n.2017-14-H.0 and INAF ``Call per interventi aggiuntivi a sostegno della 
ricerca di main stream di INAF''.
D.R. was supported by a NASA Postdoctoral Program Senior Fellowship at
the NASA Ames Research Center, administered by the Universities Space
Research Association under contract with NASA.





\begin{appendix}

\section{Mass Measurement Tests}
\label{appendix:test}


\begin{figure*}[!htb] 
 \begin{center}
  \includegraphics[scale=0.4, angle=0, clip]{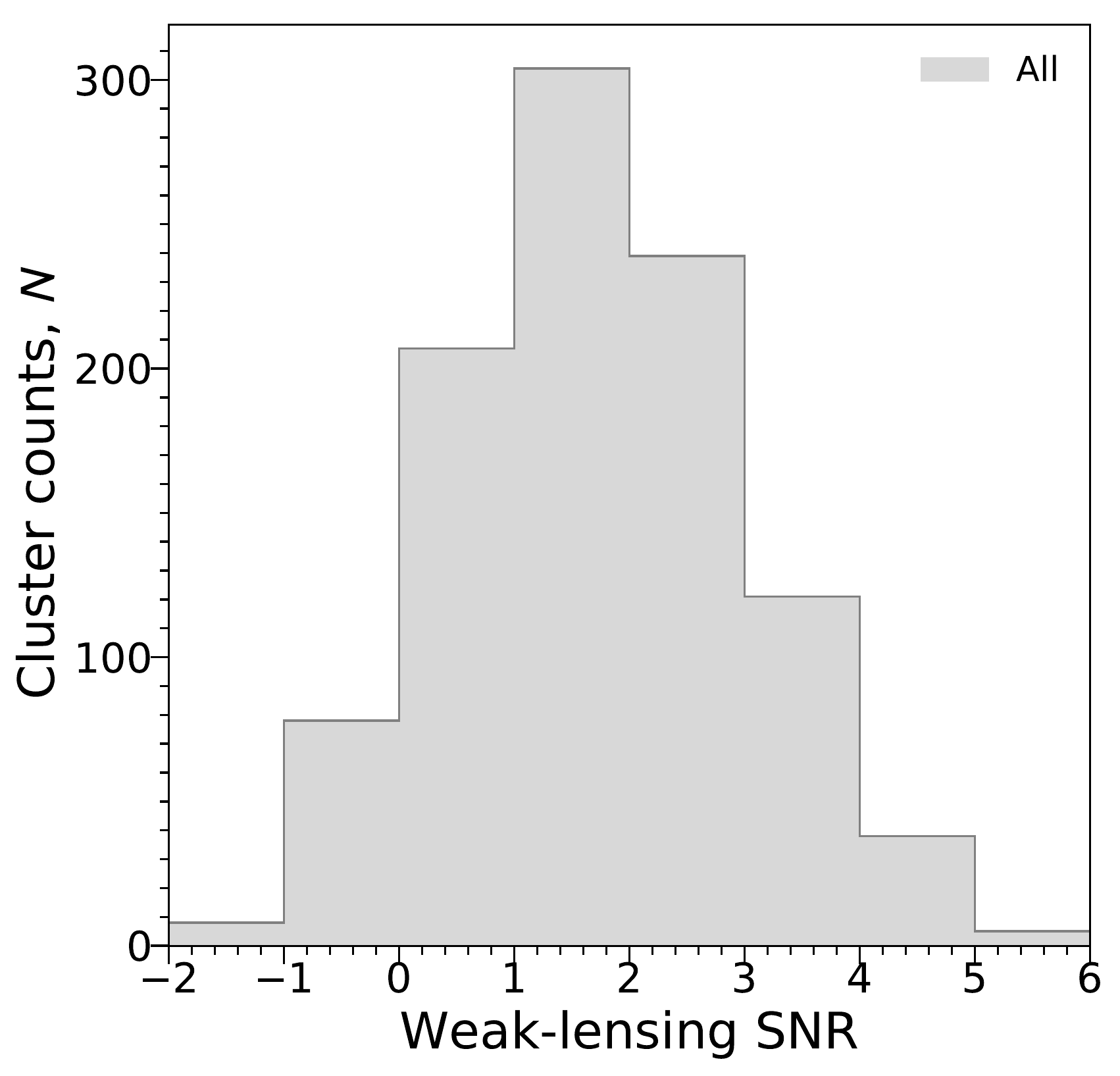}
  \includegraphics[scale=0.4, angle=0, clip]{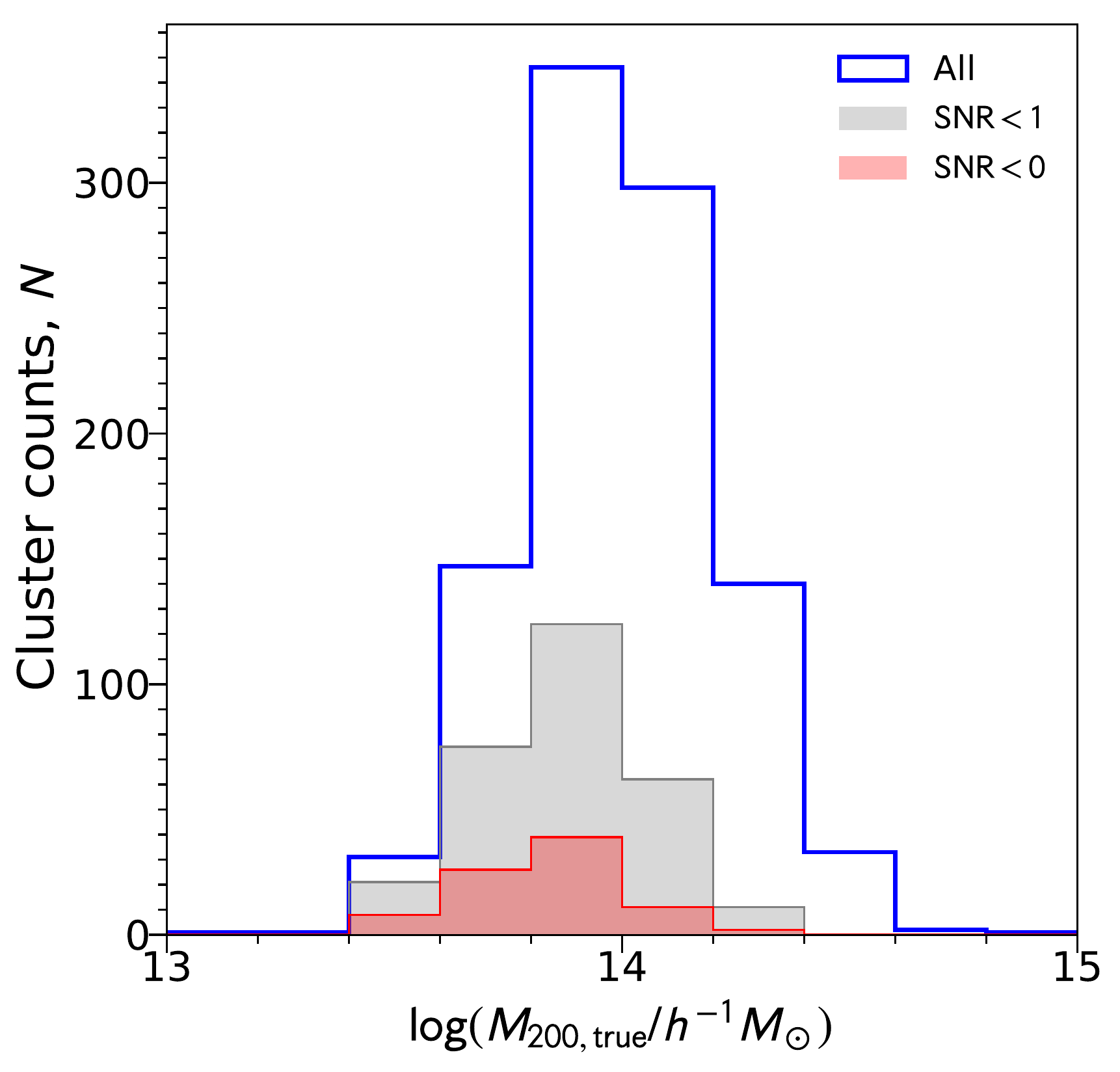}
 \end{center}
 \caption{
Synthetic weak-lensing observations from simulations of
 1000 NFW lenses at $z=0.3$.
Left panel:
 histogram distribution of the weak-lensing SNR derived
 from synthetic weak-lensing data.
Right panel: histogram distribution of the true cluster mass
 $\Mtrue$.
 The histograms are shown separately for subsamples of weak-lensing
 $\mathrm{SNR}<0$ (red shaded) and $\mathrm{SNR}<1$ (gray shaded),
 as well as for the full sample (blue).
 \label{fig:Mtrue_lognormal}
 } 
 \end{figure*}


\begin{figure*}[!htb] 
    \begin{center}    
     \includegraphics[scale=0.33, angle=0, clip]{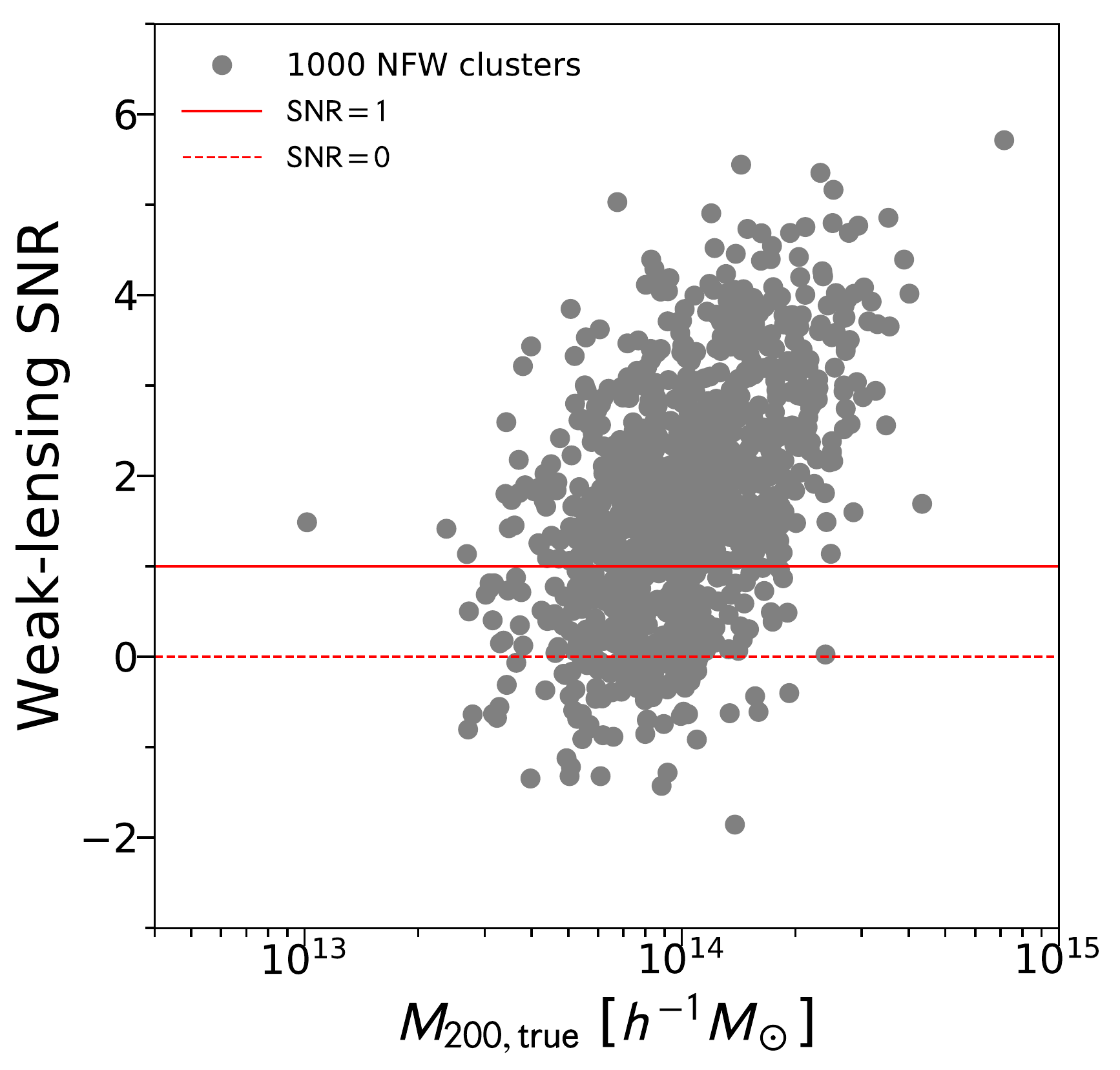}
     \includegraphics[scale=0.33, angle=0, clip]{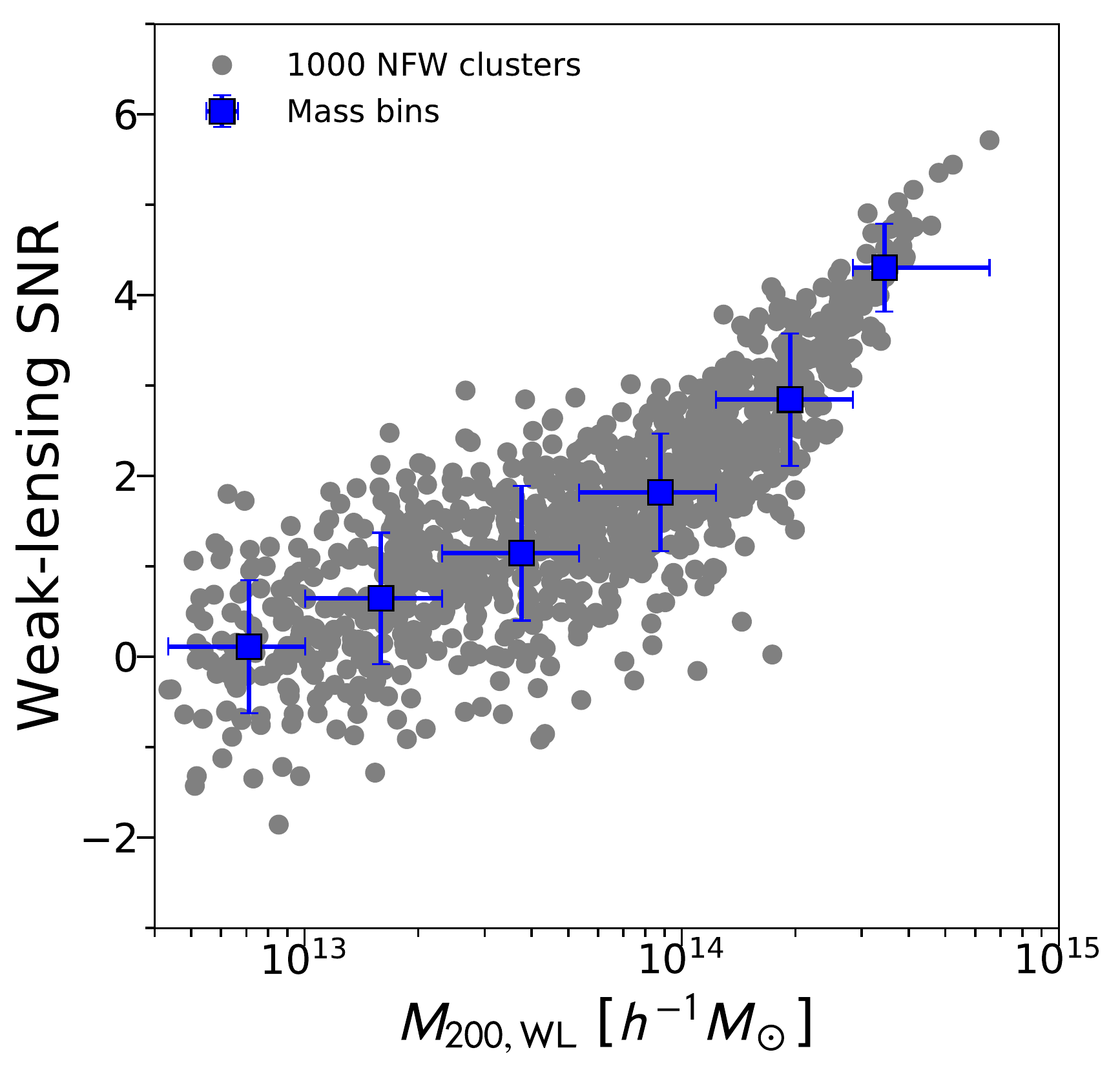}
     \includegraphics[scale=0.33, angle=0, clip]{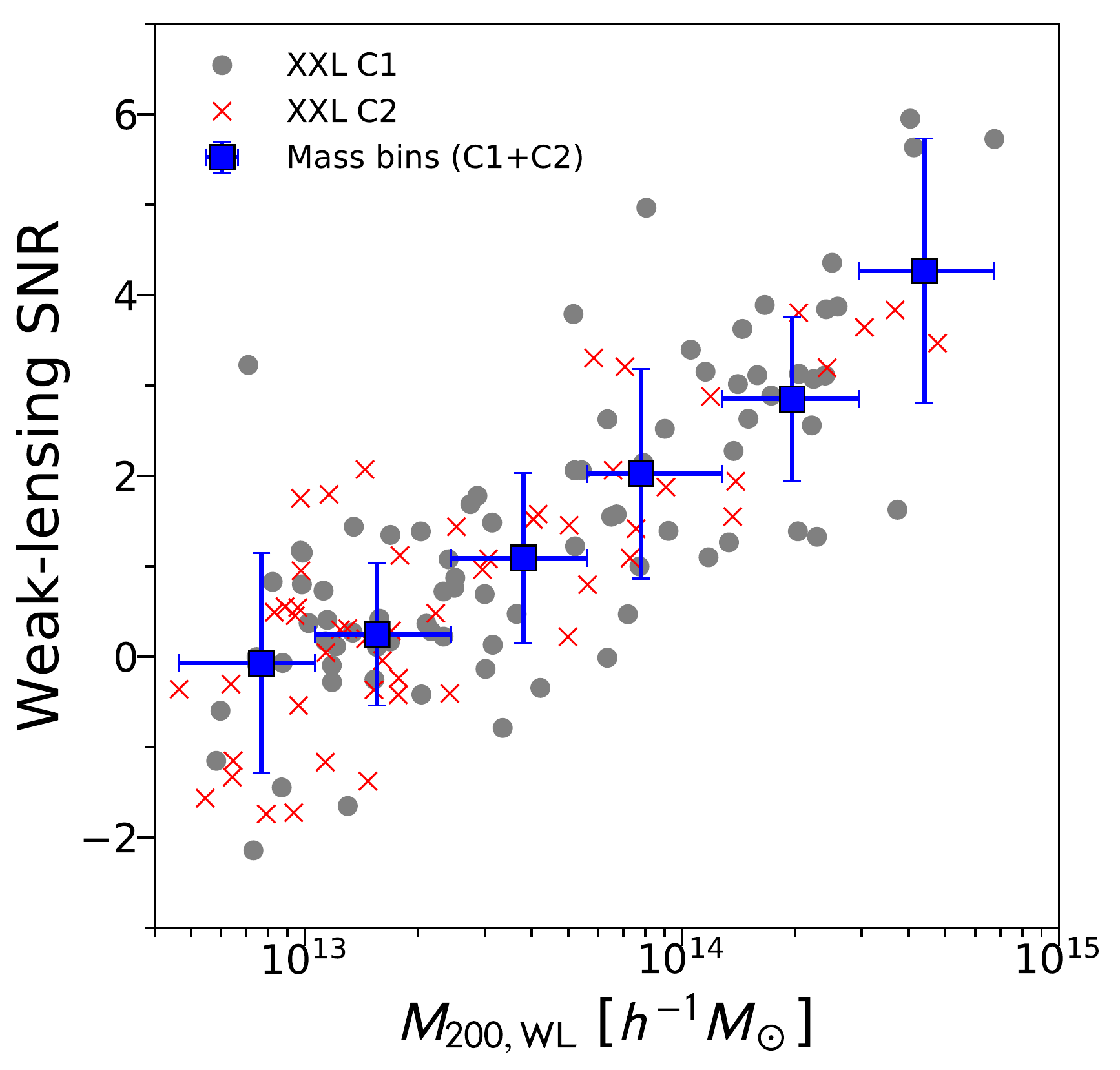}
    \end{center}
 \caption{
Left panel:
 comparison of the weak-lensing SNR and the true mass
 $\Mtrue$ (gray circles) for a synthetic sample of 1000 NFW
 lenses (see Figure \ref{fig:Mtrue_lognormal}).
 The red solid and red dashed horizontal lines represent $\mathrm{SNR}=1$
 and $\mathrm{SNR}=0$, respectively.
Middle panel: weak-lensing SNR versus weak-lensing mass $\Mwl$
 estimated from the synthetic weak-lensing data using the same analysis
 pipeline as for the real observations.
 The blue squares represent weighted geometric means in six logarithmic 
 $M_{200}$ bins, 
 where the vertical bars show the standard deviation of the
 weak-lensing SNR and the horizontal bars show the full width of each
 mass bin.
Right panel: same as the middle panel, but for the real observations
 of the XXL sample.
The gray circles and red crosses represent the C1 and C2 subsamples, 
 respectively. 
\label{fig:SNR_lognormal}
 }
\end{figure*}


\begin{figure*}[!htb] 
 \begin{center}
  \includegraphics[scale=0.33, angle=0, clip]{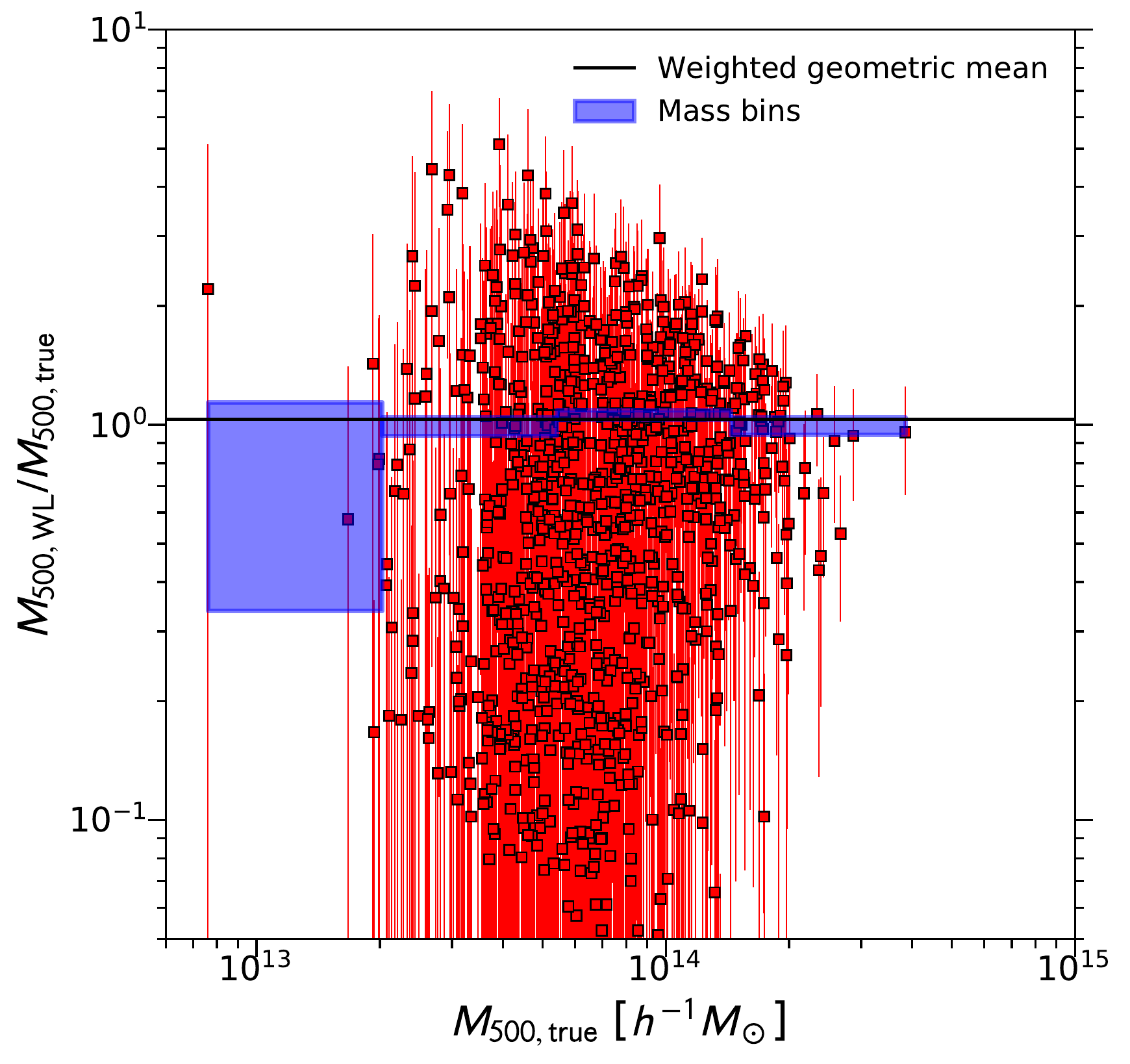}
  \includegraphics[scale=0.33, angle=0, clip]{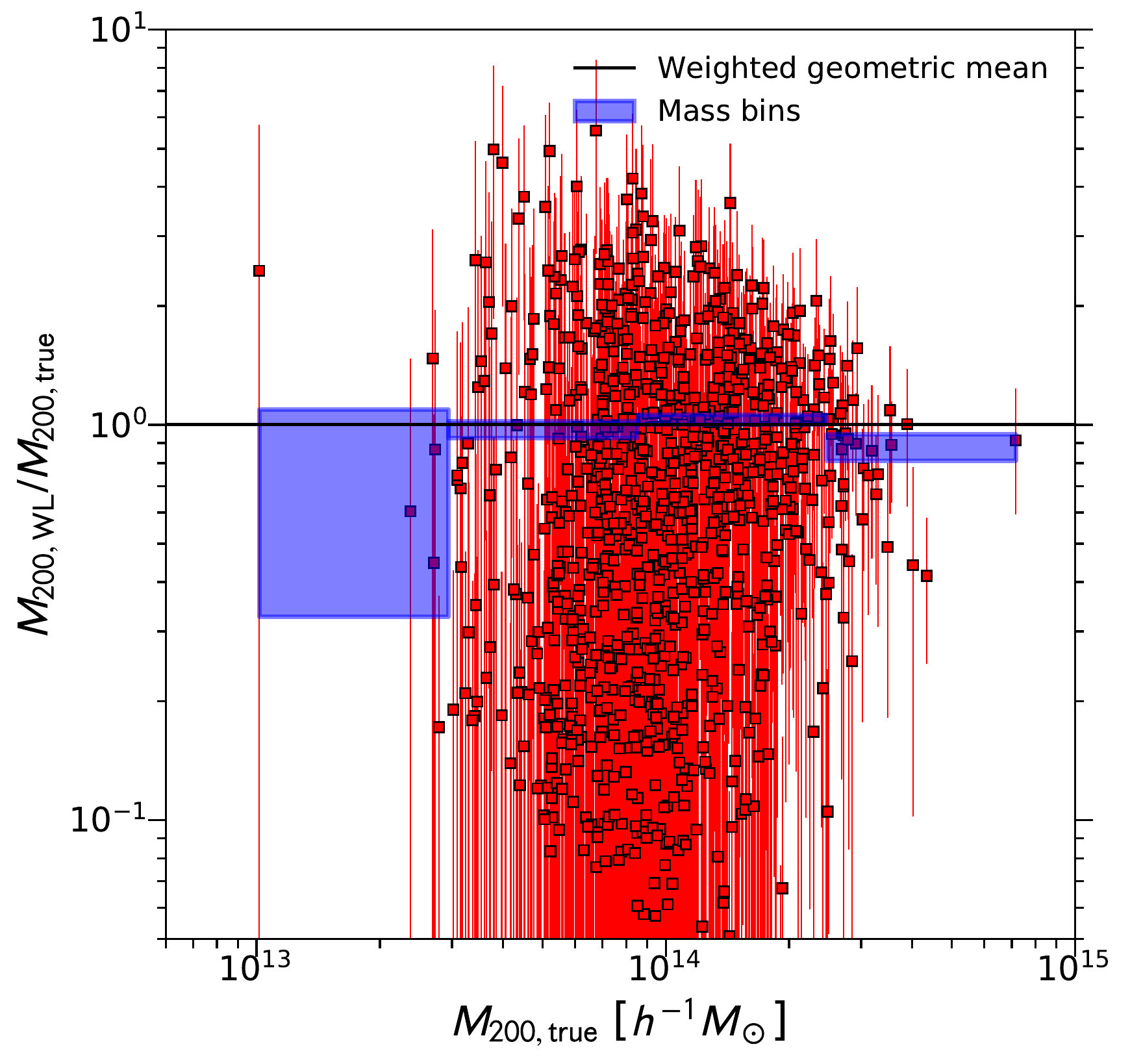}
  \includegraphics[scale=0.33, angle=0, clip]{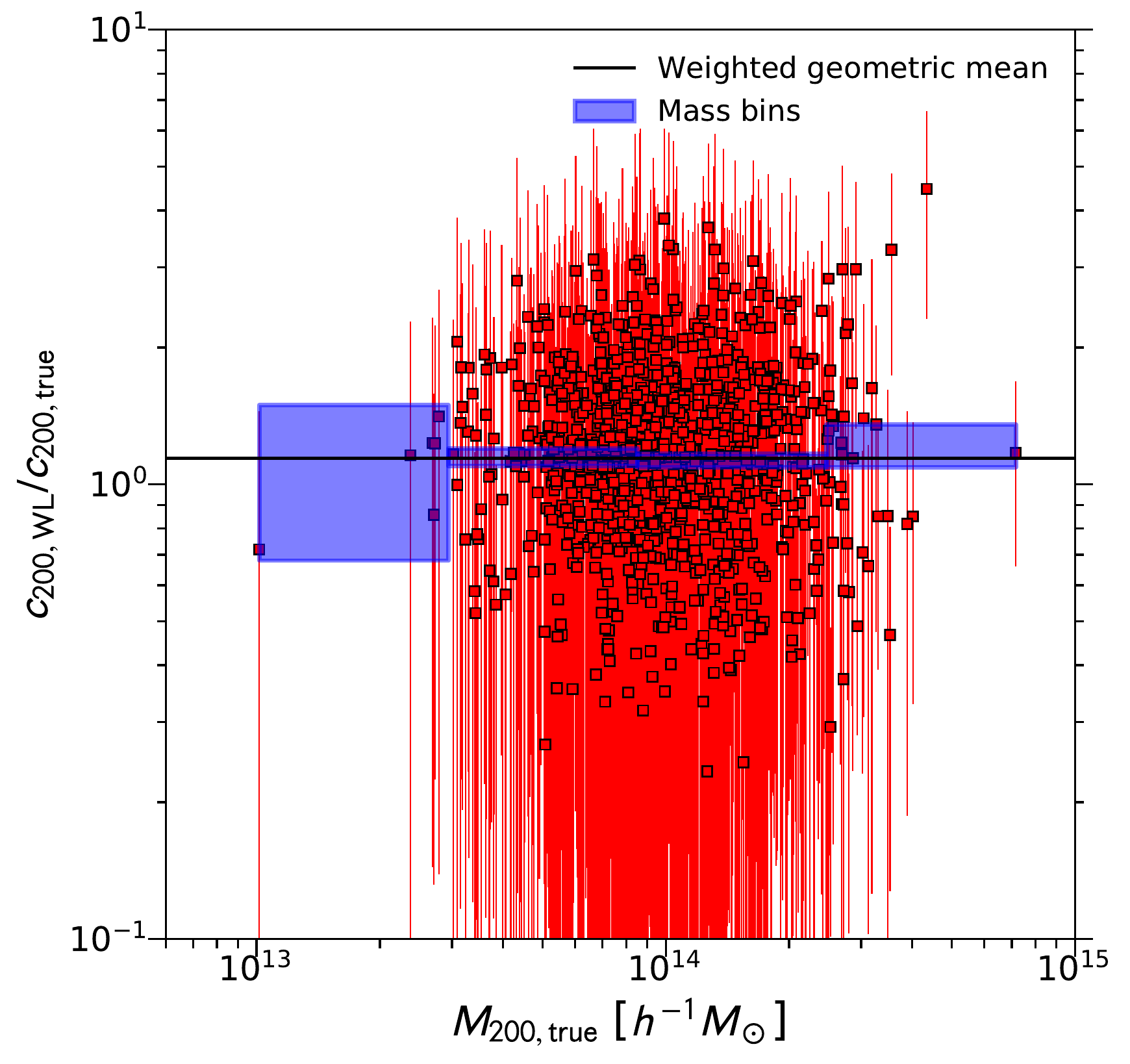}
 \end{center}
 \caption{
 Comparison of the true and estimated values of the mass
 and concentration parameters $M_{500}$ (left), $M_{200}$ (middle), and
 $c_{200}$ (right) from synthetic weak-lensing observations of 1000 NFW
 lenses at $z=0.3$ (see Figure \ref{fig:SNR_lognormal}).
In each panel, the error-weighted geometric mean ratio of the simulated sample,
 $\langle M_{\Delta,\mathrm{WL}}/M_{\Delta,\mathrm{true}}\rangle_\mathrm{g}$ or
 $\langle \cwl/\ctrue\rangle_\mathrm{g}$,
 is marked with a solid line.
Similarly, the shaded blue boxes represent weighted geometric mean
  ratios and their errors in four equally log-spaced $M_\mathrm{true}$ bins.
  The weighted average mass ratio
  $\langle M_{\Delta,\mathrm{WL}}/M_{\Delta,\mathrm{true}}\rangle_\mathrm{g}$  
  is consistent with unity to better than $2\sigma$ in all mass bins.
  Overall, $\cwl$ is biased high at a mean level of
  $(13\pm 3)\percent$,
  with no evidence of systematic mass dependence.
  \label{fig:cM_lognormal}
 }
 \end{figure*}

\subsection{Simulations of Analytical NFW Lenses}
\label{appendix:test_lognormal}

First, we test and quantify the accuracy of our cluster mass
 measurements using synthetic weak-lensing data
 that closely match the HSC survey in terms of the weak-lensing  
 SNR distribution. 
Specifically,
 the aim of this test is to assess the impact of low weak-lensing SNR
 objects on ensemble mass measurements for a sample of XXL-like
 clusters. 
To this end, we create synthetic weak-lensing data from
simulations of analytical NFW lenses at a redshift of $z=0.3$, 
 the median redshift of the full C1+C2 sample (Table \ref{tab:stack}).  
We model the weak-lensing signal of each cluster using the
 ``true'' profile shape (i.e., NFW),
 with $M_{200}$ and $c_{200}$ as fitting parameters.
We use the same analysis pipeline as done for the real
 observations.
 In this way, we can separate possible sources of systematic effects.
Hence, any significant level of mass bias,
 especially in the low-mass regime,
 would indicate systematics effects
caused by noisy mass estimates for low-SNR objects.

A synthetic sample of 1000 NFW lenses was drawn from a
 Gaussian intrinsic PDF in $Z=\log{(M_{200}/\Msunh)}$ with 
 a mean $\mu_Z=\langle Z\rangle=14$ and
 a dispersion $\sigma_Z=0.5/\ln{10}$, which closely resembles the XXL
 cluster sample
\citepalias{2016AA...592A...2P,2018AA...620A...5A}.
Concentrations were drawn from the scattered $c_{200}$--$M_{200}$
 relation of \citet{Bhatt+2013} with a lognormal intrinsic dispersion
 of $\sigma(\ln{c_{200}})=0.15\ln{10}\simeq 35\percent$.
The range of true $M_{200}$ masses for the simulated sample is
 $1.0\times 10^{13}\simlt M_{200}/(\Msunh)\simlt 7.2\times 10^{14}$
 (see the right panel of Figure \ref{fig:Mtrue_lognormal}). 
The synthetic data include 
 the cosmic noise contribution due to the projected
 uncorrelated large-scale structure, as well as the random shape noise,
 with a net intrinsic shear dispersion of
 $\sigma_g =0.4$ per shear component.
 Source galaxies are distributed over the redshift range $0.3<z_s<1.2$
 with a mean number density of $\ngal=17$ galaxies\,arcmin$^{-2}$. 
Finally, the $\Delta\Sigma(R)$ profiles were simulated in eight equally 
spaced logarithmic bins of comoving cluster radius ($R$)
from $R_\mathrm{min}=0.3\Mpch$ to $R_\mathrm{max}=3\Mpch$, to be
consistent with the observations (Section \ref{subsec:DSigma}).

The left panel of Figure \ref{fig:Mtrue_lognormal} shows the
distribution of weak-lensing SNR measured in a fixed comoving aperture
of $R\in [0.3,3]\,\Mpch$ for 1000 simulated NFW lenses.
The values of weak-lensing SNR span the range from $-1.9$ to $5.7$, with 
a median of $1.7$ and a standard deviation of $1.3$, closely mimicking
the observed SNR distribution (Figure \ref{fig:SNR}). 
About $30\percent$ ($9\percent$) of simulated NFW lenses are detected
with weak-lensing $\mathrm{SNR}<1$ (0),
as shown in the right panel of Figure \ref{fig:Mtrue_lognormal}).  
The left (middle) panel of Figure \ref{fig:SNR_lognormal} compares the
weak-lensing SNR and $\Mtrue$ ($\Mwl$) for all NFW lenses in the sample.
The resulting distribution of simulated NFW lenses in the SNR--$\Mwl$
plane reproduces the observations of the XXL sample fairly well (see the
right panel of Figure \ref{fig:SNR_lognormal}).

The weighted average weak-lensing mass
$\langle \Mwl\rangle_\mathrm{g}=(1.28\pm 0.03)\times 10^{14}\Msunh$
over the full sample (in terms of the error-weighted geometric mean;
see Equation \ref{eq:geom})
is $\simeq 30\percent$ higher than the true log-mean (or the true
median) mass, $\Mtrue = 10^{14}\Msunh$, and the true mean mass of the
population, $\Mtrue\simeq 1.13\times 10^{14}\Msunh$. 
Qualitatively, this is because the weighted geometric mean estimator
assigns higher weights to those objects with smaller measurement errors,
which are likely to be more massive objects. The degree to which
$\langle \Mwl\rangle_\mathrm{g}$ is different from the true
population mean should depend on both the shape of the intrinsic mass
PDF and the level of observational noise.

We introduce the following quantity to characterize the level of bias in
the average cluster mass estimated from weak lensing:   
\begin{equation}
\label{eq:bsim}
 1+b_{\mathrm{sim},M_\Delta} = \langle M_{\Delta,\mathrm{WL}}/M_{\Delta,\mathrm{true}}\rangle_\mathrm{g},
\end{equation}
where $M_{\Delta,\mathrm{true}}$ represents the true $M_\Delta$
mass from simulations and $M_{\Delta,\mathrm{WL}}$ represents the
$M_\Delta$ mass estimated from weak lensing. Similarly, we define the
bias parameter $b_{\mathrm{sim}, c_{200}}$ for the concentration
parameter, $c_{200}$.

Figure \ref{fig:cM_lognormal} shows that
$b_{\mathrm{sim},M_{500}}$ and $b_{\mathrm{sim},M_{200}}$
are consistent with zero to better than $2\sigma$
in all mass bins, with no significant mass dependence over the full
range of $M_{\Delta,\mathrm{true}}$.
On the other hand, $\cwl$ is biased high at a mean level of 
  $b_{\mathrm{sim},c_{200}}=(13\pm 3)\percent$,
 with no evidence of systematic mass dependence.
This systematic offset is likely because the typical scale
radius for this sample ($r_\mathrm{s}\simeq 0.21\,\Mpch$ in comoving
length units) lies below the radial range for fitting,
$R\in [0.3,3]\,\Mpch$ (comoving).

In this realization, there are a total of 86 clusters with negative
values of weak-lensing SNR.
Their weak-lensing mass estimates span the range
$\Mwl\in [0.4, 11]\times 10^{13}\Msunh$, with a median
value of $1.0\times 10^{13}\Msunh$, which is comparable to our
observations (Section \ref{subsec:single}). 
The median mass uncertainty of these clusters is
$\SBI(\Mwl)/\CBI(\Mwl)\sim 140\percent$.
This indicates that such noisy objects can reach $\Mwl/\Mtrue\sim 4$
(i.e., the boundary of the $99.7\percent$ confidence region; see Figure
\ref{fig:cM_lognormal}).  
As shown in the right panel of Figure \ref{fig:Mtrue_lognormal}
(see also the left panel of Figure \ref{fig:SNR_lognormal}),
these clusters span a fairly representative range in ``true'' mass:  
$M_{200}\in [2.7, 19]\times 10^{13}\Msunh$,
with a median value of $7.0\times 10^{13}\Msunh$
and a mean value of $7.5\times 10^{13}\Msunh$.
At a given true mass, it is expected that there is a statistical
counterpart of positively scattered clusters with apparently boosted SNR
and thus overestimated $M_{\Delta,\mathrm{WL}}$.  
In fact, we do not find any significant bias in ensemble weak-lensing
mass measurements even at low-mass scales (Figure \ref{fig:cM_lognormal}).  
In contrast, if one selects a
subsample of clusters according to their weak-lensing SNR values or mass
estimates, they are no more representative of the parent population.
In particular, such an SNR-limited selection will 
bias high the weak-lensing mass estimates at a given X-ray cut,
an effect known as the Malmquist bias
\citep[e.g.,][]{CoMaLit2,CoMaLit5,Sereno2017psz2lens}.

\subsection{BAHAMAS Simulation}
\label{appendix:test_BAHAMAS}

\subsubsection{Simulated Halos and Synthetic Weak-lensing Data} 
\label{appendix:halos_BAHAMAS}


\begin{figure*}[!htb] 
 \begin{center}
  \includegraphics[scale=0.7, angle=0, clip]{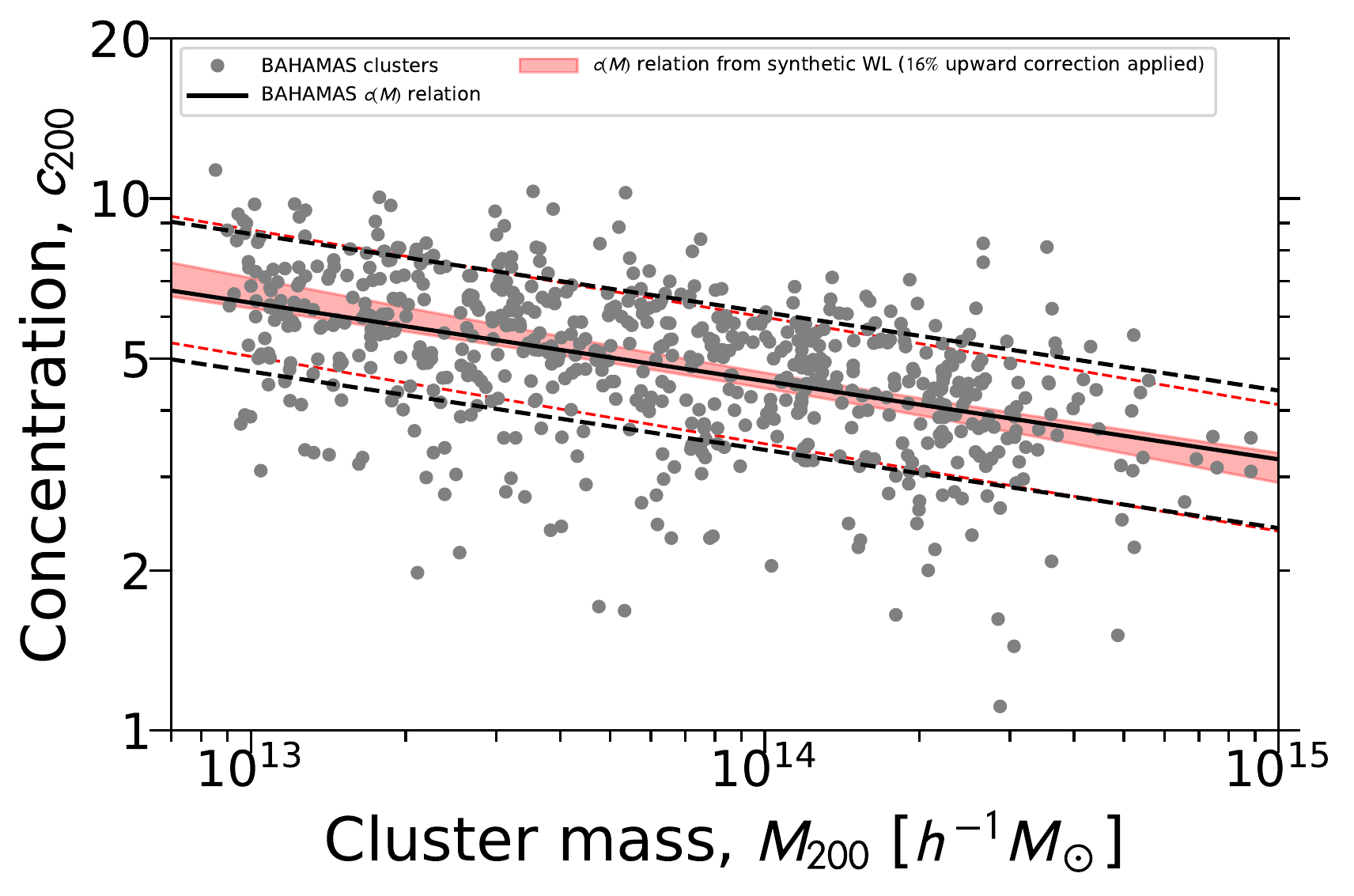}
 \end{center}
\caption{
  Halo mass and concentration of 639 $\Lambda$CDM halos (gray
  circles) selected from a DM-only realization of BAHAMAS simulations at
  $z=0.25$. The thick black line shows the $c_{200}$--$M_{200}$
  relation of the selected sample. The lognormal intrinsic scatter
  ($1\sigma$) around this relation is indicated by a pair of black
  dashed lines.  
  The red shaded region shows the $1\sigma$ range of the mean
  $c_{200}$--$M_{200}$ relation recovered from a regression analysis 
  of the synthetic weak-lensing measurements
  $(\Mwl, \cwl)$ shown in Figure \ref{fig:BAHAMAS}. 
  Here an upward correction of $16\percent$ is applied to the
  normalization inferred from the regression analysis.
  The inferred intrinsic scatter ($1\sigma$) is indicated by a pair of
  red dashed lines.   
  \label{fig:cM_BAHAMAS}
 } 
 \end{figure*}


\begin{figure*}[!htb] 
 \begin{center}
  \includegraphics[scale=0.4, angle=0, clip]{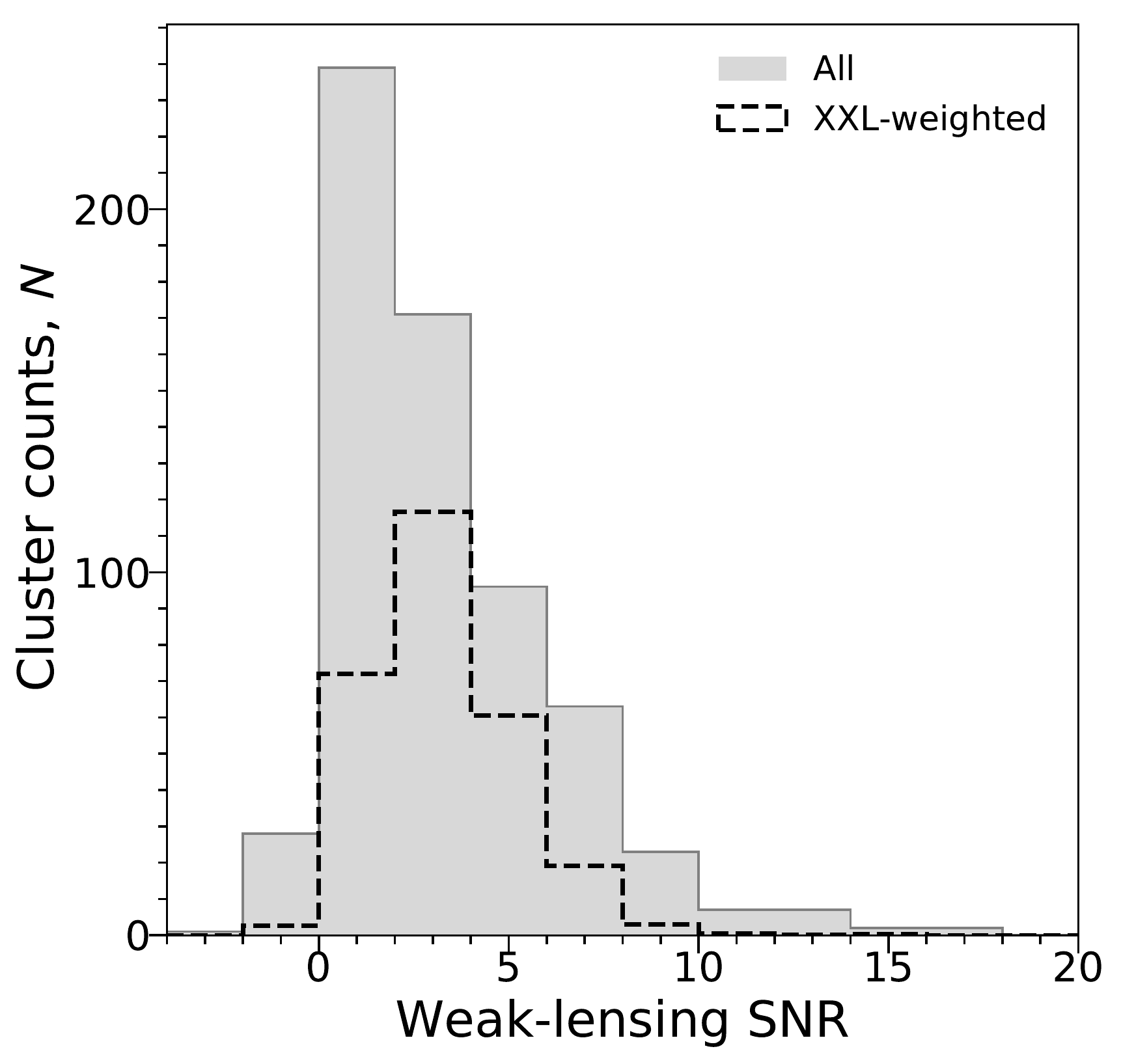}
  \includegraphics[scale=0.4, angle=0, clip]{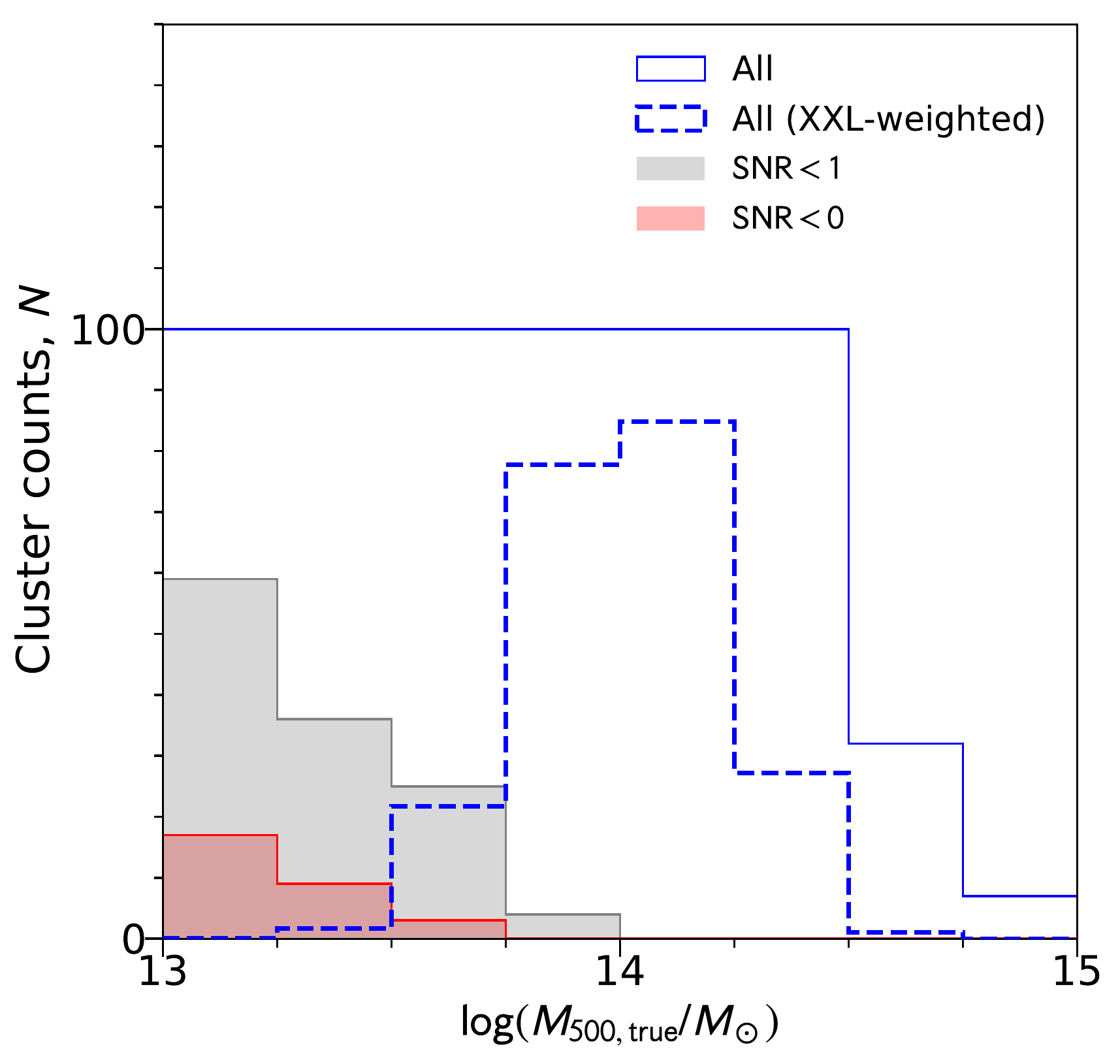}
 \end{center}
\caption{
Synthetic weak-lensing observations from a DM-only realization of
  BAHAMAS simulations at $z=0.25$.
Left panel: histogram distribution of the weak-lensing SNR derived
  from synthetic weak-lensing data.
The black dashed histogram shows an XXL-weighted
  distribution of the sample.
Right panel: histogram distribution of the true cluster mass
  $M_\mathrm{500,true}$.
The histograms are shown for subsamples of weak-lensing
  $\mathrm{SNR}<1$ (gray shaded) and $\mathrm{SNR}<0$ (red shaded),
  as well as for the full sample (blue solid).
The blue dashed histogram shows an XXL-weighted
  distribution of the sample.
  \label{fig:Mtrue_BAHAMAS}
 } 
 \end{figure*}


\begin{figure*}[!htb] 
 \begin{center}    
  \includegraphics[scale=0.4, angle=0, clip]{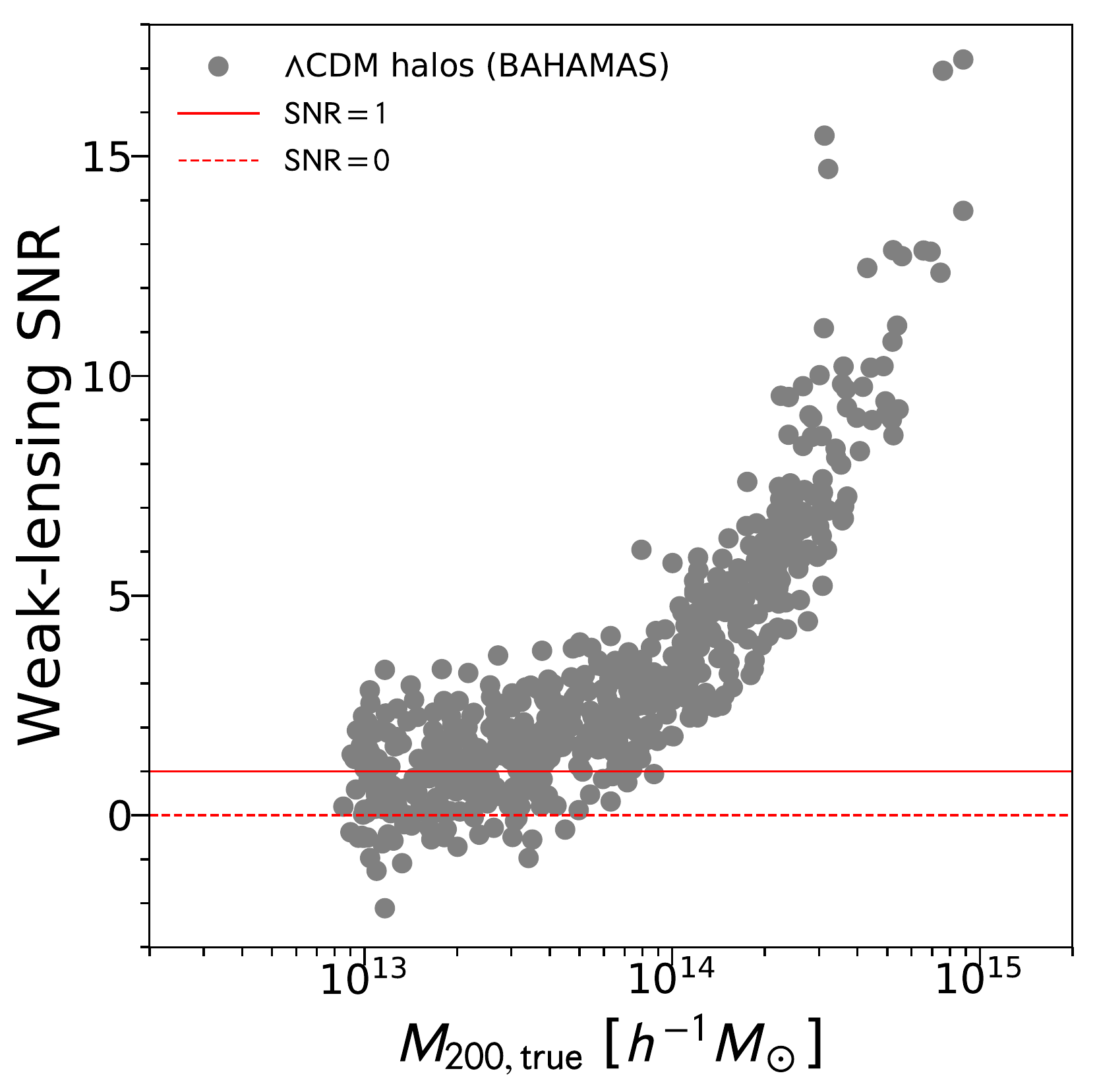}
  \includegraphics[scale=0.4, angle=0, clip]{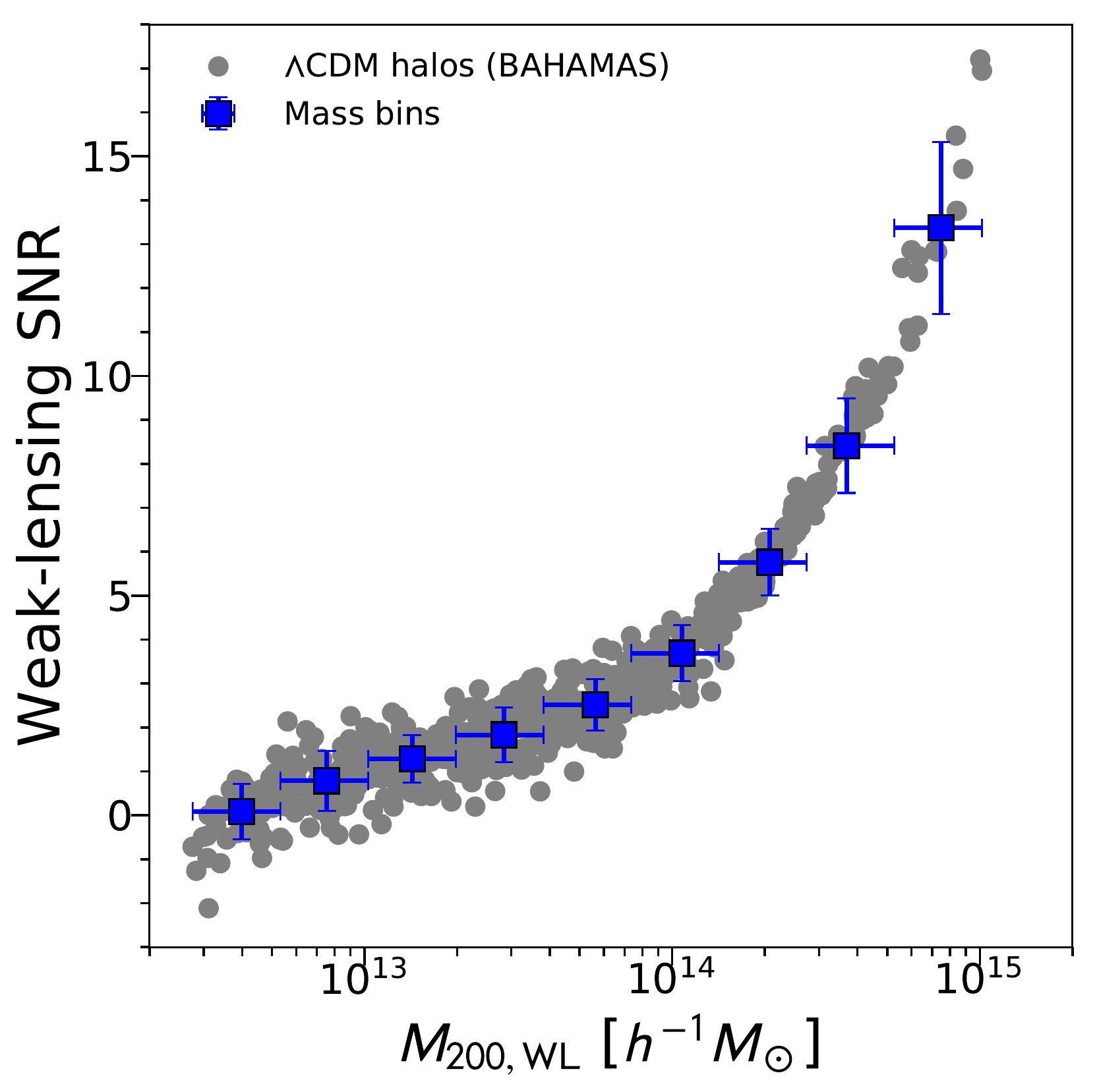}
 \end{center}
\caption{
Left panel:
 comparison of the weak-lensing SNR and the true mass $\Mtrue$ (gray
 circles) for a sample of 639 $\Lambda$CDM halos selected from a DM-only
 realization of BAHAMAS simulations.
 The red solid and red dashed horizontal lines represent
 $\mathrm{SNR}=1$ and $\mathrm{SNR}=0$, respectively. 
Right panel: 
 weak-lensing SNR versus total mass $\Mwl$ (gray circles)
 estimated from synthetic weak-lensing using the same analysis
 pipeline as done for the real observations
 (see Figure \ref{fig:SNR_lognormal}).
 The blue squares represent weighted geometric means in nine logarithmic
 $M_{200}$ bins, where the vertical bars show the standard deviation of the
 weak-lensing SNR and the horizontal bars show the full width of each
 mass bin. 
\label{fig:SNR_BAHAMAS}
 }
\end{figure*}

Next, we test and characterize the accuracy of our weak-lensing mass
measurements using synthetic observations of realistic $\Lambda$CDM
halos, selected from a DM-only run from the BAHAMAS
simulations \citep{McCarthy2017,McCarthy2018}.
The aim of this test is to assess the impact of modeling uncertainties
in the projected cluster profile shape down to low-mass group scales.
The specific simulation we use adopts a flat $\Lambda$CDM cosmology with
\WMAP\ 9\,yr cosmological parameters in a box of 400$\,\Mpch$
(comoving) on a side with $1024^3$ particles.  The particle mass is
$3.85\times10^9 \Msunh$ and the softening length is $4\kpch$ (physical).   

To efficiently survey any mass-dependent bias in our methodology, we
randomly select 100 halos per logarithmic mass bin 
$\Delta\log{M_{500}} = 0.25$ over the mass range
$\log{(M_{500}/M_\odot)} \in [13,15]$ from the simulation (i.e., a total
of eight logarithmic mass bins),
at a redshift of $z=0.25$.  We note that given the finite size of the
simulation volume, the two highest-mass bins have fewer than 100 unique
halos (they have 32 and 7, respectively).  For these bins, we select all 
halos for analysis, yielding a total sample of 639 halos.

Figure \ref{fig:cM_BAHAMAS} shows the distribution of selected halos in
the $c_{200}$--$M_{200}$ plane.
The $c_{200}$--$M_{200}$ relation of the selected sample is described by
a power law of the form
$c_{200}=4.5\times (M_{200}/10^{14}\Msunh)^{-0.15}$ with a lognormal
intrinsic dispersion of $\sigma(\ln{c_{200}}) = 30\percent$.  
The right panel of Figure \ref{fig:Mtrue_BAHAMAS} shows the distribution
of the selected halos in $M_\mathrm{500,true}$ (blue solid histogram),
along with an XXL-weighted distribution (blue dashed histogram) where
the counts are weighted by the mass PDF expected for the XXL sample 
(Appendix \ref{appendix:test_lognormal}).

Around each selected halo, we extract all particles in a cube of length
30\,Mpc (physical) centered on the most bound particle of each selected
halo.  The particle distribution is then projected along the $z$-axis and
interpolated to a regular two-dimensional grid using a triangular-shaped
cloud algorithm to produce an image of surface mass density.
We compute convergence and reduced shear maps from the surface mass
density map following the methods described in
\citet[][see their Section 3.4.3]{McCarthy2018},
assuming a single source redshift plane at $z_s=0.829$.
We randomly sample the reduced shear maps to obtain a mean background
source density of $\ngal=20$ galaxies\,arcmin$^{-2}$.  We then add
shape noise to the selected shear values, drawing from a normal
distribution with a dispersion of $\sigma_g=0.28/\sqrt{2}\simeq 0.20$ 
per shear component.

\subsubsection{NFW Modeling}
\label{appendix:NFW_BAHAMAS}

\begin{deluxetable*}{cccc|ccc|ccc}
\tablecolumns{10}
\tablewidth{0pt}
\tabletypesize{\scriptsize}
\tablecaption{\label{tab:BAHAMAS}
Systematic Bias in Weak-lensing Mass Modeling}
\tablehead{
\multicolumn{4}{c}{} &
\multicolumn{3}{c}{NFW Model} &
\multicolumn{3}{c}{Halo Model} \\
\multicolumn{1}{c}{$M_\mathrm{500,true}$
\tablenotemark{a}} &
\multicolumn{1}{c}{$M_\mathrm{200,true}$\tablenotemark{a}} &
\multicolumn{1}{c}{$c_\mathrm{200,true}$\tablenotemark{a}} &
\multicolumn{1}{c}{$N_\mathrm{cl}$} &
\multicolumn{1}{c}{$b_{\mathrm{sim},M_{500}}$} &
\multicolumn{1}{c}{$b_{\mathrm{sim},M_{200}}$} &
\multicolumn{1}{c}{$b_{\mathrm{sim},c_{200}}$} &
\multicolumn{1}{c}{$b_{\mathrm{sim},M_{500}}$} &
\multicolumn{1}{c}{$b_{\mathrm{sim},M_{200}}$} &
\multicolumn{1}{c}{$b_{\mathrm{sim},c_{200}}$} \\
\multicolumn{1}{c}{($10^{13}\Msunh$)} &\multicolumn{1}{c}{($10^{13}\Msunh$)} &
\colhead{} &
\colhead{} &
\multicolumn{1}{c}{(\percent)} &
\multicolumn{1}{c}{(\percent)} &
\multicolumn{1}{c}{(\percent)} &
\multicolumn{1}{c}{(\percent)} &
\multicolumn{1}{c}{(\percent)} &
\multicolumn{1}{c}{(\percent)} }
\startdata
$0.8$ & $1.2$ & $6.1$ & $100$ & $-19\pm 8$ & $-15\pm 8$ & $-14\pm 8$ & $-15\pm 8$ & $-12\pm 9$ & $-13\pm 8$\\ 
$1.5$ & $2.2$ & $5.7$ & $100$ & $-24\pm 7$ & $-22\pm 7$ & $-18\pm 7$ & $-20\pm 7$ & $-19\pm 7$ & $-16\pm 8$\\ 
$2.8$ & $3.9$ & $5.6$ & $100$ & $-21\pm 6$ & $-18\pm 6$ & $-13\pm 8$ & $-16\pm 6$ & $-14\pm 6$ & $-10\pm 8$\\ 
$4.9$ & $7.4$ & $5.0$ & $100$ & $-11\pm 5$ & $-9\pm 5$ & $-24\pm 6$ & $-7\pm 5$ & $-5\pm 5$ & $-24\pm 6$\\ 
$8.8$ & $12.4$ & $4.6$ & $100$ & $-5\pm 3$ & $-2\pm 3$ & $-28\pm 4$ & $-2\pm 3$ & $2\pm 3$ & $-30\pm 4$\\ 
$15.8$ & $23.1$ & $4.1$ & $100$ & $8\pm 2$ & $8\pm 2$ & $-13\pm 4$ & $10\pm 2$ & $11\pm 3$ & $-16\pm 4$\\ 
$25.5$ & $37.0$ & $4.0$ & $32$ & $3\pm 3$ & $6\pm 3$ & $-24\pm 5$ & $4\pm 3$ & $8\pm 4$ & $-26\pm 5$\\ 
$55.5$ & $74.5$ & $3.2$ & $7$ & $1\pm 4$ & $9\pm 6$ & $-2\pm 8$ & $2\pm 4$ & $10\pm 5$ & $-3\pm 8$ 
\enddata
\tablecomments{We characterize the accuracy of our weak-lensing mass measurements using synthetic observations of 639 $\Lambda$CDM halos at $z=0.25$ selected from a DM-only realization of BAHAMAS simulations. We quantify the level of bias in the average cluster mass from weak lensing as $1+b_{\mathrm{sim}, M_\Delta}(M_{\Delta,\mathrm{true}})=\langle M_{\Delta,\mathrm{WL}}/M_{\Delta,\mathrm{true}}\rangle_\mathrm{g}$, where $M_{\Delta,\mathrm{true}}$ is the true $M_\Delta$ mass, $M_{\Delta,\mathrm{WL}}$ is the $M_\Delta$ mass estimated from weak lensing, and those quantities in brackets with subscript "g"e denote error-weighted geometric means (Equation (\ref{eq:geom})). Similarly, we define the bias parameter $b_{\mathrm{sim}, c_{200}}$ for the concentration parameter.}
\tablenotetext{a}{True median value in each logarithmic mass bin.}
\end{deluxetable*}


\begin{figure*}[!htb] 
 \begin{center}
  \includegraphics[scale=0.33, angle=0, clip]{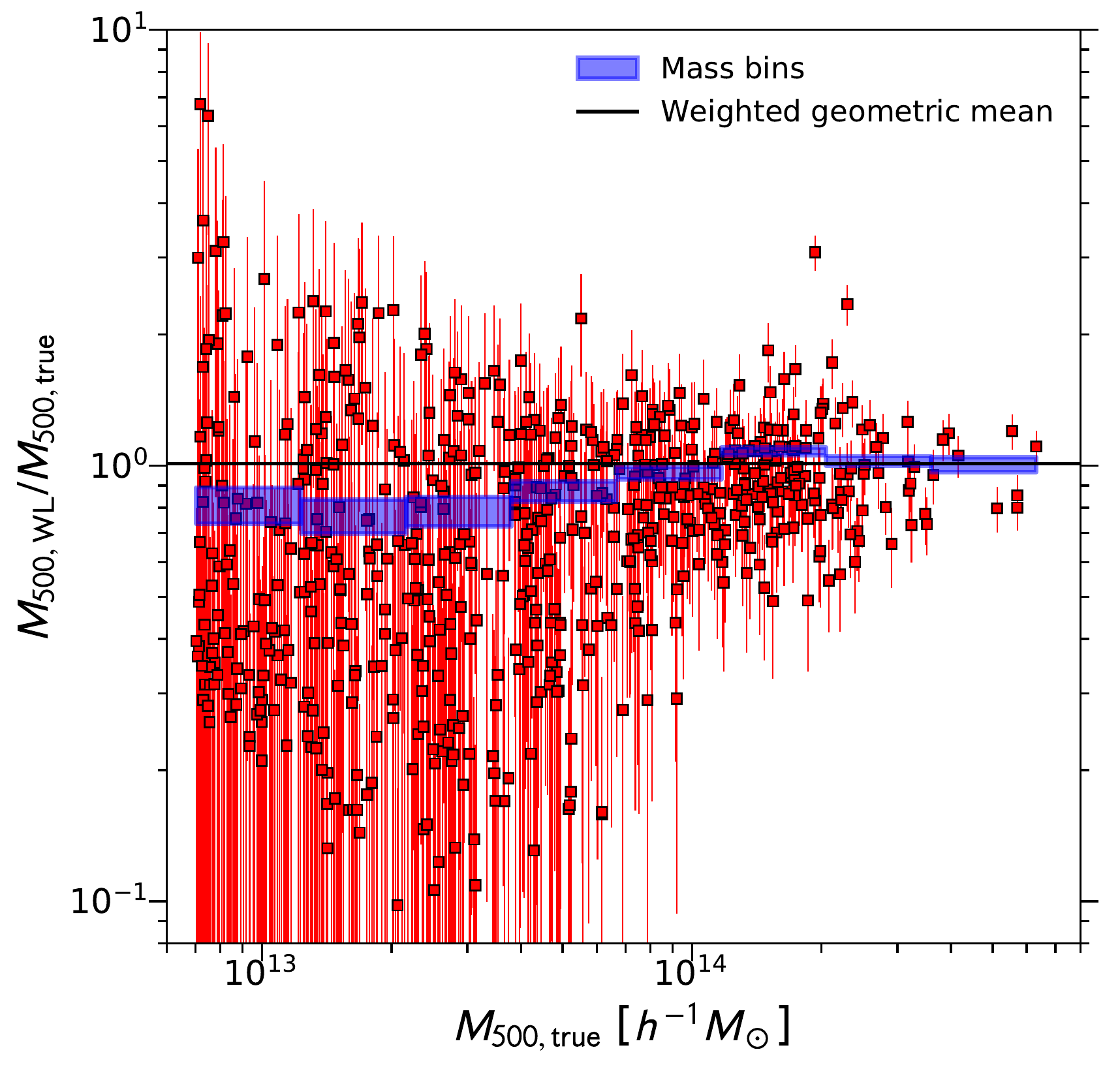}
  \includegraphics[scale=0.33, angle=0, clip]{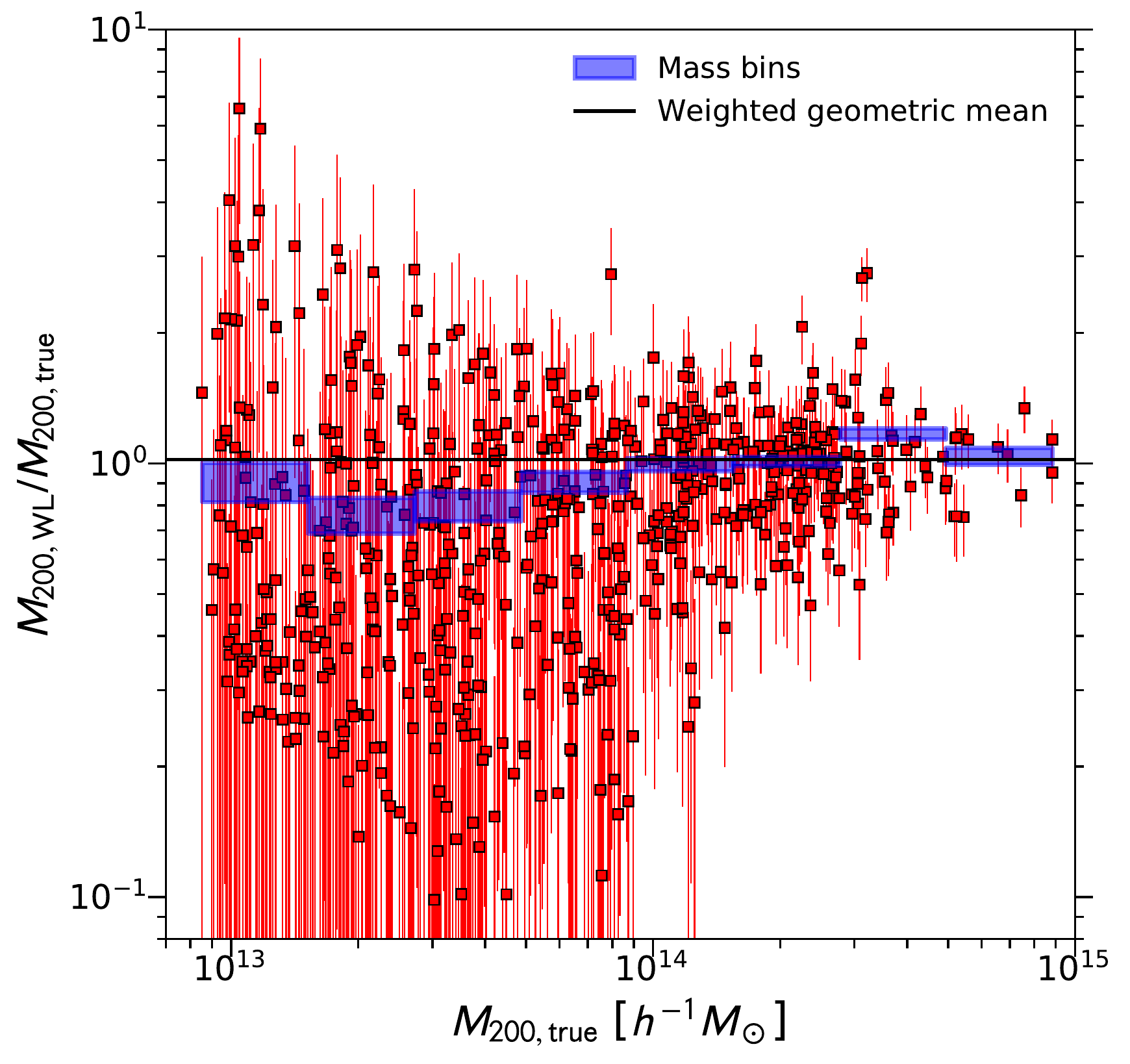}
  \includegraphics[scale=0.33, angle=0, clip]{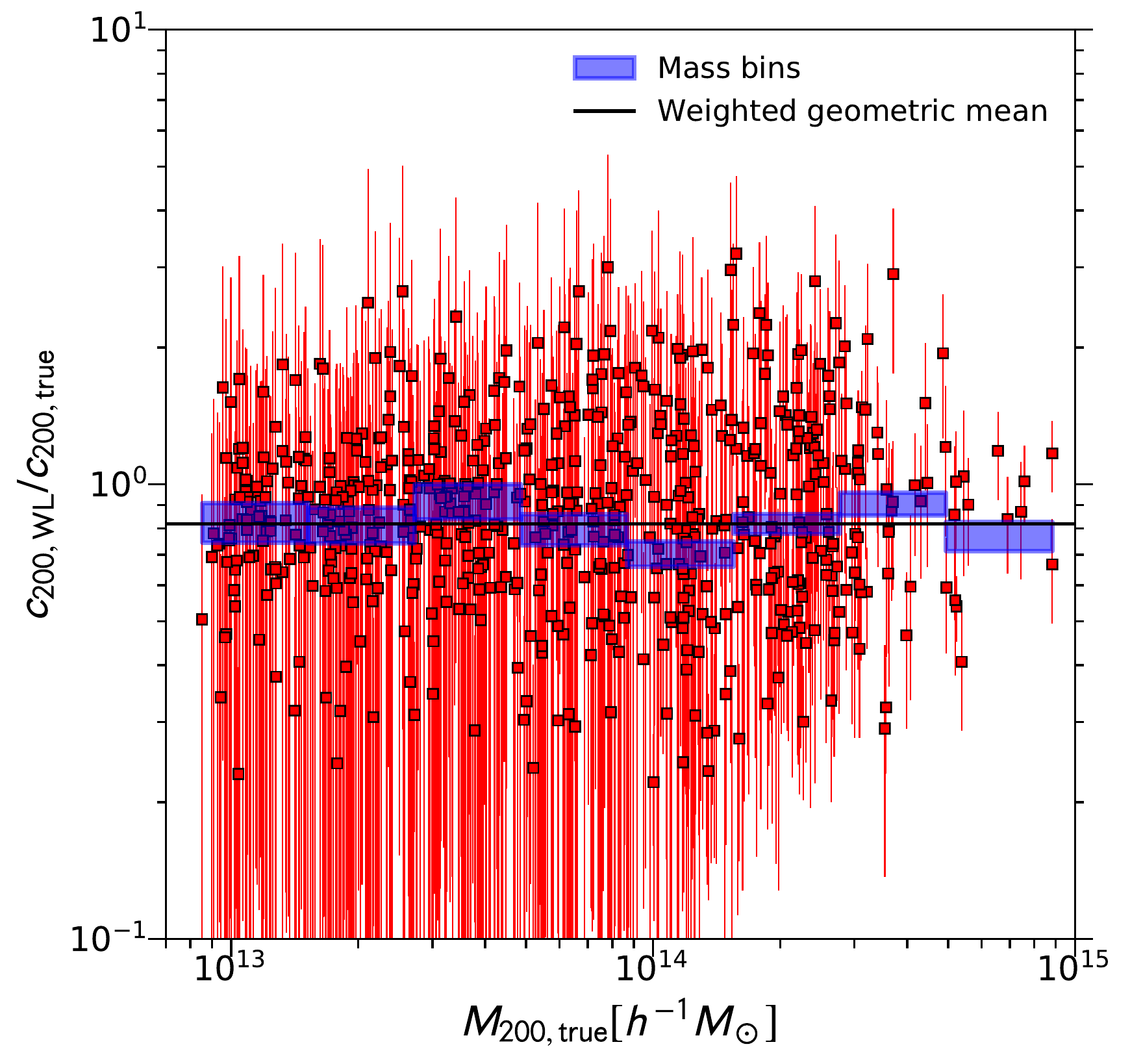}
 \end{center}
 \caption{
 Comparison of the true and estimated values of
 $M_{500}$ (left),
 $M_{200}$ (middle), and
 $c_{200}$ (right) from synthetic weak-lensing observations of 639
 $\Lambda$CDM halos at $z=0.25$ selected from a DM-only realization of
 BAHAMAS simulations. 
We model the weak-lensing signal of each individual cluster assuming an
 NFW profile.
In each panel, the weighted geometric mean ratio of the simulated sample,
 $\langle M_{\Delta,\mathrm{WL}}/M_{\Delta,\mathrm{true}}\rangle_\mathrm{g}$ or
 $\langle \cwl/c_{200,\mathrm{true}}\rangle_\mathrm{g}$,
 is marked with a solid line.
Similarly, the shaded blue boxes represent weighted geometric mean
 ratios and their errors in eight equally log-spaced $M_\mathrm{true}$ bins.
 \label{fig:BAHAMAS}
 }  
\end{figure*}

We analyze the synthetic weak-lensing data using the same analysis
pipeline as for the real observations (Section \ref{sec:wl}).    
We compute for each cluster halo the synthetic $\Delta\Sigma_+$ profile
(Equation (\ref{eq:DSigma})) and model the weak-lensing signal assuming
a spherical NFW profile with $M_{200}$ and $c_{200}$ as fitting
parameters, following the procedures laid down in Section
\ref{sec:mass}.

The left panel of Figure \ref{fig:Mtrue_BAHAMAS} shows the
distribution of weak-lensing SNR measured in a fixed comoving aperture
of $R\in [0.3,3]\,\Mpch$ for our sample of 639  halos.
The values of weak-lensing SNR span the range from
$-2.1$ to $17$, with a median of $2.5$ and a standard deviation of $2.9$.
About $20\percent$ ($5\percent$) of simulated halos are detected
with weak-lensing $\mathrm{SNR}<1$ ($0$), as shown in the right panel of 
Figure \ref{fig:Mtrue_BAHAMAS}.   
The left (right) panel of Figure \ref{fig:SNR_BAHAMAS} compares the
weak-lensing SNR and $\Mtrue$ ($\Mwl$) for all halos in the BAHAMAS
sample. 
The shape noise level assumed in this set of synthetic data is about a
factor of $2$ smaller than that in the NFW-based simulations (Appendix  
\ref{appendix:test_lognormal}).

In Figure \ref{fig:BAHAMAS}, we compare the true and estimated values of
($M_{500}, M_{200}, c_{200}$) for our simulated sample of 639 halos.
For each quantity, we compute the weighted geometric mean ratio between
the estimated and true values over the full sample, finding
$b_{\mathrm{sim}, M_{500}}=(0.9\pm1.3)\percent$,
$b_{\mathrm{sim}, M_{200}}=(2.1\pm 1.5)\percent$, and
$b_{\mathrm{sim}, c_{200}}=(-18\pm2)\percent$.

We also quantify the levels of bias in the average weak-lensing mass and 
concentration as a function of $M_\mathrm{true}$.
Table \ref{tab:BAHAMAS} lists the values of bias in 
our weak-lensing measurements of ($M_{500}, M_{200}, c_{200}$) estimated 
in eight equally log-spaced $M_{500}$ bins (see Figure
\ref{fig:Mtrue_BAHAMAS}).
We find a significant level of mass bias of $\sim -20\percent$ for
low-mass group halos with $\Mtrue\simlt 4\times 10^{13}\Msunh$, or
$M_\mathrm{500,true}\simlt 3\times 10^{13}\Msunh$.
However, such a low-mass population is expected to be subdominant in the 
XXL sample (Figures \ref{fig:MTR} and \ref{fig:Mtrue_BAHAMAS}).
At the typical mass scale
$M_{500}\simeq 7\times 10^{13}\Msunh = 10^{14}\Msun$
of the XXL sample, we find the level of bias in
$M_\mathrm{200, WL}$ and $M_\mathrm{500, WL}$
to be $\simeq -11\percent$.

\subsubsection{Recovery of the $c$--$M$ Relation}
\label{appendix:cM_BAHAMAS}

Here we test how well the parameters describing the $c_{200}$--$M_{200}$
relation can be recovered from cluster weak-lensing observations. To
this end, we perform a \texttt{LIRA} regression analysis of our synthetic
weak-lensing measurements $(\Mwl, \cwl)$ for the BAHAMAS sample.
We adopt slightly different priors to model
the BAHAMAS sample in the \texttt{LIRA} framework (Section
\ref{subsec:regression}). 
First, we set
$\gamma_{Y|Z}=\gamma_{\mu_Z,{\cal D}}=\gamma_{\sigma_Z,{\cal D}}=0$
because all halos are sampled at a single redshift of $z=0.25$.
Next, we need $\sigma_{Z,0}\simgt 1$ to approximate
$P(Z)$ of the BAHAMAS sample
(see Figure \ref{fig:Mtrue_BAHAMAS})
with a lognormal distribution in $Z=\log{(M_{200}/10^{14}\Msunh)}$.
Since such a solution is strongly disfavored by the assumed prior for
$\sigma_{Z,0}$ (see Equation (\ref{eq:Pr_Gamma})), we fix the center and 
width of the lognormal mass PDF to 
$\mu_{Z,0}=0$ and $\sigma_{Z,0}=1$, respectively.
We thus have a total of four regression parameters,
($\alpha_{Y|Z},\beta_{Y|Z},\sigma_{Y|Z},\sigma_{X|Z}$).

The results are shown in Figures \ref{fig:cM_BAHAMAS} and
\ref{fig:cMfit_BAHAMAS}.
The $c_{200}$--$M_{200}$ relation recovered from the synthetic data is 
summarized as  
$c_{200}=(3.7\pm 0.1)\times (M_{200}/10^{14}\Msunh)^{-0.14\pm 0.02}$
with a logarithmic intrinsic dispersion of
$\sigma(\ln{c_{200}})=\ln{10}\sigma_{Y|Z}=(27\pm 3)\percent$.
We thus accurately recover the true input values of
$\beta_{Y|Z}=0.15$ and
$\sigma(\ln{c_{200}})=30\percent$
(see Appendix \ref{appendix:halos_BAHAMAS})
within the statistical uncertainties. On the other hand, we
underestimate the normalization of the $c_{200}$--$M_{200}$ relation by  
$(18\pm 2)\percent$, as found in Appendix \ref{appendix:NFW_BAHAMAS}.
The intrinsic dispersion of the $\Mwl$--$M_{200}$ relation is
found to be
$\sigma(\ln{\Mwl})=\ln{10}\sigma_{X|Z}=(6.5\pm 4.8)\percent$,
with an upper limit of
$<23\percent$ ($33\percent$) at the $95\percent$ ($99.7\percent$) CL
(see Figure \ref{fig:cMfit_BAHAMAS}).
In Figure \ref{fig:cM_BAHAMAS},
we have applied an upward correction of $16\percent$ to the
normalization inferred from the regression analysis.

As a sanity check, we perform a similar test for lower fiducial values 
of the intrinsic dispersion $\sigma(\ln{c_{200}})$ in the
$c_{200}$--$M_{200}$ relation. To this end, 
we select a subsample of BAHAMAS halos that lie within $1\sigma$
scatter from the mean $c_{200}$--$M_{200}$ relation, which leaves us
with 459 halos. This subsample virtually has a lower dispersion of
$\sigma(\ln{c_{200}})=16\percent$, which matches the level 
expected for X-ray regular clusters  \citep{Meneghetti2014clash}.
For this subsample, we obtain
$c_{200}=(3.7\pm 0.1)\times (M_{200}/10^{14}\Msunh)^{-0.14\pm 0.02}$
with $\sigma(\ln{c_{200}})=(21\pm 4)\percent$
from the synthetic data (Figure \ref{fig:cMfit_BAHAMAS}).
We thus recover the true input values of 
$\beta_{Y|Z}=0.15$ and $\sigma(\ln{c_{200}})=16\percent$
within the statistical uncertainties.
We checked that even lower fiducial values of $\sigma(\ln{c_{200}})\sim
10\percent$ can be accurately recovered.

\subsubsection{Halo Modeling}
\label{appendix:halomodel}


\begin{figure*}[!htb] 
 \begin{center}
  \includegraphics[scale=0.33, angle=0, clip]{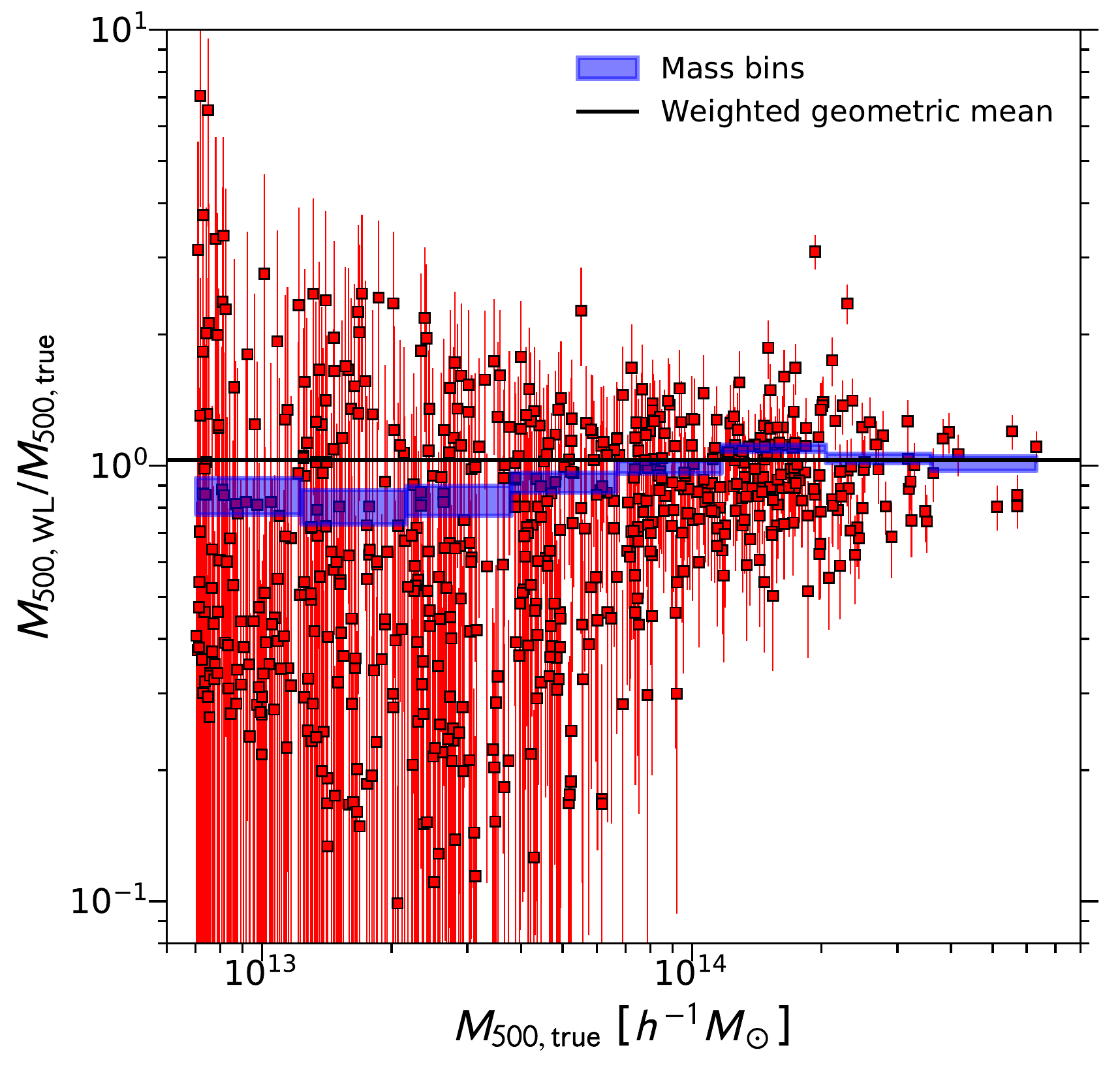}
  \includegraphics[scale=0.33, angle=0, clip]{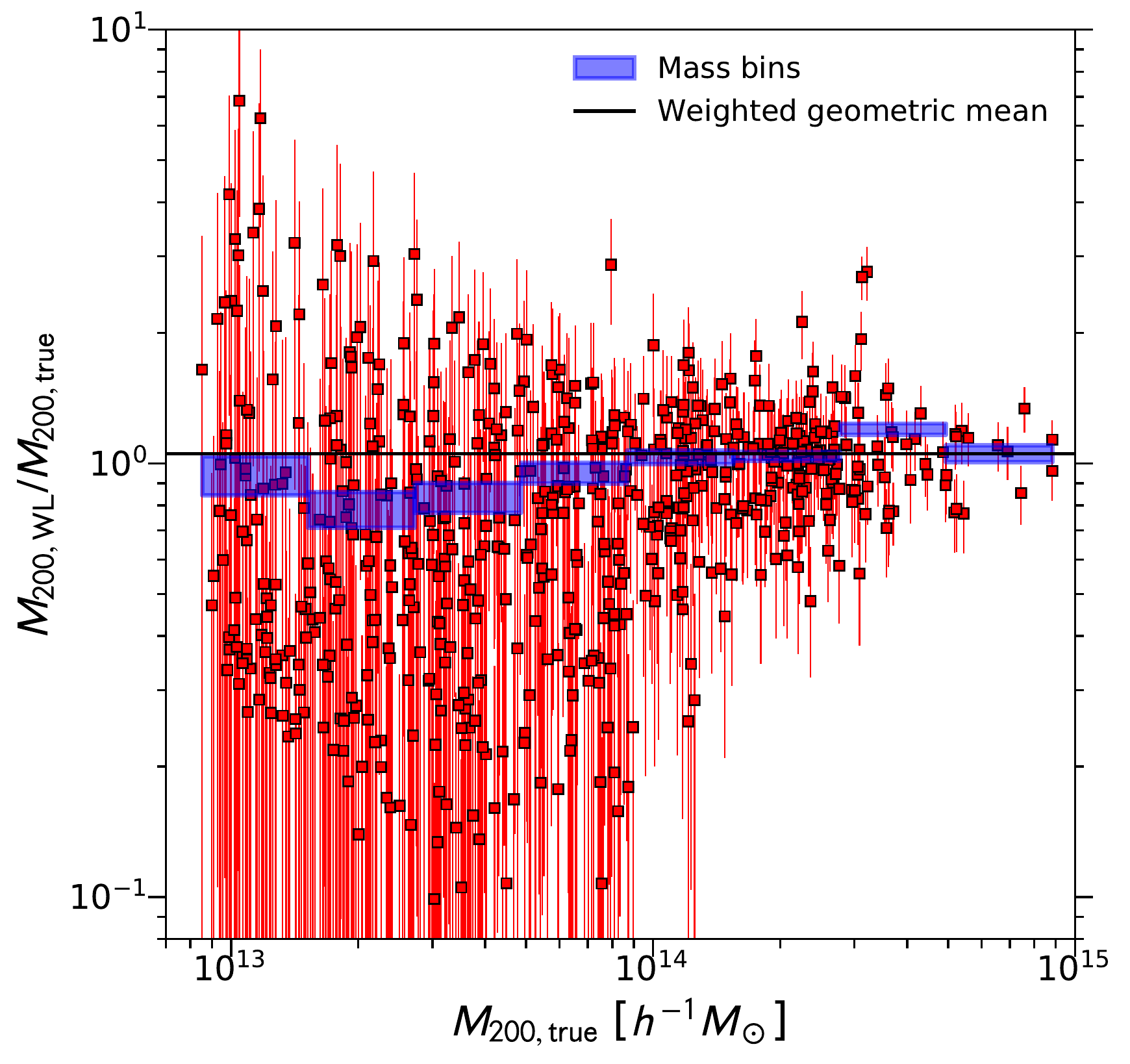}
  \includegraphics[scale=0.33, angle=0, clip]{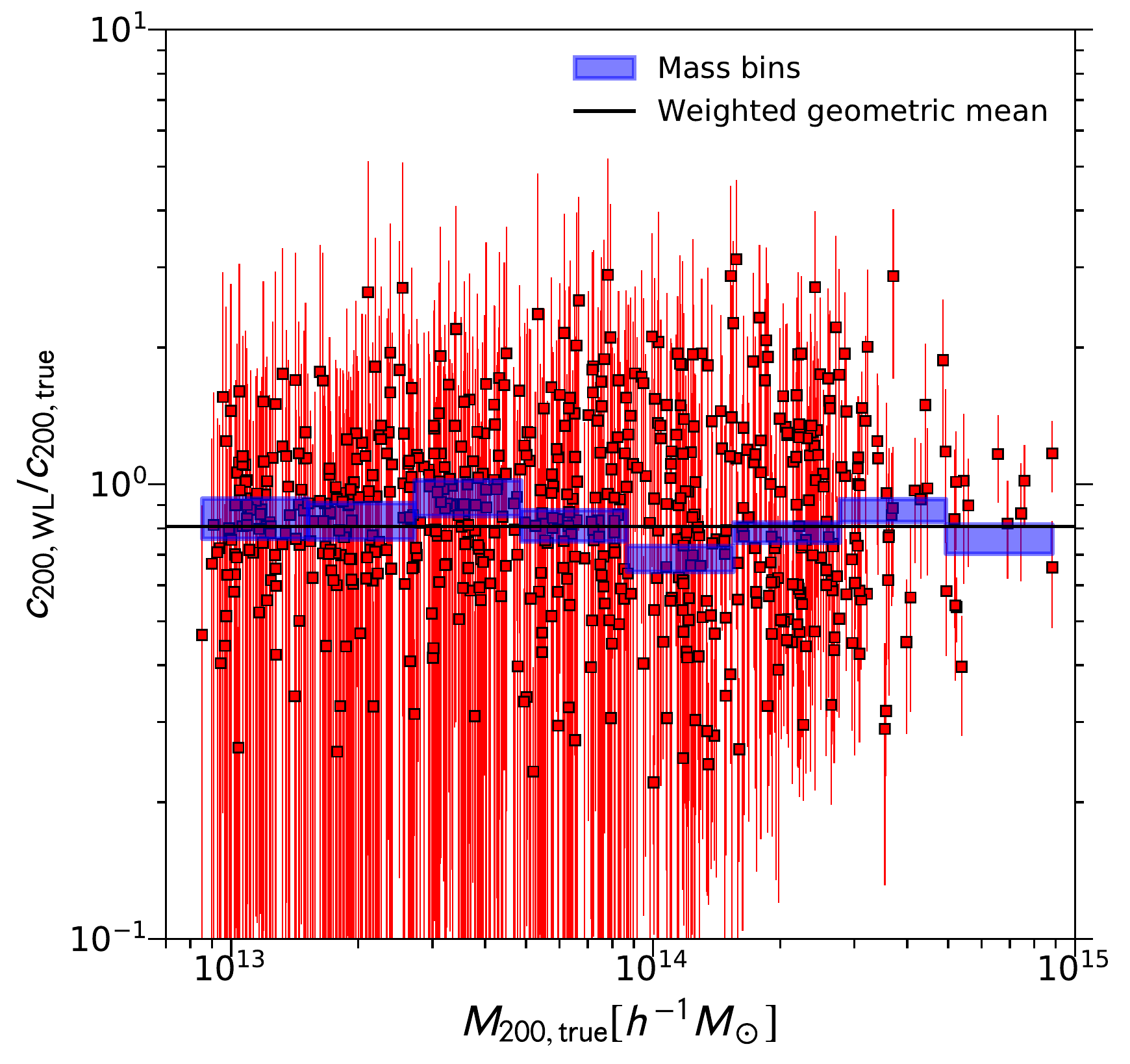}
 \end{center}
 \caption{
 Same as in Figure \ref{fig:BAHAMAS}, but fitting each individual
 cluster with the two-parameter halo model. 
 \label{fig:BAHAMAS_halomodel}
 }
\end{figure*}

Furthermore, we have tested our shear fitting procedures and pipeline 
using the standard halo model including the effects of surrounding
large-scale structure as a 2-halo term (Equation (\ref{eq:2ht})). We
describe the projected halo model with $M_{200}$ and $c_{200}$ as
fitting parameters and use the same priors as for the NFW model. 
As demonstrated in Figure \ref{fig:DSigma}, the 2-halo term
$\Delta\Sigma_\mathrm{2h}(R)$ is negligibly small in the comoving radial
range $R\in [0.3,3]\,\Mpch$.
When the 2-halo term is neglected, the halo model  reduces to the BMO
model that describes a smoothly truncated NFW profile (Section
\ref{subsec:stack}). 

The results are summarized in Figure \ref{fig:BAHAMAS_halomodel} and 
Table \ref{tab:BAHAMAS}.
Overall, the two-parameter halo modeling of each individual cluster does
not significantly improve the accuracy of weak-lensing measurements of
cluster mass and concentration, although it yields slightly
($<1\sigma$) improved levels of accuracy in the determination of
$M_{500}$ and $M_{200}$.

\section{Marginalized Posterior Constraints on Regression Parameters}
\label{appendix:PDF}


\begin{figure*}[!htb] 
  \begin{center}
   \includegraphics[scale=0.33, angle=0, clip]{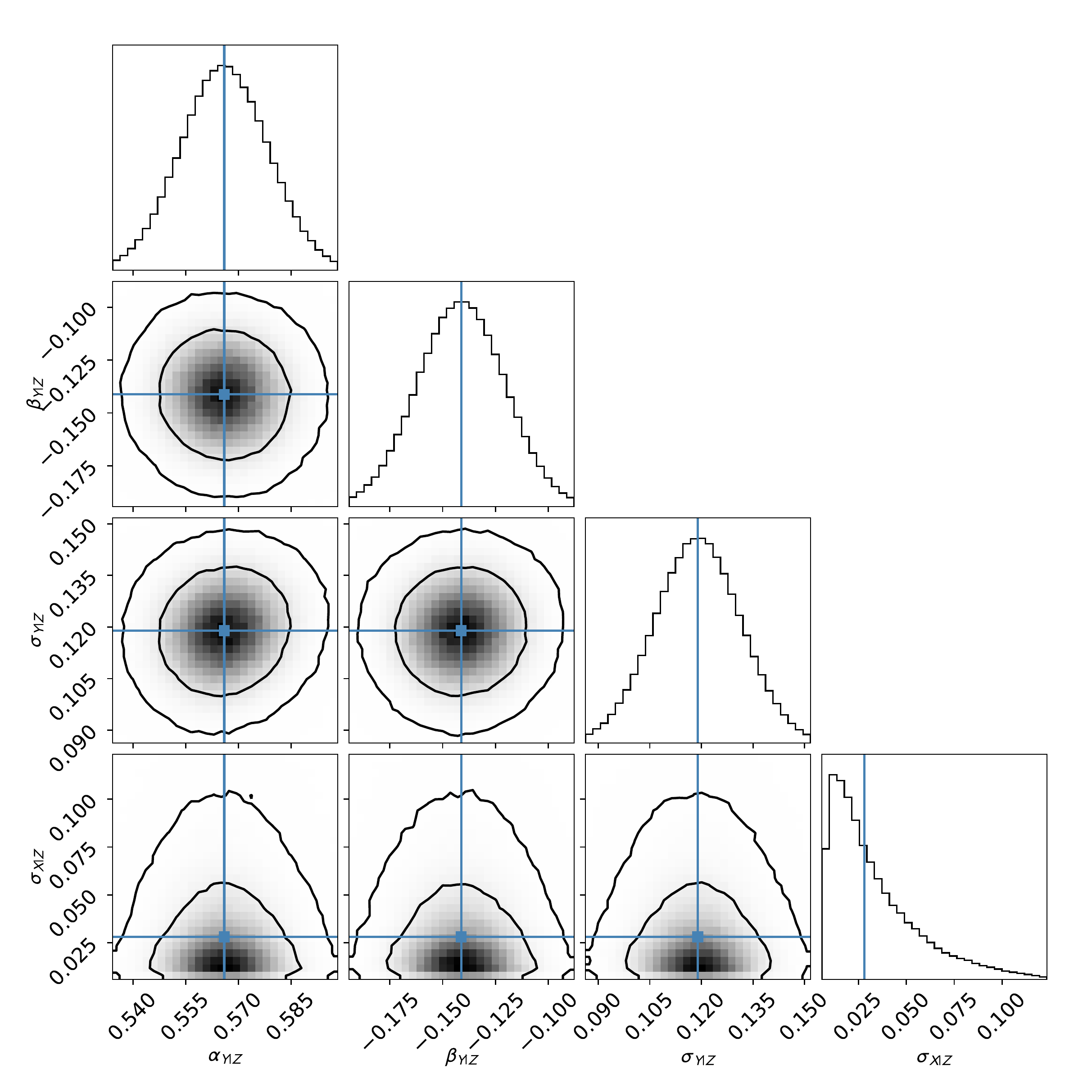}
   \includegraphics[scale=0.33, angle=0, clip]{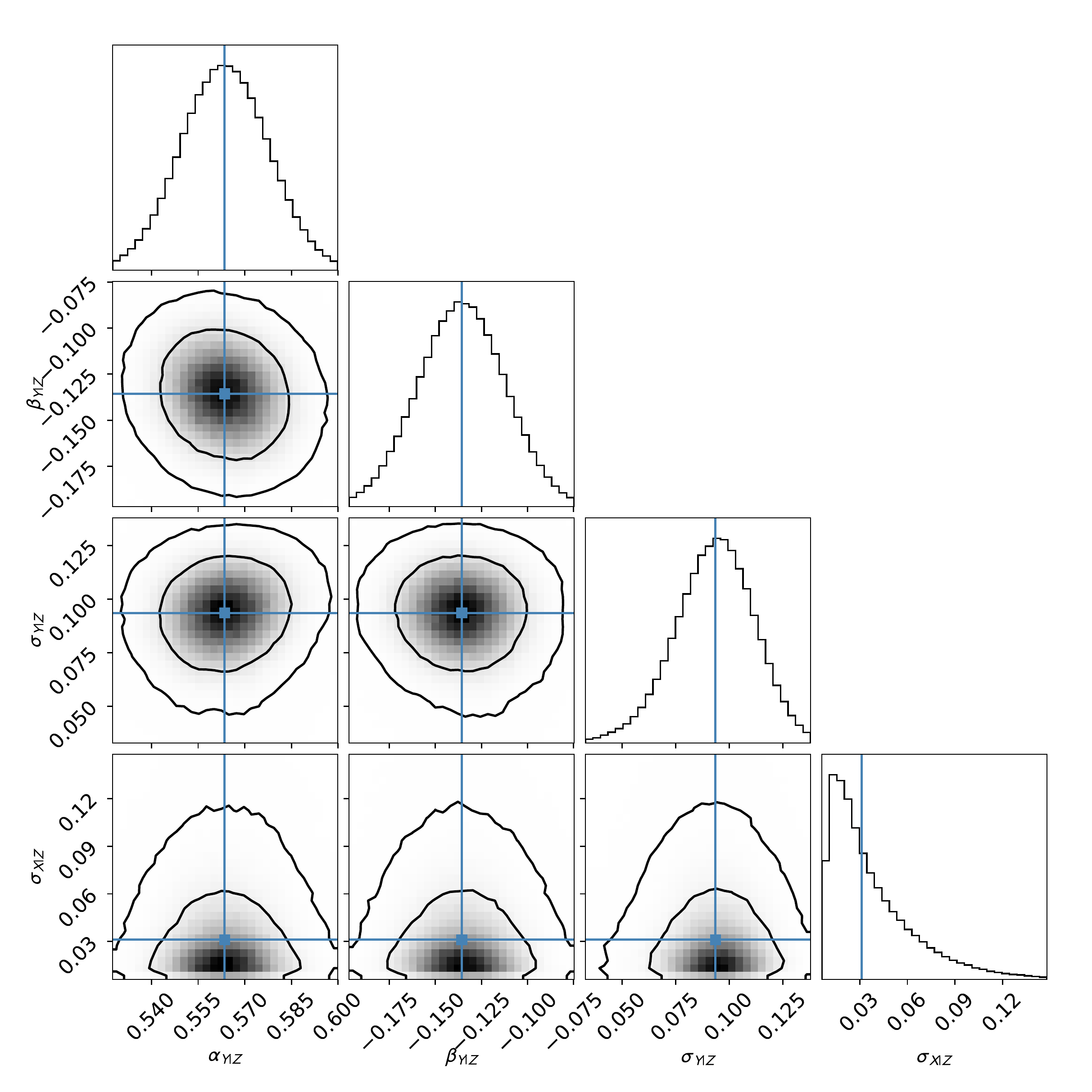}
  \end{center}
\caption{
 \label{fig:cMfit_BAHAMAS}
Marginalized one-dimensional (histograms) and two-dimensional
 ($68\percent$ and $95\percent$ confidence level contour plots)
 posterior distributions of the regression parameters 
 for the $c_{200}$--$M_{200}$ relation of the BAHAMAS sample at
 $z=0.25$, recovered from synthetic weak-lensing measurements (see
 Figure \ref{fig:BAHAMAS}).      
For each parameter, the blue solid line shows the biweight central
 location ($\CBI$) of the marginalized one-dimensional distribution.
The left panel is for the full BAHAMAS sample of 639 halos (Figure
 \ref{fig:Mtrue_BAHAMAS}), and the right panel is for a subsample of 459
 halos that lie within $1\sigma$ scatter from the mean
 $c_{200}$--$M_{200}$ relation.
 }
\end{figure*}


\begin{figure*}[!htb] 
  \begin{center}
   \includegraphics[scale=0.35, angle=0, clip]{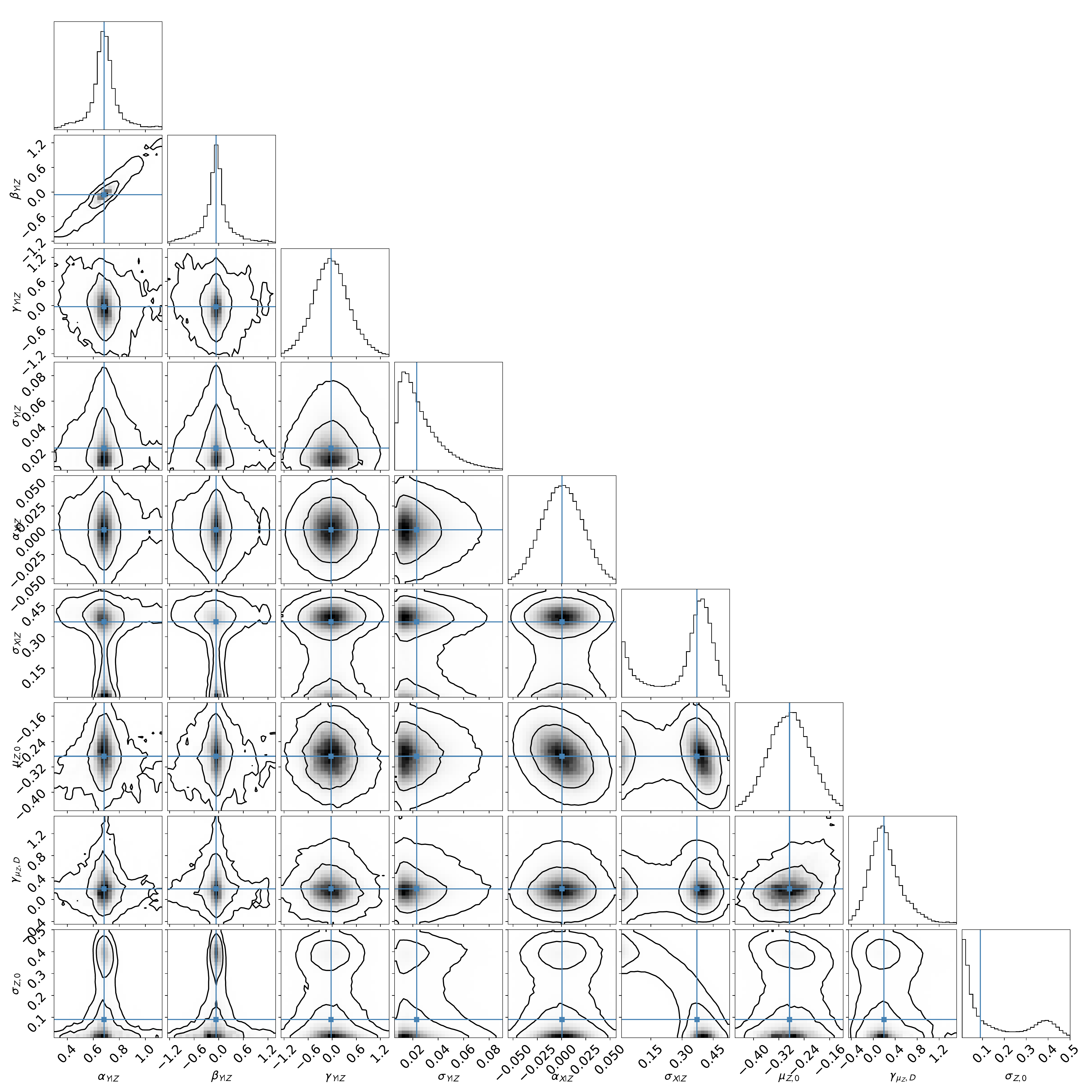} 
  \end{center}
\caption{
 \label{fig:cMfit}
 Constraints on the regression parameters for the $c_{200}$--$M_{200}$
 relation of the XXL sample, 
 showing marginalized one-dimensional (histograms) and two-dimensional
 ($68\percent$ and $95\percent$ confidence level contour  plots) posterior
 distributions. For each parameter, the blue solid line shows the
 biweight central location ($\CBI$) of the marginalized one-dimensional
 distribution.  
 }
\end{figure*}


\begin{figure*}[!htb] 
  \begin{center} 
   \includegraphics[scale=0.32, angle=0, clip]{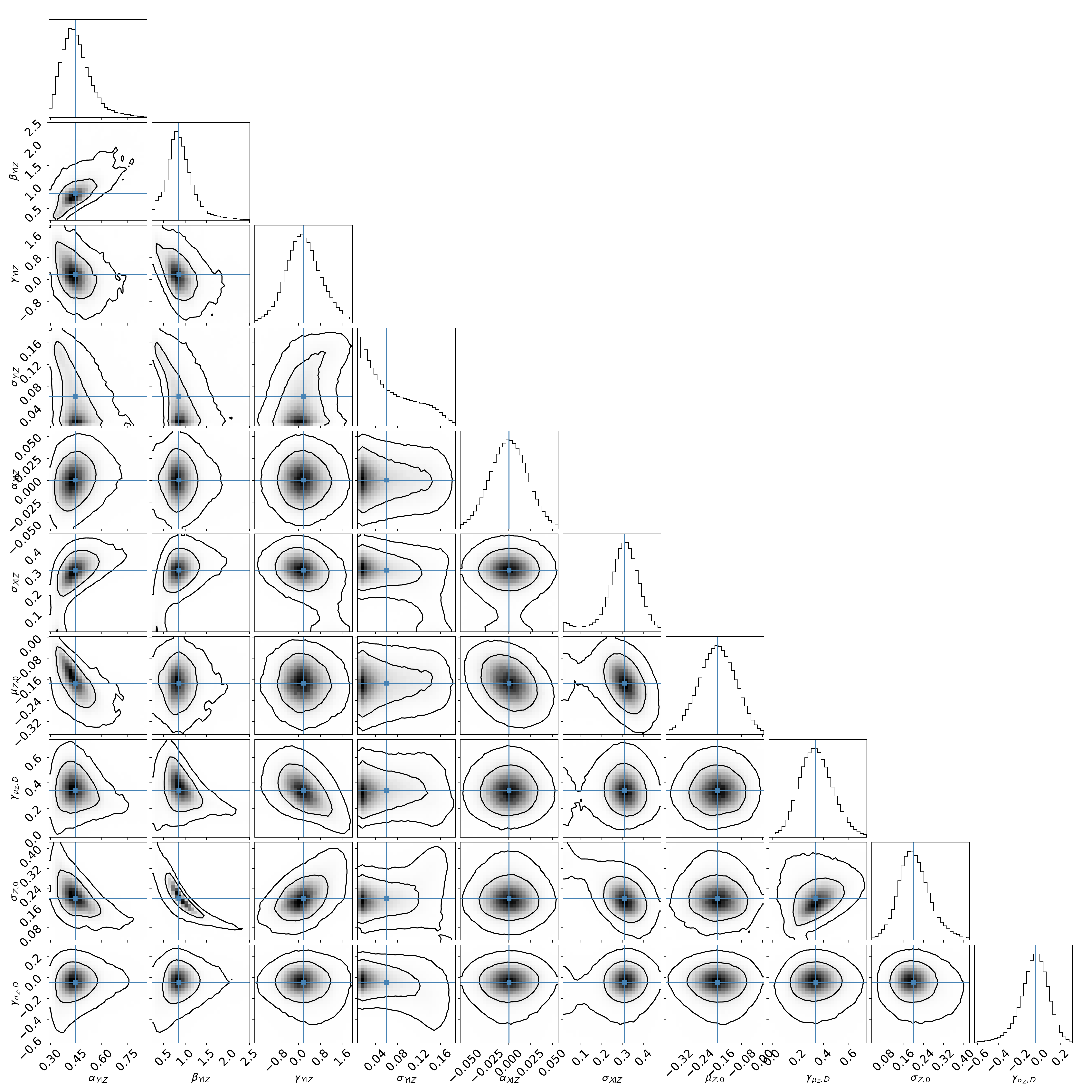} 
  \end{center}
\caption{
 \label{fig:TMfit}
 Constraints on the regression parameters for the $\Tx$--$M_{500}$
 relation of the XXL sample, 
 showing marginalized one-dimensional (histograms) and two-dimensional
 ($68\percent$ and $95\percent$ confidence level contour  plots) posterior
 distributions. For each parameter, the blue solid line shows the
 biweight central location ($\CBI$) of the marginalized one-dimensional
 distribution.   
 }
\end{figure*}

In Figure \ref{fig:cMfit_BAHAMAS}, we show marginalized one- and
two-dimensional posterior probability distributions of the regression
parameters $(\alpha_{Y|Z}, \beta_{Y|Z}, \sigma_{Y|Z}, \sigma_{X|Z})$
for the $c_{200}$--$M_{200}$ relation of the BAHAMAS sample
(Figure \ref{fig:cM_BAHAMAS}) recovered from the synthetic weak-lensing 
measurements shown in Figure \ref{fig:BAHAMAS} (for details, see
Appendix \ref{appendix:test_BAHAMAS}).    

In Figures \ref{fig:cMfit} and \ref{fig:TMfit},
we show marginalized posterior probability distributions of the
regression parameters for the $c_{200}$--$M_{200}$ relation (Section
\ref{subsec:cMR}) and the $\Tx$--$M_{500}$ relation (Section
\ref{subsec:TMR}), respectively, derived for the XXL cluster
population.

\end{appendix}

\end{document}